\documentclass[aps,prd,superscriptaddress,reprint]{revtex4-1}

\usepackage[usenames,dvips]{color}
\usepackage{graphicx}
\usepackage{amsmath}

\begin{document}

\setlength{\pdfpageheight}{\paperheight}
\setlength{\pdfpagewidth}{\paperwidth}

\title{Constraints on the Dark Matter Particle Mass from the Number of Milky Way Satellites}

\author{Emil Polisensky}
\affiliation{Naval Research Laboratory, Washington, D.C. 20375, USA}
\affiliation{Department of Astronomy, University of Maryland, College Park, Maryland 20745, USA}
\email{emilp@astro.umd.edu}

\author{Massimo Ricotti}
\affiliation{Department of Astronomy, University of Maryland, College Park, Maryland 20745, USA}                                                                     
\date{\today}

\begin{abstract}

  We have conducted $N$-body simulations of the growth of Milky
  Way-sized halos in cold and warm dark matter cosmologies. The number
  of dark matter satellites in our simulated Milky Ways decreases with
  decreasing mass of the dark matter particle. Assuming that the
  number of dark matter satellites exceeds or equals the number of
  observed satellites of the Milky Way we derive lower limits on
  the dark matter particle mass. We find with $95\%$ confidence $m_s > 13.3$~keV for a
  sterile neutrino produced by the Dodelson and Widrow mechanism, $m_s > 8.9$~keV for the Shi and Fuller mechanism, $m_s > 3.0$~keV for the Higgs decay mechanism, and $m_{WDM} > 2.3$~keV for a thermal dark matter
  particle. The recent discovery of many new dark matter dominated
  satellites of the Milky Way in the Sloan Digital Sky Survey allows
  us to set lower limits comparable to constraints from the
  complementary methods of Lyman-$\alpha$ forest modeling and X-ray observations of the
  unresolved cosmic X-ray background and of dark matter halos from dwarf galaxy to cluster scales. 
  Future surveys like LSST, DES,
  PanSTARRS, and SkyMapper have the potential to discover many more
  satellites and further improve constraints on the dark matter
  particle mass.

\end{abstract}

\pacs{}

\maketitle

\section{Introduction}\label{sec:1}

Cold dark matter (CDM) is extremely successful at describing the large
scale features of matter distribution in the Universe but has problems
on small scales. Below the Mpc scale CDM predicts numbers of satellite
galaxies for Milky Way-sized halos about an order of magnitude in
excess of the number observed. This is the `missing satellites'
problem \cite{kly1999, moo1999}. One proposed solution is that, due to
feedback mechanisms, some dark matter satellites do
not form stars and are nonluminous dark halos \cite{efs1992,tho1996,bul2001,ric2004,ric2005A}. 
Another solution is the power spectrum of density fluctuations may be truncated which may arise if the
dark matter is `warm' (particle mass $\sim 1$~keV) instead of
`cold' (particle mass $\sim 1$~GeV). Warm dark matter (WDM) particles decouple from 
the other particle species in the early Universe with relativistic 
velocities and only become nonrelativistic when about a Galactic mass ($\sim 10^{12} M_{\odot}$) 
is within the horizon. Streaming motions while the particles are still relativistic 
can erase density fluctuations on
sub-Galactic scales and reduce the number of satellites. WDM models have been studied by a number
of authors \cite{col2000, avi2001, bod2001, kne2002, kne2003, zen2003, mac2009} in relation to the missing satellites problem and other issues with CDM such as the apparent density cores in spiral and dwarf galaxies \cite{van2001,swa2003,wel2003,don2004,gen2005,sim2005,gen2007,sal2007,kuz2010}.

\begin{figure}[!th]
\includegraphics*[width=3in]{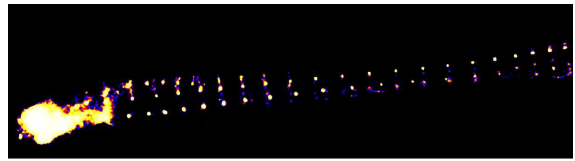}
\caption{Nonphysical halos formed along a filament and accreting
 onto a larger halo at $z=1$ in a WDM simulation
 ($m_{WDM}=1$~keV). These halos are numerical
 artifacts.\label{figI1}}
\end{figure}

$N$-body simulations of WDM
cosmologies are complicated by numerical artifacts produced by the discrete sampling
of the gravitational potential with a finite number of particles (see
\cite{mel2007} for a review).
Matter perturbations collapse and form filaments with
nonphysical halos separated by a distance equal to the mean particle
spacing (see Fig.~\ref{figI1}) \cite{wan2007,mel2007}. 
These halos are numerical artifacts. The ability of
these halos to survive disruption as they accrete from filaments onto 
Milky Way-sized halos has not been studied and they may contaminate
the satellite abundances and distributions in WDM simulations.

In the past few years, 16 new dwarf spheroidal galaxies have been discovered in the
Sloan Digital Sky Survey (SDSS) \cite{cas1998} (see Table 3 and references
therein). After correcting for completeness the
estimated number of Milky Way (MW) satellites is $>60$ (see Sec.\ \ref{sec:4}). 
These new dwarfs have low luminosities, low surface
brightnesses, and most appear to be dark matter dominated. Since the number of dark
matter halos must be greater than or equal to the number of observed
satellites, the new data from the SDSS may provide improved
limits on the mass of the dark matter particle independent of complementary techniques.

Motivated by the recent increase in the number of observed Milky Way
satellites, 
we have performed new simulations of the growth of Milky Way-like galaxies in
CDM and WDM cosmologies for a variety of WDM particle masses. Our
goal is to constrain the dark matter particle mass by comparing the
number of satellite halos in the simulated Milky Ways to the observed number
of luminous satellites for the actual Milky Way. 
Macci{\`o} and Fontanot \cite{mac2009} combined $N$-body simulations with semianalytic 
models of galaxy formation to compare the simulated and observed Milky Way satellite luminosity
functions for CDM and WDM cosmologies.
In this work, we do not make
any assumptions on how we populate dark matter halos with luminous
galaxies. We simply impose that the number of observed satellites is
less than or equal to the number of dark matter halos for a range of
Galactocentric radii. This guarantees a robust lower limit on the dark
matter particle mass.

\section{Simulations}

All our simulations were conducted with the $N$-body cosmological
simulation code GADGET-2 \cite{spr2005} assuming dark matter only. 
We adopted values for cosmological parameters from the
third year release of the WMAP mission \cite{spe2007}, 
($\Omega_m$, $\Omega_{\Lambda}$, $h$, $\sigma_8$, $n_s$) = (0.238, 0.762, 0.73, 
0.751, 0.951) to facilitate comparison with the Via~Lactea~II simulation \cite{die2008}. For each simulation set we produced a single
realization of the density field in the same periodic, comoving volume
but varied the power spectrum of fluctuations appropriate for CDM and
WDM cosmologies. Our initial conditions were generated on a cubic
lattice using the GRAFIC2 software package \cite{ber2001}. The power
spectra for CDM and WDM are given by
\begin{eqnarray}
P_{CDM}(k) &\propto& k^{n_s} T_{CDM}^2,\\
P_{WDM}(k) &=& P_{CDM} T_{WDM}^2,
\end{eqnarray}
respectively, with the normalization of $P_{CDM}$ determined by
$\sigma_8$. For our first two sets of simulations (\textit{set A} and \textit{B} described below) we used the transfer function for CDM adiabatic
fluctuations given by Bardeen et al.~(BBKS) \cite{bbks}:
\begin{eqnarray}
T_{CDM}(k)=\frac{\text{ln} (1 + 2.34q)}{2.34q} \text{\hspace{1.6in}}\nonumber \\*
 \left[1 + 3.89q + (16.1q)^2 + (5.46q)^3 + (6.71q)^4\right]^{-0.25},
\end{eqnarray}
where $q = k/(\Omega_mh^2)$.  A potential problem with the BBKS
transfer function is that it underestimates power on large scales. In
the Appendix we investigate the effect that our choice for the CDM
transfer function may have on the number of Milky Way satellites. We
run one of our simulations adopting the transfer functions from
Eisenstein and Hu \cite{eis1997} and we find that this does not affect
our results on the number of satellites. We also ran additional CDM
simulations (\textit{set C}) using the transfer function
calculated by the LINGER program in the GRAFIC2 package 
($\Omega_b=0.04$~was used for calculating the effects of baryons on the matter transfer function). 
LINGER integrates the linearized equations of general relativity, the Boltzmann equation, 
and the fluid equations governing the evolution of scalar metric perturbations, photons, neutrinos, baryons, and CDM.
The mass and circular velocity functions for satellites are consistent across both transfer functions.

Assuming the WDM to be a thermal particle, a particle that was in
thermal equilibrium with the other particle species at the time of its
decoupling, we used the transfer function valid for thermal particles
given by Bode, Ostriker, and Turok \cite{bod2001}:
\begin{equation}
T_{WDM}(k) = \left[1 + (\alpha k/h)^{\nu}\right]^{-\mu},
\end{equation}
where $\nu=2.4$, $\mu=4.167$ and
\begin{eqnarray}
\alpha = 0.0516\left(\frac{m_{WDM}}{\text{1~keV}}\right)^{-1.15} \text{\hspace{1.3in}} \nonumber \\*
         \left(\frac{\Omega_m}{0.238}\right)^{0.15}\left(\frac{h}{0.73}\right)^{1.3}\left(\frac{g_X}{1.5}\right)^{-0.29}.
\end{eqnarray}
The parameter $g_X$ is the number of degrees of freedom for the WDM
particle, conventionally set to the value for a light neutrino
species: $g_X=1.5$. The parameter $k$ is the spatial wavenumber in
$\text{Mpc}^{-1}$ and $m_{WDM}$ is the mass of the WDM particle in
keV.

A candidate for a thermal WDM particle is the gravitino, the
superpartner of the graviton in supersymmetry theories. The lightest
stable particle (LSP) in supersymmetry theories is a natural dark
matter candidate. If the scale where supersymmetry is spontaneously
broken is $\lesssim 10^6$~GeV, as predicted by theories where
supersymmetry breaking is mediated by gauge interactions, then the
gravitino is likely to be the lightest stable particle and can have a
mass reaching into the keV regime \cite{gor2008}.

In general the dark matter particle may not have been in thermal
equilibrium when it decoupled. This is the case for a sterile neutrino
(see \cite{kus2009}
and references therein), a theoretical particle added to standard electroweak
theory, the only matter it interacts with (except through gravity) is
left-handed neutrinos.
Sterile neutrinos have been proposed \cite{gni2010a,gni2010b,kar2009,sor2004,mel2009,mal2007,pas2005,akh2010} 
as an explanation for the anomalous excess of oscillations observed 
between muon and electron neutrinos and antineutrinos \cite{ath1995,ath1996,ath1998a,ath1998b,aa2007,aa2009,aa2010}.
There are several mechanisms by which sterile neutrinos can be produced. In the 
standard mechanism proposed by Dodelson and Widrow (DW) \cite{dw1994}, 
sterile
neutrinos are produced when oscillations convert some of the more familiar active neutrinos into 
the sterile variety. 
The amount produced depends on the sterile neutrino mass and the mixing angle
but we will not consider such details here and when considering sterile neutrinos we simply assume 
they compose the entirety of the dark matter. 
The transfer function for DW sterile neutrinos with mass $m_s$ is given by \cite{aba2006}
\begin{equation}
T_{s}(k) = \left[1 + (\alpha k/h)^{\nu}\right]^{-\mu},
\end{equation}
where $\nu=2.25$, $\mu=3.08$, and
\begin{equation}
\alpha = 0.1959\left(\frac{m_{s}}{\text{1~keV}}\right)^{-0.858}\left(\frac{\Omega_m}{0.238}\right)^{-0.136}\left(\frac{h}{0.73}\right)^{0.692}.
\end{equation}
Viel et al.~\cite{vie2005} give a scaling relationship between the
mass of a thermal particle and the mass of the DW sterile neutrino for
which the transfer functions are nearly identical:
\begin{equation}
m_s = 4.379 \text{ keV}\left(\frac{m_{WDM}}{\text{1~keV}}\right)^{4/3}\left(\frac{\Omega_m}{0.238}\right)^{-1/3}\left(\frac{h}{0.73}\right)^{-2/3}. \label{sneqn}
\end{equation}
Other sterile neutrino production mechanisms include that of Shi and Fuller (SF) \cite{sf1999} 
who showed the DW mechanism is 
altered in the presence of a universal lepton asymmetry where production 
can be enhanced by resonance effects.
Sterile neutrinos can also be produced from decays of 
gauge-singlet Higgs bosons at the electroweak scale \cite{kus2006}. 
The momentum distribution of the sterile neutrinos depends on the production mechanism. In the absence of 
transfer function calculations we use the expressions in 
\cite{kus2009} for the free streaming length and average momentum to derive approximate scaling factors for the SF and
Higgs produced sterile neutrinos: $m_{DW}/m_{SF}=1.5$, $m_{DW}/m_{Higgs}=4.5$. 

There are also ways other than WDM to reduce small scale
power. Broken-scale invariance inflation models \cite{kam2000} have a
cutoff length below which power is suppressed. Particle theories where
the LSP dark matter particle arises from the decay of the next
lightest supersymmetric particle (NLSP) can also suppress small scale
power if the NLSP is charged and coupled to the photon-baryon plasma
\cite{sig2004} or if the NLSP decay imparts a large velocity to the
LSP \cite{kap2005}. 
Further possibilities include composite dark matter models where stable charged heavy leptons
and quarks bind to helium nuclei by Coulomb attraction and can play the role of dark matter
with suppression of small scale density fluctuations \cite{khl2005,khl2006,bel2006A,bel2006B,khl2007,khl2008A,khl2008B}.
Our method can potentially be applied to constrain these
models as well. However, in the rest of this paper we will not discuss
further the consequences that our work has on these theories.

In our simulations we assume the dark matter is thermal and scale the
results to the standard sterile neutrino mass using Eq.~(\ref{sneqn}). Our initial
conditions include particle velocities due to the gravitational
potential using the Zeldovich approximation but we do \textit{not} add
random thermal velocities appropriate for WDM to the simulation
particles. Bode, Ostriker, and Turok \cite{bod2001} argue that for warm
particle masses greater than $1$~keV thermal motions are unimportant for
halos on scales of a kiloparsec and above. Regardless, we expect
thermal motions, if anything, would reduce the number of small mass
halos and by not including thermal motions the mass limits derived from
our simulations will be more conservative.

\begin{figure}[!t]
\includegraphics*[scale=0.33,angle=270]{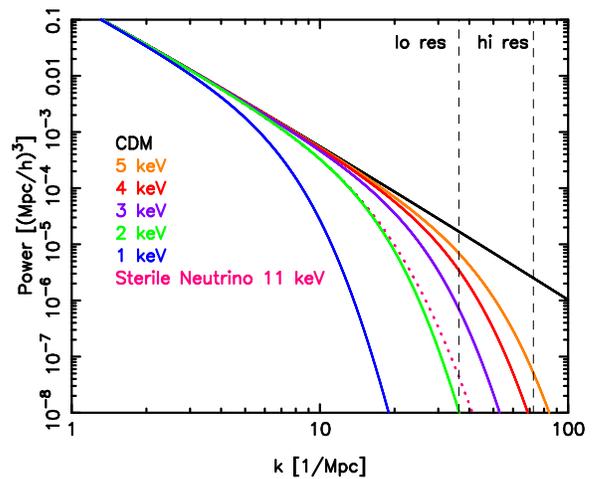}
\caption{Power spectra for our simulations. The dotted line is
  the power spectrum for an $11$~keV standard sterile neutrino from Abazajian
  \cite{aba2006}. The neutrino spectrum is approximately the same as a
  $2$~keV thermal particle, validating the scaling relation of Viel et
  al.~\cite{vie2005}. The vertical dashed lines indicate the lattice
  cell size in our high and low resolution refinement
  levels.\label{figPS1}}
\end{figure}

We ran simulations for CDM and WDM cosmologies with particle
masses of $m_{WDM}=$~1, 2, 3, 4, and $5$~keV ($m_s=$~4.4, 11.0, 18.9,
27.8, $37.4$~keV). Figure~\ref{figPS1} shows the power spectra for
these cosmologies along with the spectrum for an $11$~keV standard sterile
neutrino using Eq.~(\ref{sneqn}). We ran two separate sets of
simulations both consisting of a comoving cubic box $90$~Mpc on a
side. \textit{Set A} consisted of $204^3$ particles giving a `coarse'
particle mass of $3.0\times10^9$~M$_{\odot}$ and a force resolution of
$8.8$~kpc (all our force resolutions were fixed in comoving
coordinates). We ran the HOP halo finding software \cite{eis1998} at
$z=0$ and identified Milky Way-sized halos with masses
$1-2\times10^{12} M_{\odot}$. Halos were examined visually, one was
chosen that was at least several Mpc away from clusters and other
large structures so as to be relatively isolated. Its particles were
identified in the initial conditions and a cubic refinement level,
$6.2$~Mpc on a side, was placed on the region. For the refinement
region in our low resolution simulations we used $11,239,424$ ($224^3$) 
particles with mass and force resolutions of
$7.3\times10^5 M_{\odot}$ and $550$~pc, respectively. For CDM and
WDM particle masses of 1, 2, and $4$~keV we also conducted higher
resolution simulations with $89,915,392$ ($448^3$) particles in the
refinement region and mass and force resolutions of
$9.2\times10^4 M_{\odot}$ and $275$~pc, respectively. The simulated
Milky Way halo had a neighbor halo with mass $0.23 M_{MW}$ at a distance of
$700$~kpc in the low resolution simulations. The real Milky Way has a massive
neighbor in M31, the Andromeda galaxy ($M_{M31}\sim 1-3 M_{MW}$), at a
distance $\sim 700$~kpc. Being nonlinear and chaotic systems, small perturbations to the trajectories of dark matter halos can be amplified exponentially and in the higher resolution simulation this
satellite is merging with the Milky Way at $z=0$. Such a merger may disrupt
the equilibrium of the halo and make it nonrepresentative of the
actual Milky Way. The difference between the high and the low resolution
simulations is significant and complicates the comparison between the
resolutions; however, this merger is not a violation of the selection method
used for the \textit{set C} halos described below and
we find excellent agreement across all simulation sets.

The need to explore the scatter between the subhalo distributions of
different realizations of Milky Way-type halos, in addition to the
complications arising with the high and low resolution simulations of
{\it set A}, prompted us to conduct a second set of simulations. \textit{Set B}
consisted of $408^3$ particles giving a coarse particle mass of $3.8
\times 10^8 M_{\odot}$ and a force resolution of $4.4$~kpc. We ran
HOP and identified halos with masses $0.8-2.2 \times
10^{12} M_{\odot}$. For each halo we also found the nearest
neighboring halo with mass $> 0.8 \times 10^{12} M_{\odot}$. We
selected a halo whose nearest massive neighbor was at least $5$~Mpc away
and visually verified the halo was indeed isolated. A rectangular
refinement level $6.1 \times 7.0 \times 7.9$~Mpc was placed over this
halo's particles in the initial conditions. Low and high resolutions
were conducted with the same mass and force resolutions as {\it set
  A}. The low resolution simulations used $16,515,072$ ($\sim255^3$)
particles in the refinement level while high resolution used
$132,120,576$ ($\sim510^3$) particles in the refinement level.

A third set of CDM only, low resolution simulations was run to further explore 
the scatter between the subhalo distributions of
different realizations of Milky Way-type halos and to explore the possibility of a bias introduced 
by the use of the BBKS transfer function.
\textit{Set C} consisted of $408^3$ particles but the CDM transfer function
was generated from the LINGER software in GRAFIC2 \cite{ber2001}
after correcting a bug where the power spectrum for baryons was used for dark matter when calculating the transfer function. 
We used the AMIGA's Halo Finder (AHF) software \cite{kno2009} to find MW-sized halos with no equal sized
neighbor within two virial radii (defined below). Nine halos were selected for refinement at low resolution from a variety of environments, 
low density with few large halos to high density with many large halos. The rectangular refinement regions had lengths $7.5-15.8$~Mpc 
and $31,752,192-69,009,408$ ($316^3-410^3$) particles. 

Table~\ref{tabMW} summarizes the properties calculated by AHF for all of our simulated Milky Way halos
at $z=0$. We write $R_\Delta$ to mean the
radius enclosing an overdensity $\Delta$ times the critical value. The
mass and number of particles inside $R_\Delta$ are $M_\Delta$ and
$N_\Delta$, respectively; $v_{\Delta}$ is the circular velocity
$v_\Delta^2 \equiv GM_\Delta/R_\Delta$ at $R_\Delta$, and $v_{max}$ is
the maximum circular velocity of the halo. We use the value $\Delta=178\Omega_m^{0.4}=100$ \cite{eke1996} (which is also very close to the value using the definition from \cite{bry1998}) for the virial radius of
our MWs and consider subhalos within $R_{100}$ when comparing to other published work. 
The mass, radius, and velocity at $\Delta=50$ are also used in the literature and these values are also listed in Table~\ref{tabMW}.

\begin{figure}[!th]
\includegraphics*[scale=0.35,angle=270,clip=true]{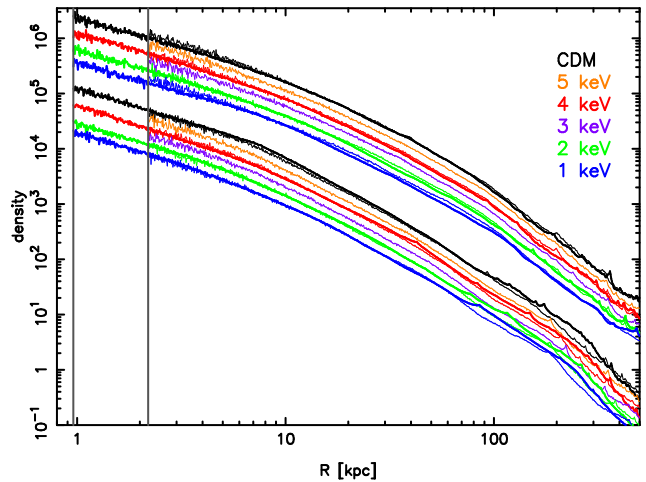}
\caption{Density profile of Milky Way halos in the \textit{set A} and \textit{set B} 
  CDM and WDM simulations. Thick
  lines are the high resolution simulations. \textit{Set B} simulations are at top, the {\it set A}
  and the WDM cosmologies in each set have been offset
  downward for clarity. The profiles are plotted starting from the convergence
  radius of Power et al.~\cite{pow2003} for both
  resolutions (vertical lines). \label{figden2}}
\end{figure}

\begin{figure*}[!pth]
 \begin{minipage}[b]{0.45\textwidth}
   \includegraphics*[width=2.5in]{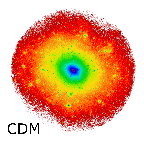}
 \end{minipage}
 \hspace{0.04\textwidth}
 \begin{minipage}[b]{0.45\textwidth}
   \includegraphics*[width=2.5in]{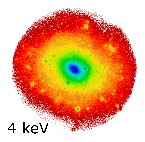}
 \end{minipage}\\
 \begin{minipage}[b]{0.45\textwidth}
   \includegraphics*[width=2.5in]{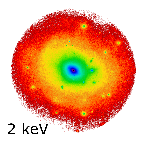}
 \end{minipage}
 \hspace{0.04\textwidth}
 \begin{minipage}[b]{0.45\textwidth}
   \includegraphics*[width=2.5in]{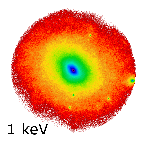}
 \end{minipage}
 \caption{Portraits of the simulated Milky Way halo at $z=0$ in the
   {\it set B} high resolution simulations. Structure within $500$~kpc 
   of the MW centers is shown.\label{figPORT}}
\end{figure*}

\begin{table*}[!th]
\caption{Properties of simulated Milky Way halos.\label{tabMW}}
\begin{center}
\begin{tabular}{l c c c c c c r r}    \hline
 Simulation & $M_{100}$ & $R_{100}$ & $M_{50}$ & $R_{50}$ & $v_{50}$ & $v_{max}$ & $N_{100}$ & $N_{50}$ \\
 & [$10^{12}M_{\odot}$] & [kpc] & [$10^{12} M_{\odot}$] & [kpc] & [km/s] & [km/s] & & \\
\hline
\hline
\multicolumn{9}{c}{\textit{Set A}}\\
\hline
\hline
 CDM lo & 1.4867 & 288.25 & 1.6786 & 378.47 & 138.11 & 183.02 & $2,026,414$ & $2,287,923$\\
 5 keV lo & 1.4964 & 288.86 & 1.6825 & 378.75 & 138.22 & 183.38 & $2,039,597$ & $2,293,239$\\
 4 keV lo & 1.5060 & 289.45 & 1.6833 & 378.82 & 138.24 & 183.87 & $2,052,643$ & $2,294,398$\\
 3 keV lo & 1.5141 & 290.00 & 1.6850 & 378.95 & 138.29 & 182.98 & $2,063,747$ & $2,296,714$\\
 2 keV lo & 1.5100 & 289.74 & 1.6702 & 377.84 & 137.88 & 181.59 & $2,058,104$ & $2,276,518$\\
 1 keV lo & 1.4983 & 289.00 & 1.6615 & 377.18 & 137.64 & 180.04 & $2,042,264$ & $2,264,672$\\
\hline
 CDM hi & 1.8403 & 309.49 & 2.0331 & 403.43 & 147.22 & 191.94 & $20,067,182$ & $22,169,072$\\
 4 keV hi & 1.8261 & 308.70 & 2.0383 & 403.77 & 147.34 & 189.69 & $19,911,999$ & $22,225,367$\\
 2 keV hi & 1.8326 & 309.06 & 2.0266 & 402.99 & 147.06 & 183.82 & $19,982,705$ & $22,098,268$\\
 1 keV hi & 1.8373 & 309.33 & 2.0244 & 402.85 & 147.01 & 179.39 & $20,033,935$ & $22,073,940$\\
\hline
\hline
\multicolumn{9}{c}{\textit{Set B}}\\
\hline
\hline
 CDM lo & 1.9005 & 312.84 & 2.1325 & 409.89 & 149.58 & 195.87 & $2,590,475$ & $2,906,549$\\
 5 keV lo & 1.8862 & 312.04 & 2.1254 & 409.44 & 149.41 & 195.76 & $2,570,982$ & $2,896,920$\\
 4 keV lo & 1.8863 & 312.04 & 2.1212 & 409.16 & 149.32 & 195.84 & $2,570,992$ & $2,891,165$\\
 3 keV lo & 1.8800 & 311.70 & 2.1185 & 409.00 & 149.25 & 195.75 & $2,562,445$ & $2,887,566$\\
 2 keV lo & 1.8479 & 309.92 & 2.0936 & 407.38 & 148.67 & 195.24 & $2,518,690$ & $2,853,610$\\
 1 keV lo & 1.8263 & 308.70 & 2.0752 & 406.19 & 148.23 & 192.33 & $2,489,258$ & $2,828,485$\\
\hline
 CDM hi & 1.7533 & 304.53 & 1.9948 & 400.88 & 146.29 & 194.01 & $19,117,720$ & $21,751,717$\\
 4 keV hi & 1.7426 & 303.92 & 1.9781 & 399.75 & 145.88 & 188.52 & $19,001,776$ & $21,569,680$\\
 2 keV hi & 1.7288 & 303.11 & 1.9640 & 398.81 & 145.53 & 185.18 & $18,850,480$ & $21,415,983$\\
 1 keV hi & 1.6230 & 296.80 & 1.8655 & 392.01 & 143.06 & 179.59 & $17,697,389$ & $20,341,369$\\
\hline
\hline
\multicolumn{9}{c}{\textit{Set C}}\\
\hline
\hline
 CDM lo 1 & 2.4814 & 342.19 & 2.8071 & 449.22 & 163.93 & 214.42 & $3,351,495$ & $3,761,164$\\
 CDM lo 2 & 2.3512 & 336.10 & 2.8287 & 450.37 & 164.35 & 213.86 & $3,204,746$ & $3,855,526$\\
 CDM lo 3 & 1.9846 & 317.63 & 2.2133 & 415.01 & 151.45 & 203.23 & $2,705,093$ & $3,016,787$\\
 CDM lo 4 & 2.2587 & 331.63 & 2.6486 & 440.60 & 160.79 & 199.95 & $3,078,658$ & $3,610,116$\\
 CDM lo 5 & 1.7665 & 305.53 & 1.9226 & 345.21 & 154.77 & 193.06 & $2,382,645$ & $2,589,405$\\
 CDM lo 6 & 1.6004 & 295.64 & 1.8977 & 394.26 & 143.88 & 187.76 & $2,174,733$ & $2,567,693$\\
 CDM lo 7 & 1.8704 & 311.41 & 2.7754 & 447.52 & 163.31 & 187.56 & $2,549,352$ & $3,782,898$\\
 CDM lo 8 & 1.9858 & 317.70 & 2.3401 & 422.78 & 154.29 & 194.43 & $2,706,610$ & $3,189,609$\\
 CDM lo 9 & 1.6881 & 300.95 & 1.8936 & 393.97 & 143.77 & 201.23 & $2,300,887$ & $2,581,006$\\
\hline
\end{tabular}
\end{center}
\end{table*}
Figure~\ref{figden2} shows the density profiles of the \textit{A} and \textit{B} Milky
Way halos calculated by breaking the halos into spherical shells. Small
differences between the high and low resolution {\it set A} halos caused by
the merging neighbor are apparent but generally the profiles are very
similar across all simulations of each set. We do not see an inner
flattening of the halos in the WDM simulations because we did not add
thermal motions to our particles. Figure~\ref{figPORT} shows portraits
of the Milky Way in our {\it set B} high resolution simulations.

\subsection{Identification of Satellites}

We used the AHF halo finding software \cite{kno2009} to find the
gravitationally bound dark matter halos in our $N$-body simulations.
Unbound particles were iteratively removed and we selected
gravitationally bound halos with ten or more particles.

AHF calculates properties of the halos it finds such as the total
mass and the maximum circular velocity. For our purposes the maximum
circular velocity is a better characteristic of a subhalo than the
mass because quantifying the outer boundary of a subhalo embedded in
a larger halo is somewhat arbitrary and can introduce systematic errors. The maximum circular velocity however typically occurs at a radius well inside the
subhalo outskirts.

\section{Results}

\subsection{Satellite Distribution Functions}\label{sec:3}

\begin{figure*}[!th]
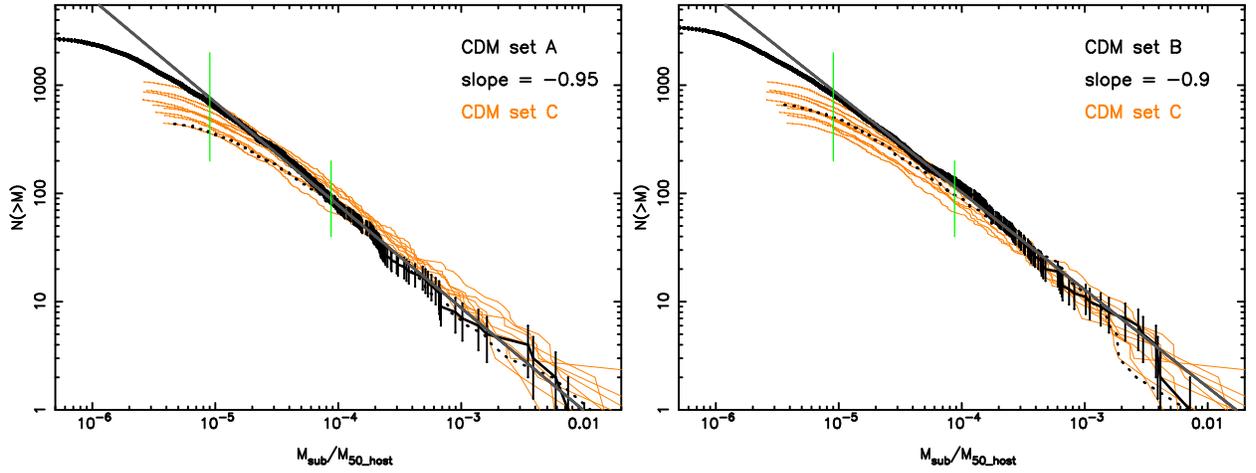

\includegraphics*[scale=0.34,angle=270]{fig5a.eps}
\includegraphics*[scale=0.34,angle=270]{fig5b.eps}
\caption{Cumulative mass functions for subhalos within $R_{50}$ for the CDM {\it set A} (left) and {\it set B} (right) simulations. Subhalo masses have been normalized by the $M_{50}$ mass of the host. Poisson error bars have been added to the high resolution simulations and fit with a straight line. Both high and low resolution (dotted lines) turn away from the straight fit below about 200 times the mass resolutions of the simulations (short vertical lines). Mass functions for the \textit{set C} halos have also been added (thin lines) and show the \textit{A} and \textit{B} abundances are within the halo to halo scatter.\label{figNMV}}
\end{figure*}

We first compare our CDM simulations to other CDM simulations in the literature.
Figure~\ref{figNMV} shows the cumulative mass functions, $N(>M_{sub})$, for subhalos within
$R_{50}$ for the \textit{set A} and \textit{B} MWs. Poisson statistic error bars have been added to the 
high resolution simulations and fit by $N \propto M^{-\beta}$. The values of $\beta$ ($0.9$ and $0.95$) agree
with other published work that find values of $0.7-1.0$ \cite{moo1999,ghi2000,hel2002,gao2004,del2004,van2005,die2007,gio2008,spr2008}. 
At both high and low resolution
the simulated mass functions turn away from the fit at masses below about 200 times
the mass resolution of the simulation; this was also seen in the Via~Lactea simulation \cite{die2007}. 
Also plotted are the mass functions for the \textit{set C} simulations. The subhalo
abundances of the \textit{set A} and \textit{B} halos are within the halo to halo scatter and are consistent with the \textit{set C} simulations.

\begin{figure}[!th]
\includegraphics*[scale=0.35,angle=270]{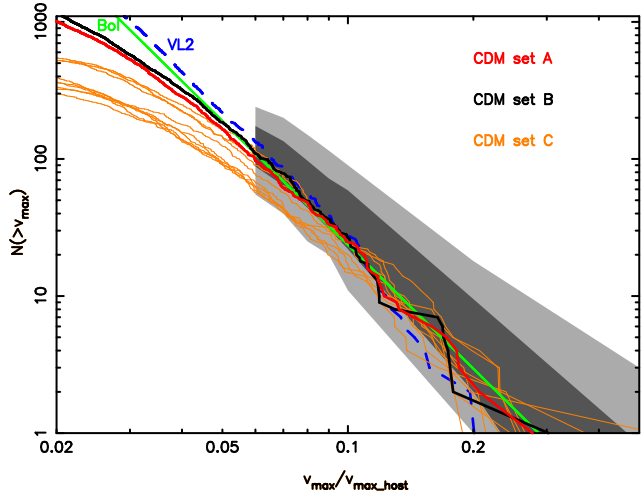}
\caption{Cumulative velocity functions for subhalos within $R_{100}$ in our low
  resolution \textit{set C} (thin lines) and high resolution {\it set A}
  and {\it set B} (thick lines) MW halos. Subhalo circular
  velocities have been normalized to the maximum circular velocity of the
  host halo. The dashed line is the subhalo velocity function of Via~Lactea~II, the straight solid line is the fitting formula from the Bolshoi simulation applied to our \textit{A} and \textit{B} halos. The shaded regions show the minimum and maximum and $\pm 1 \sigma$ from the mean of the $68$ MW-sized halos of Ishiyama et al.\ \cite{ish2009}\label{figISHI}}
\end{figure}

The cumulative maximum circular velocity functions for subhalos within $R_{100}$ are plotted in Figure~\ref{figISHI}. 
The maximum velocities of the subhalos have been normalized by the maximum circular velocity of their host MW. 
The shaded region shows the minimum and maximum (lighter) and $\pm1\sigma$ (darker) from the mean of the 
$68$ halos with masses $1.5-3 \times 10^{12} M_{\odot}$ in the simulation of Ishiyama et al.~\cite{ish2009}.
We use the fit to the density profile of the Via~Lactea~II halo \cite{die2008} to estimate its $R_{100}$ ($298$~kpc) and from the published subhalo catalog we calculate and plot the Via~Lactea~II velocity function as the dashed line. 
The solid straight line is the 
fitting formula from the Bolshoi simulation \cite{kly2010},
\begin{eqnarray}
N(>x) &=& 1.7 \times 10^{-3} v_{max,host}^{1/2} x^{-3},\\
x &\equiv& v_{max}/v_{max,host},
\end{eqnarray}
applied to our high resolution halos which provides an excellent fit (the difference between the fit for the \textit{set A} and \textit{B} $v_{max,host}$ is less than the thickness of the line).
Via~Lactea~II used the same cosmological parameters as our simulations and their subhalo abundance agrees well with our simulations.  
Our simulations are consistent with the Ishiyama et al.\ simulation but are systematically on the low end of their distribution. 
This is likely due to the different cosmology used in the Ishiyama et al.\ simulation (discussed below).

\begin{figure}[!th]
\includegraphics*[scale=0.35,angle=270]{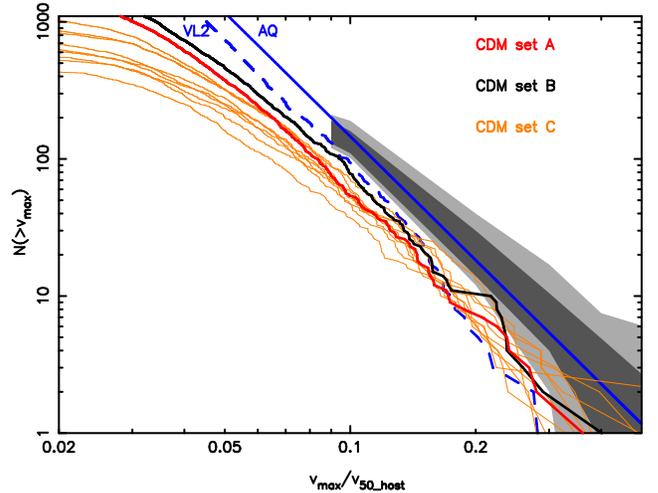}
\caption{Cumulative velocity functions for subhalos within $R_{50}$ in our low
  resolution \textit{set C} (thin lines) and
  high resolution {\it set A}
  and {\it set B} (thick lines)
  simulations. Subhalo circular
  velocities have been normalized to the circular velocity of the
  host halo at a radius enclosing an overdensity of $\Delta=50$. The dashed line is the velocity function for Via~Lactea~II and the straight
  solid line is the average abundance from the Aquarius simulations \cite{spr2008}.
  The shaded region is the minimum/maximum range and $\pm1\sigma$ about the mean
  for halos from the simulations of Ishiyama et al.\ \cite{ish2009}.
\label{figAQ}}
\end{figure}

Figure~\ref{figAQ} also plots the cumulative velocity functions but includes all subhalos within
$R_{50}$ and the subhalo velocities have been normalized by the circular velocity at $R_{50}$ of their host MW.
The Ishiyama et al.\ halos are again plotted as in Figure~\ref{figISHI} as well as Via~Lactea~II. 
The solid straight line is the result from the Aquarius simulations \cite{spr2008}. 
Again there is good agreement between our simulations and Via~Lactea~II but an offset between our simulations and those of Ishiyama et al.\ and Aquarius. 
To first order, the abundance of halos of any size depends on the power spectrum of density perturbations which depends on the normalization, $\sigma_8$, and the tilt of the power spectrum, $n_s$. Larger values of either parameter increases the power on small scales and leads to a larger number of satellies for a given mass and $v_{max}$ of the host. 
The values ($\sigma_8=0.9$, $n_s=1$) were used in the Aquarius simulations and ($0.8$, $1$) were used by Ishiyama et al. 
Both are significantly greater than our adopted values of ($0.74$, $0.951$), and this is the likely cause of the abundance offset.

We adopted a WMAP3 cosmology to facilitate comparison to the Via~Lactea~II simulation. 
The WMAP3 values of $n_s$, $\sigma_8$, and $\Omega_m$ are $1.0$, $2.9$, and $2.5$ standard deviations below the latest WMAP7 
values \cite{jar2010}. 
The Bolshoi simulation used parameters in agreement with WMAP7 and constraints from other cosmology projects. 
A comparison of the subhalo abundances of $4960$ Bolshoi halos with circular velocities and masses comparable to the Via~Lactea~II halo
indicated Bolshoi has more subhalos by about $10\%$. Although Via~Lactea~II is just one halo and may not be representative of the average for a WMAP3 cosmology, this agrees with expectations from the $10\%$ smaller value of $\sigma_8$ used by Via~Lactea~II. We used the same value of $\sigma_8$ as Via~Lactea~II but the Bolshoi fitting formula applied to our
high resolution simulations in Figure~\ref{figISHI} provides an excellent fit with no indication of an offset. This could be because, as we show in the Appendix, the BBKS power spectrum used in our high resolution simulations has about $10\%$ more power at sub-Galactic scales. Below we argue that an intrinsic scatter in subhalo abundance of $30\%$ ($1\sigma$) is reasonable to adopt and we conclude this can account for variations in the adopted cosmology without the need for a separate correction.

From our simulations and those of Ishiyama et al.\ in Figures~\ref{figISHI} and~\ref{figAQ} it is clear
that the subhalo abundances of halos of similar sizes have a scatter. 
For a given cosmology the scatter in abundance includes an intrinsic scatter and 
a statistical scatter from the number of subhalos. 
The Aquarius simulation suite included $6$ MW-sized halos simulated at very high resolution.
For subhalos within $R_{50}$, at high values of the abundances where the statistical scatter is small, the $1\sigma$ intrinsic scatter was determined to be $10\%$.
In Figure~\ref{figAQ} the scatter in the Ishiyama et al.\ abundances decreases for increasing $N$ and appears to be converging to the $10\%$ found in Aquarius. 
However the variation in the Ishiyama et al.\ abundances in Figure~\ref{figISHI} is clearly converging to a larger value.
As argued in \cite{ish2009}, the smaller scatter in Figure~\ref{figAQ} can be explained by the inclusion of subhalos at distances up to $R_{50}$ which are outside the virial radius and, hence, their evolution has not been affected by the structure of the host halo. Using $v_{50}$ to normalize the subhalo velocities can also reduce scatter since, unlike $v_{max}$, it is less dependent on the central concentration. 
We will be interested in the number of subhalos in the inner regions of the host MW which are expected to be sensitive to the host concentration so we will adopt the higher value for the intrinsic variation in the number of subhalo from Figure~\ref{figISHI} which we estimate to be about $30\%$ ($1\sigma$) after subtracting the Poissonian statistical scatter expected from the number of subhalos.

\begin{figure}[!th]
\includegraphics*[scale=0.35,angle=270]{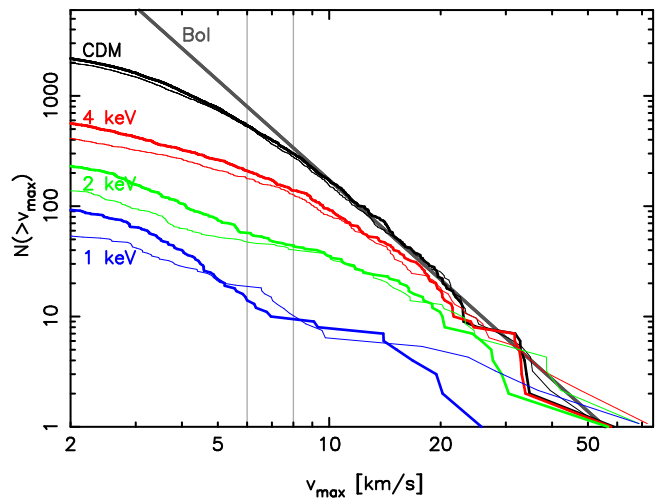}
\caption{Cumulative velocity functions for satellites in our
  high resolution {\it set A}
  and {\it set B} CDM and WDM
  simulations. The {\it set A} abundances (thin lines) have been increased by $7\%$ to normalize the CDM abundances to those of the {\it set B} and show that the relative suppression of halos in WDM cosmologies compared to CDM is similar across both simulations. The straight gray line is the Bolshoi fitting formula applied to the {\it set B} halo.
\label{figVCA}}
\end{figure}

In Figure~\ref{figVCA} the cumulative circular velocity functions for subhalos within $R_{100}$ for the high resolution
{\it set A} and {\it set B} CDM and WDM simulations are plotted. The {\it set A} abundances have been increased $7\%$ to normalize the CDM abundances to those of the {\it set B} simulation and illustrate that the relative suppression of subhalo abundances for each WDM simulation compared to CDM
is the same across both simulation sets. The straight line is the Bolshoi fitting function applied to the {\it set B} CDM halo.
The vertical lines in Figure~\ref{figVCA} show where $v_{max} = 6$ and $8$~km/s. 
Below $8$~km/s the high resolution CDM simulations begin to fall away 
from the Bolshoi line due to the resolution limits of the simulations. 
For $v_{max} > 8$~km/s our simulations are reasonably complete within 
$R_{100}$ of each Milky Way although numerical destruction of a small fraction of 
satellites in the inner Milky Way would not be apparent in Figure~\ref{figVCA}, 
especially for the CDM and $4$~keV cosmologies. Before comparing our 
simulations to observations we need to
determine to what distance our simulations are convergent.

\subsection{Convergence Study}

Satellites
orbiting in the halo of a larger galaxy are destroyed by tidal
stripping and heating through encounters with other
satellites. Satellites in simulations are also destroyed artificially
by numerical effects that become dominant for poorly resolved halos in
the inner halo region. There will therefore be a radius inside of
which our simulations will not converge to a realistic representation
of the actual Milky Way.

To determine the convergence of our simulations and have an
idea of the variance of the results, we performed simulations at
lower and higher resolution of two different
realizations of a Milky Way-sized galaxy. We performed convergence
studies following the argument elucidated below, in combination with
results of published high resolution simulations found in the
literature.  Using the work of Moore et al., De Lucia et al., and
Ishiyama et al.~\cite{moo1999, del2004, ish2009}, we will assume that
the shape of the cumulative satellite velocity function for host halos of
different masses is nearly constant and the total number of satellites
scales linearly with the host mass. If the simulations are convergent,
the cumulative circular velocity function for satellites, $N(R)$,
within a given Galactocentric radius, $R$, should be proportional to
the enclosed mass, $M(R)$, and a function of $R$ that represents the
fraction of satellites that survive destruction from physical effects:
\begin{equation}
N(R) \equiv f(R) M(R),
\end{equation}
where $f(R) \propto R^{\alpha}$.  The normalization of $f(R)$ can be
set using values of $N(R)$ and $M(R)$ at a distance $R_0$:
\begin{equation}
  N(R) \left( \frac{R_0}{R} \right)^{\alpha} \left( \frac{M_0}{M(R)} \right) = N_0 = const. \label{conveqn}
\end{equation}
The velocity functions normalized in this way will be constant with radius
where the simulations are convergent. Where numerical effects destroy
satellites the velocity functions will normalize to a lower value. We
expect that $\alpha$ is constant because there is no
characteristic scale for the destruction rate in dark matter only
simulations. Hence, $\alpha$ can be determined at large radii where
convergence is certain.
    
\begin{figure*}[!th]
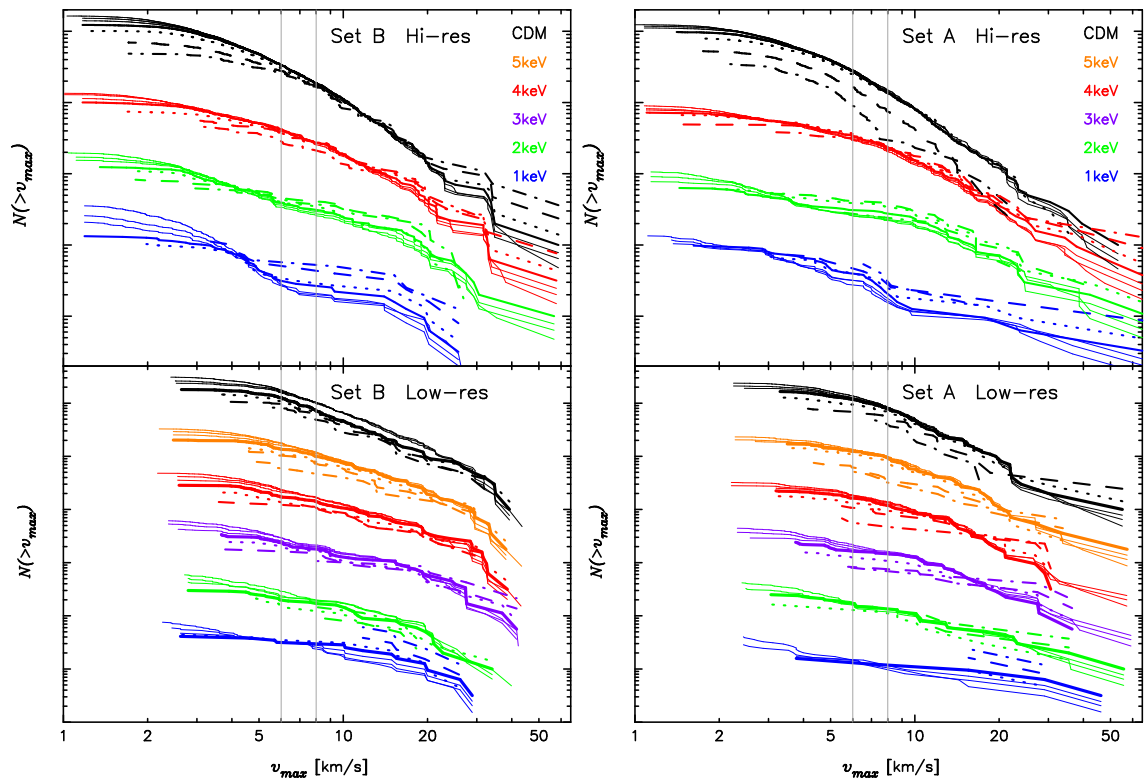

\includegraphics*[scale=0.55,angle=270]{fig9a.eps}
\includegraphics*[scale=0.55,angle=270]{fig9b.eps}
\caption{{\it (left)} Velocity functions for {\it set B} high
(\textit{top}) and low (\textit{bottom}) resolution simulations normalized with Eq.~(\ref{conveqn}). Solid lines 
are $R = 400$, $300$, $250$, and $200$~kpc (thick), dotted line is $R = 150$~kpc, dashed line is $R = 100$~kpc,
dot-dashed line is $R = 75$~kpc. The WDM cosmologies have been shifted
down vertically for clarity. The value $\alpha=0.55$ was set by the
high resolution simulation and provides good normalization for the low
resolution as well but the effects of incompleteness become apparent
at much larger radii ($150-200$~kpc compared to $75-100$~kpc for high
resolution). {\it (right)} Same as the left panel, but for the {\it set A}
high (\textit{top}) and low (\textit{bottom}) resolution
simulations. The value $\alpha=0.55$ also provides good normalization
for this set of halos. \label{figVCdistcompB}}
\end{figure*}

Figure~\ref{figVCdistcompB} shows the normalized velocity functions for
the simulations. The normalization constants $M_0, R_0$ have been
chosen at $200$~kpc and the value of $\alpha$ ($0.55$) was adjusted by hand
until a good fit was achieved for the {\it set B} velocity functions above
$200$~kpc in the high resolution CDM cosmology at circular velocities
$> 6$~km/s (vertical lines).  This $\alpha$ also provides a good
fit for the WDM cosmologies and for the {\it set A} simulations, although
the $1$ and $2$~keV velocity functions have a wider scatter due to the smaller
numbers of satellites in these simulations. The $m_{WDM}=4$~keV
simulation is convergent for $v_{max}>6$~km/s to distances $>100$~kpc.
At $75$~kpc the effects of numerical resolution are apparent. The
same value of $\alpha$ has been used in the normalization of the low
resolution sets and appears to provide a good fit for the velocity
functions $> 200-250$~kpc. The effects of numerical
resolution on the destruction of satellites are apparent at larger
distances in these simulations: $< 200$~kpc for CDM and $< 150$~kpc for WDM.

The $1$ and $2$~keV WDM velocity functions in Figures~\ref{figVCA} and~\ref{figVCdistcompB}
show a flattening when going from high to low velocities until about $6$~km/s, below which the
number of subhalos increases greatly. This is a generic feature of WDM simulations \cite{bod2001,bar2001} and is usually explained as top-down fragmentation of matter filaments. 
Given that WDM simulations are known to form nonphysical halos along filaments \cite{mel2007,wan2007}, 
it is likely the low velocity upturn in the velocity function of subhalos is actually caused by
these numerical artifacts accreting onto the MW.
Since these nonphysical halos form with separations typical of the mean particle distances in the initial conditions, the number of halos should increase with the mass resolution of the simulation. Our low resolution simulations do not show clear evidence of upturns in the velocity functions, we require simulations with resolution higher than our high resolution sets to confirm this effect.
Regardless, we consider only satellites with velocities greater than $6$~km/s in our high resolution simulations when deriving our constraints on the dark matter particle mass. 

\subsection{Comparison to Observations}\label{sec:4}

Before the Sloan Digital Sky Survey there were only 12
classically known satellite galaxies to the Milky Way. Sixteen new
satellites have been discovered in the SDSS, currently in Data Release
7. We list all known Milky Way satellites in Table~\ref{tabDwarfs}. We use the 
satellite distances given in Table~\ref{tabDwarfs} as their Galactocentric 
distances. When
comparing the observed satellites to the simulations, we must correct
the SDSS dwarfs for completeness. The primary incompleteness of the
SDSS is its sky coverage, 28.3\% ($11663$~deg$^2$). Second, being a
magnitude limited survey, the SDSS has a luminosity bias. 
The detection efficiency of dwarfs in the SDSS is a function of dwarf size, 
luminosity, distance, and Galactic
latitude as shown by Walsh et al.~\cite{wal2009}.
An approximate expression is given in Tollerud et al.~\cite{tol2008} (using the work of Koposov
et al.~\cite{kop2008}) for the distance which
galaxies of luminosity $>L$ are completely detected: $d \approx 66
\text{kpc}(L/1000~L_\odot)^{1/2}$. 
Galaxies with $L > 10^4$~L$_\odot$ should be approximately complete to $200$~kpc, 
with $L > 2300$~L$_\odot$ to $100$~kpc.
The distance range $100-200$~kpc is thus suited
for comparisons because our simulations are convergent and the
observations are nearly, but not quite, complete. 
For our analysis we will only use satellites with distances $< 200$~kpc.
\begin{table}[!th]
\caption{Summary of known Milky Way satellites.\label{tabDwarfs}}
\begin{center}
\begin{tabular}{l c c c r}\hline
 Name & $dist$ & $\sigma_{star}$ & $M_V$ & References \\ 
 & [kpc] & [km/s] & & \\ \hline
\hline
\multicolumn{5}{c}{Classical (pre-SDSS)}\\ 
\hline
 Sagittar & $24\pm2$ & $11.4\pm0.7$ & -13.4 & \cite{mat1998}\\
 LMC & $49\pm2$ & ... & -18.4 & \cite{mat1998}\\
 SMC & $58\pm2$ & ... & -17.0 & \cite{mat1998}\\
 Ursa Minor & $66\pm3$ & $9.3\pm1.8$ & -8.9 & \cite{mat1998}\\
 Draco & $79\pm4$ & $9.5\pm1.6$ & -8.8 & \cite{mat1998}\\
 Sculptor & $79\pm4$ & $6.6\pm0.7$ & -11.1 & \cite{mat1998}\\
 Sextans & $86\pm4$ & $6.6\pm0.7$ & -9.5 & \cite{mat1998}\\
 Carina & $94\pm5$ & $6.8\pm1.6$ & -9.3 & \cite{mat1998}\\
 Fornax & $138\pm8$  & $10.5\pm1.5$ & -13.2 & \cite{mat1998}\\
 Leo II & $205\pm12$ & $6.7\pm1.1$ & -9.6 & \cite{mat1998}\\
 Leo I & $270\pm30$ & $8.8\pm0.9$ & -11.9 & \cite{mat1998}\\
 Phoenix & $405\pm15$ & ... & -10.1 & \cite{mat1998}\\
\hline
\multicolumn{5}{c}{SDSS discovered}\\
\hline
 Segue I & $23\pm2$  & $4.3\pm1.2$  & -1.5 & \cite{geh2009}\\
 Ursa Major II & $30\pm5$  & $6.7\pm1.4$ & -3.8 & \cite{mar2007, sim2007}\\
 Segue II & $\sim35$ & $3.4\pm2.0$  & -2.5 & \cite{bel2009}\\
 Willman I & $38\pm7$  & $4.3^{+2.3}_{-1.3}$ & -2.5 & \cite{wil2005, mar2007}\\
 Coma Berenics & $44\pm4$  & $4.6\pm0.8$  & -3.7 & \cite{bel2007, sim2007}\\
 Bootes II & $60\pm10$  & ...  & -3.1 & \cite{wal2007} \\
 Bootes I & $62\pm3$ & $6.5^{+2.0}_{-1.4}$  & -5.8 & \cite{mar2007} \\
 Pisces I & $80\pm14$ & ... & ... & \cite{wat2009, kol2009}\\
 Ursa Major I & $106^{+9}_{-8}$  & $7.6\pm1.0$  & -5.6 & \cite{sim2007}\\
 Hercules & $140^{+13}_{-12}$  & $5.1\pm0.9$ & -6.0 & \cite{bel2007, sim2007}\\
 Canes Venatici II & $150^{+15}_{-14}$  & $4.6\pm1.0$  & -4.8 & \cite{bel2007, sim2007} \\
 Leo IV & $160^{+15}_{-14}$ & $3.3\pm1.7$ & -5.8 & \cite{bel2007, sim2007}\\
 Leo V & $175\pm9$  & $2.4\pm1.8$  & -5.2 & \cite{dej2009, bel2008}\\
 Pisces II & $\sim180$ & ... & -5.0 & \cite{bel2010}\\
 Canes Venatici I & $220^{+25}_{-16}$  & $7.6\pm0.4$  & -7.9 & \cite{zuc2006, sim2007} \\
 Leo T & $\sim420$  & $7.5\pm1.6$ & -7.1 & \cite{irw2007, sim2007}\\
\hline
\end{tabular}
\end{center}
\end{table}
For comparison with the simulations, we do a first order correction for
the sky coverage of the SDSS assuming an 
isotropic distribution
of satellites. The SDSS covers $0.283$ of the sky, so for every
satellite detected we assume there are actually $3.54$ satellites at that 
distance. 
Combined with the classic Milky Way satellites this forms our observed data set. 
We also implemented a conservative luminosity correction for the SDSS dwarfs 
using the formulas in Walsh et al.~\cite{wal2009}. This adds two satellites 
making a total of $61\pm13$ satellites within $200$~kpc where the error only 
includes Poisson statistics of the SDSS dwarfs: $3.54\sqrt{N_{SDSS}}$. The 
extra two satellites are at distances $150-200$~kpc and do not affect our 
conclusions. We note that we get the same result if we only include dwarfs 
detected in the Data Release 5 footprint and correct for the smaller sky 
coverage ($8000$~deg$^2$). We also note the formulas in Walsh et al.~\cite{wal2009} 
assume the size-luminosity distribution of known dwarfs is representative of 
all satellites. There may be a population of dwarfs with surface brightnesses 
below the detection limit of the SDSS~\cite{ric2005, ric2008, ric2010, bov2009}.

Willman~1 is an exceptional case in that it may not be a dark matter dominated
dwarf galaxy but a globular cluster undergoing tidal disruption. Its stellar velocity dispersion
implies a large mass to light ratio like other dwarf spheroidals and it has a
size and luminosity intermediate between MW dwarfs and
globular clusters~\cite{wil2005}, but unresolved binaries and tidal heating may contaminate the velocity dispersion and lead to an overestimated mass. Although it has a large metallicity spread unlike the
stellar population of a globular cluster~\cite{mar2007}, follow-up spectroscopy~\cite{sie2008} 
suggests there may be contamination by foreground stars and when these are excluded the metallicity 
spread can be consistent with a metal-poor globular cluster.
The detection of an X-ray emission line from decaying dark matter in Willman~1 has been claimed \cite{loe2010,kus2010} but is provisional \cite{boy2010}. If confirmed this would indicate a dark matter halo in Willman~1 and confirm its status as a MW satellite.
When deriving constraints
on the dark matter particle mass we will consider both including and excluding Willman~1
as a Milky Way satellite.

When comparing observations and simulations we
will apply cuts to the simulated subhalos and consider only those with velocites above $6$ and $8$~km/s.
As discussed in the previous section this is to avoid potential contamination from numerical effects in the WDM simulations.
That these velocity cuts are a reasonable  
estimate of the minimum
$v_{max}$ of the dark matter halos the observed galaxies are 
presumably embedded in can be shown as follows. 
Ricotti and Gnedin~\cite{ric2005} found in simulations that the
maximum circular velocities of satellites are at least
twice the velocity dispersion of the stellar component. 
Assuming the stellar velocity dispersions of the observed
dwarfs are $\sqrt3$ times the line-of-sight velocity dispersions 
($\sigma_{star}$ in Table~\ref{tabDwarfs}), then all dwarfs with measured 
velocity dispersions have maximum circular velocities greater than $8$~km/s. 
We assume dwarfs without measured velocity dispersions are similar to the 
other known dwarfs and conservatively conclude all dwarfs reside in dark matter halos with 
$v_{max}$ greater than $8$~km/s. An alternative approach
is the work of Wolf et al.~\cite{wol2009} relating the circular velocity
at half light radius to $\sigma_{star}$: $v_c(r_{1/2})=\sqrt{3} \sigma_{star}$. 
All the observed dwarfs except Leo V have circular velocities at half light radius about
$6$~km/s or greater. Since the maximum circular velocity must be greater than or equal to the 
half light circular velocity, it is also reasonable to consider that all observed dwarfs
reside in halos with $v_{max} > 6$~km/s. We stress that these cuts reflect the need to reduce the numerical effects
of the nonphysical halos in WDM simulations that dominate our high resolution simulations
at subhalo velocities below $6$~km/s rather than an assumption on the relationship between luminous satellites and dark matter halos.

\begin{figure*}[!th]
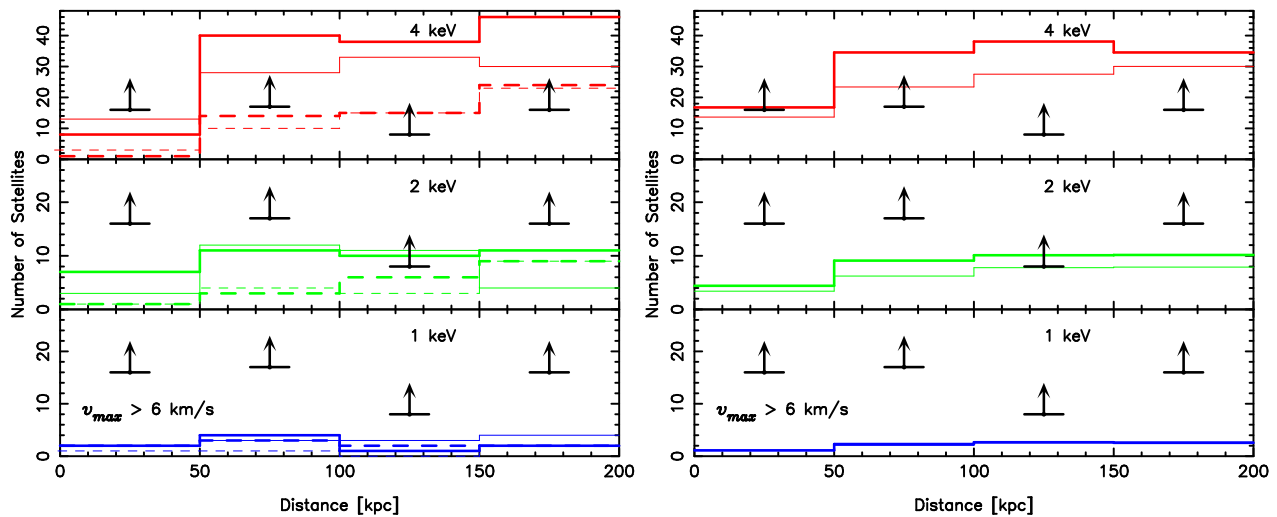

\includegraphics*[scale=0.34,angle=270]{fig10a.eps}
\includegraphics*[scale=0.34,angle=270]{fig10b.eps}
\caption{\textit{(left)} Number of satellites with distance from the Milky Way grouped into $50$~kpc bins. The
  simulation satellites have been cut by circular
  velocity $> 6$~km/s. Solid lines show the data from the high
  resolution and dashed lines show the low resolution simulations,
  thick lines are the {\it set B} simulations. The bars with
  arrows are the observed satellites after correcting for
  the sky coverage of the SDSS but not including Willman~1. Observations are incomplete at
  distances greater than about $50-100$~kpc (depending on the
  luminosity and surface brightness of the dwarf), while simulations
  have not converged for less than about $100$~kpc for high resolution and
  $150$~kpc for low resolution. \textit{(right)} Number of satellites with distance from the Milky Way like
  the plot at left but calculated from the convergence
  equation [Eq.~\ref{conveqn}]. The convergence correction affects mainly the $4$~keV $0-50$~kpc bin.\label{figSDSScomp3}}
\end{figure*}

In the left panel of Figure~\ref{figSDSScomp3} we plot histograms of the number of
satellites with distance for the observed and simulation data sets with a $6$~km/s 
maximum circular velocity cut.
The upward arrows on the observed data bars indicate these are only lower limits 
due to the surface brightness limits of the SDSS; it is possible there are more dwarfs yet to be discovered.
Willman~1 has not been included as a MW satellite in this figure. 
The $6$~km/s cut to the simulation data assures the high
resolution simulations are convergent to at least $r=100$~kpc. The low resolution
simulations are also plotted in these figures but they are convergent
only to $150$~kpc.
Focusing on the $100-200$~kpc bins it is clear the $1$~keV has far too
few satellites to match the observations. 
The $2$~keV simulations are consistent with the observations in the $100-150$~kpc bin but to be generally consistent with the observations would require the
simulations to be significantly incomplete below $100$~kpc or the sky correction to have overestimated the 
number of satellites in the inner Milky Way. 
The $4$~keV simulations can be consistent with the observations although 
they may require some of the dark matter halos to not host luminous galaxies. 
Strong conclusions cannot be drawn from this plot because it is not clear how variance in the satellite abundances for the simulated halos may affect the results.

The number of satellites in the simulations can be corrected for
completeness using the convergence equation, Eq.~(\ref{conveqn}). We
used the mass and number of satellites inside $R_{50}$ for the
normalization and calculated the number of satellites in $50$~kpc
bins for the high resolution simulations. The results are shown in the 
right panel of Figure~\ref{figSDSScomp3}.
The results are very similar across cosmologies, a nearly constant number 
of satellites per bin from $50-200$~kpc with about half as many in the 
$0-50$~kpc bin. The plots are also very similar to the simulation data plots 
except for the $0-50$~kpc bin in the $4$~keV cosmology where about 
ten satellites were destroyed by numerical effects in the \textit{set B} halo. 
The $0-50$~kpc bin is most important for constraining the dark matter 
particle mass because the observations are most complete in this bin.

\begin{figure}[!th]
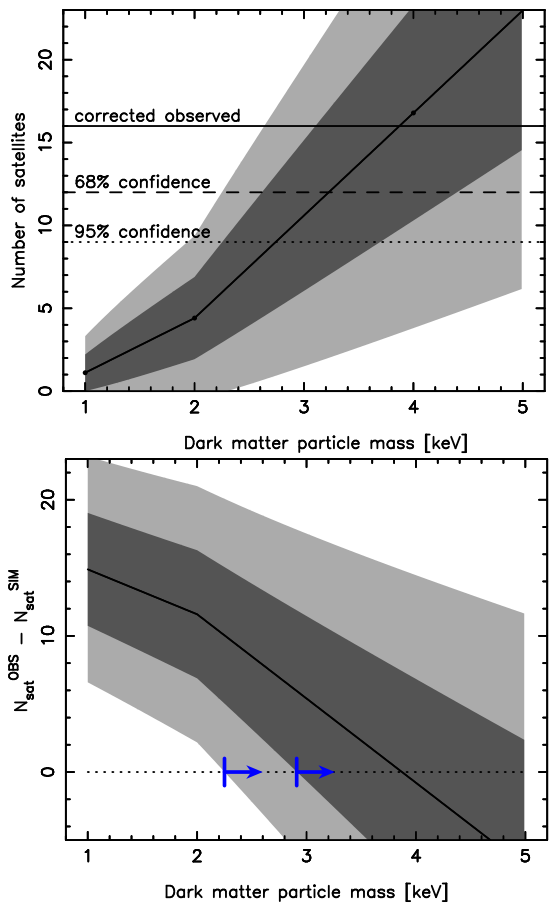

\includegraphics*[scale=0.32,angle=270]{fig11a.eps}
\includegraphics*[scale=0.32,angle=270]{fig11b.eps}
\caption{\textit{top} Number of satellites from 0-50 kpc calculated from the convergence equation in 
the \textit{set B} high resolution simulation for $v_{max} > 6$~km/s (sloped solid line) with $1\sigma$ and $2\sigma$ limits (shaded regions) compared to 
the observed number of Milky Way satellites, excluding Willman~1, 
with correction for sky coverage (solid horizontal line) and 
at 68\% and 95\% confidence (dashed and dotted horizontal lines). 
\textit{bottom} The observed number of satellites minus the number of satellites in simulation with $1\sigma$ and $2\sigma$ limits (shaded regions). The number of satellites in simulation must be greater than or equal to the number of observed satellites, therefore where this quantity equals zero sets a lower limit to the dark matter particle mass (arrows).
\label{figSDSScomp1bin}}
\end{figure}

In the top panel of Figure~\ref{figSDSScomp1bin} we plot the number of satellites in the $0-50$~kpc bin 
calculated from the convergence equation for the high resolution 
\textit{set B} Milky Way as a function of the particle mass with a $6$~km/s velocity cut to the subhalos. 
We set the variance in satellite abundances equal to that for a $30\%$ intrinsic rms scatter plus
a Poissonian variance in the number of satellites. The dark and light shaded regions in the plot show the $1\sigma$ and $2\sigma$ ranges, respectively.
The solid horizontal line shows the number of observed satellites after 
applying sky coverage corrections to the SDSS data (excluding Willman~1) while 
the dashed and dotted horizontal lines show the $68\%$ and $95\%$ confidence lower 
limits from Poisson statistics. 
In the bottom panel of Figure~\ref{figSDSScomp1bin} we plot the difference in the number of observed and simulated satellites with $1\sigma$ and $2\sigma$ limits (shaded regions) from combining the variances of the observed and simulated data sets.
The number of satellites in simulation must be at least equal to the number of observed satellites, therefore where this quantity equals zero defines a lower limit on the dark matter particle mass. The arrowed lines indicate the lower limits at $1\sigma$ and $2\sigma$ for this case of the \textit{set B} simulation with a $6$~km/s cut to the subhalos and excluding Willman~1 from the observed set. 

We repeated the same analysis using a $v_{max} > 8$~km/s cut to the simulation
data. We also considered the effects when Willman~1 was included in the observed data set
for both the $6$~km/s and $8$~km/s analysis. We present the results in Table \ref{tabResults}.
The \textit{set B} halo is slightly more abundant in subhalos but both the \textit{set A} and \textit{B} simulations give the same results to within about $10\%$. Rather than take the average of the two simulations we will simply adopt the more conservative of the two constraints.
\begin{table}[!th]
\caption{Dark matter particle mass constraints (in keV) from the high resolution \textit{set A} and \textit{B} MW halos. Constraints for simulated subhalo $v_{max}$ 
cuts of $8$ and $6$~km/s and including or excluding Willman~1 from the observed data set are given.\label{tabResults}}
\begin{center}
\begin{tabular}{l |c c |c c |c c |c c}\hline
\multicolumn{1}{c}{} & \multicolumn{4}{|c}{ $v_{max} > 8$ km/s} & \multicolumn{4}{|c}{$v_{max} > 6$ km/s}\\ 
\hline
\multicolumn{1}{l}{Will 1?} & \multicolumn{2}{|c}{Included} & \multicolumn{2}{c}{Excluded} & \multicolumn{2}{|c}{Included} & \multicolumn{2}{c}{Excluded} \\
\hline
MW & A & B & A & B & A & B & A & B\\
\hline 
 $2\sigma$ & $>3.6$ & $>3.3$ & $>2.9$ & $>2.7$ & $>3.0$ & $>2.6$ & $>2.5$ & $>2.3$\\
 $1\sigma$ & $>5$ & $>4.6$ & $>4.2$ & $>3.8$ & $>3.9$ & $>3.4$ & $>3.3$ & $>2.9$\\
\hline
\end{tabular}
\end{center}
\end{table}
In the most conservative case, where Willman~1 is not a dark matter dominated dwarf
galaxy and all observed satellites correspond to dark matter halos with
$v_{max} > 6$~km/s, we can say $m_{WDM}>2.3$~keV with $95\%$ confidence.
We adopt this as our formal limit for this work.

\section{Discussion}

We found that a model with $m_{wdm}=4$~keV produces the best fit to observations at $< 50$~kpc, i.e.\ this model has a number of
dark matter satellites equal to the number of observed luminous satellites. However, due to the large uncertainties in
the number of observed satellites due to partial sky coverage and on the number of simulated satellites due to Poisson and intrinsic scatter, that partially reflects
observational uncertainties on the mass and $v_{max}$ of the Milky Way, we find much weaker lower limits on $m_{wdm}$ than $4$~keV. In the future however, the lower limit on $m_{wdm}$ will improve as observations of MW satellites become more complete.
The scatter of the simulation can also be reduced using constrained simulations of the Local Group (also including the
effect of baryons) in combination with more accurate determination of the mass, rotation curve, and concentration of the
Milky Way.

Considering the various uncertainties in the number of observed and simulated satellites, we found a conservative lower limit of $m_{WDM} > 2.3$~keV ($2\sigma$)
on the dark matter particle mass.
We also found the $1$~keV WDM simulations have too few satellites to match
the Milky Way observations.
This agrees with the
semianalytic modeling and Milky Way satellite luminosity functions 
in WDM cosmologies work of
Macci{\`o} and Fontanot \cite{mac2009}; however, we only apply a cut to the simulated halos to avoid numerical effects and do not make assumptions on how the dark matter halos are populated by luminous galaxies.

Our result can also be compared to limits on the particle mass from the
Lyman-$\alpha$ forest in high redshift quasars. Lyman-$\alpha$
absorption by neutral hydrogen along the line of sight to distant
quasars over redshifts 2--6 probes the matter power spectrum in the
mildly nonlinear regime on scales 1--80 Mpc/$h$. Viel et al.~\cite{vie2005, vie2006, vie2008} 
have numerically modeled the
Lyman-$\alpha$ forest flux power spectra for varied cosmological
parameters and compared to observed quasar forests to obtain lower
limits on the dark matter particle mass. Their 2006 work \cite{vie2006} used low
resolution spectra for 3035 quasars ($2.2 < z < 4.2$) from the SDSS
\cite{mcd2006} and found a 2$\sigma$ lower limit of $2$~keV for a
thermal WDM particle. This limit agrees with our results that a
$2.3$~keV particle is the lower limit that can reproduce the observed
number of Milky Way satellites and agrees with the Lyman-$\alpha$ work of Seljak et al.~\cite{sel2006} who find a 2$\sigma$ limit $> 2.5$~keV for a thermal particle. The latest work of Viel et al.\
\cite{vie2008} uses high resolution spectra for 55 quasars ($2.0 < z <
6.4$) from the Keck HIRES spectrograph in addition to the SDSS
quasars. With the new data they report a lower limit of $4$~keV
(2$\sigma$). A caveat arises in Viel et al.~\cite{vie2009}, who show
the flux power spectrum from the SDSS data prefer larger values of the
intergalactic medium (IGM) temperature at mean density than expected
from photoionization. The flux power spectrum temperature is also
higher than that derived from an analysis of the flux probability
distribution function of 18 high resolution spectra from the Very Large Telescope and
also higher than constraints from the widths of thermally broadened
absorption lines \cite{ric2000, sch2000}. This could be explained by
an unaccounted for systematic error in the SDSS flux power spectrum
data which may also affect the derived dark matter particle mass
limits.

Using the scaling relation for sterile neutrinos we find a lower
limit $m_s > 13.3$~keV with $95\%$ confidence 
for a DW produced sterile neutrino particle. Scaling to the other production mechanisms
we get $m_s > 8.9$~keV for the SF mechanism and $m_s > 3.0$~keV for Higgs decay sterile neutrinos; however, we note this is not based on transfer function calculations for the SF and Higgs mechanisms but assumes a simple scaling for the average momentum for the different production mechanisms \cite{kus2009}.
The Lyman-$\alpha$ forest observations discussed above in the context of a thermal particle also set limits on the sterile neutrino mass. The 2006 work of Viel et al.~\cite{vie2006} sets $m_s > 11$~keV and is similar to the Seljak et al.~\cite{sel2006} limit $m_s > 14$~keV. The 2008 work of Viel et al.~\cite{vie2008} sets the highest limit of $m_s > 28$~keV but is subject to the caveats mentioned above.

Sterile neutrinos are expected to radiatively decay to a lighter
mass neutrino and a X-ray photon with energy $E_{\gamma}=m_s/2$. 
X-ray observations of the diffuse X-ray background \cite{boy2006A} and dark matter halos in clusters \cite{aba2006A,boy2006,rie2007,boy2008}, M31 \cite{wat2006}, dwarf spheroidal galaxies \cite{boy2007,rie2009,boy2009,loe2009}, and the halo of the Milky Way \cite{rie2006,aba2007,boy2007} have all been used to set constraints on the sterile neutrino mass. 
Observations of the diffuse X-ray background have set $m_s < 9.3$~keV~\cite{boy2006A},
while the Virgo and Coma clusters have been used to set $m_s < 6.3$~keV~\cite{aba2006A} which also agrees with limits from the
bullet cluster, 1E~0657-56, $m_s < 6.3$~keV~\cite{boy2008} and is close to results from the Milky Way halo $m_s < 5.7$~keV~\cite{aba2007}.
Tighter constraints have been determined from
M31 observations $m_s < 3.5$~keV~\cite{wat2006} and from the dwarf spheroidal
Ursa Minor $m_s < 2.5$~keV~\cite{loe2009}. 

These upper limits are well below the lower limits derived in this work and from Lyman-$\alpha$ 
observations and seem to rule out the DW and SF production mechanisms. However, all of these 
mass limits, including the constraints set in this work, are model dependent and make certain assumptions.
In general X-ray constraints depend on the sterile neutrino mass, the mixing angle with active 
neutrinos $\theta$, and the cosmic matter density of sterile neutrinos $\Omega_s$. There are also assumptions about 
the initial conditions, that there were no sterile neutrinos in the early Universe at temperatures $> 1$~GeV, there was no entropy dilution after creation, and no coupling to other particles. There are also uncertainties with the calculation of production rates because these occur at temperatures where the plasma is neither well described by hadronic nor quark models \cite{asa2006,boy2006A}. Depending on the assumptions made and the adopted production model the relationship between $m_s$, $\theta$, and $\Omega_s$ changes so that robust constraints cannot be placed on any one model parameter.

There has also been a report of a detection of a dark matter X-ray emission line from Willman~1 consistent with $m_s=5.0 \pm 0.2$~keV~\cite{loe2010,kus2010}. This detection is provisional \cite{boy2010} but if confirmed the limits derived in this work imply the sterile neutrinos are not produced by the DW mechanism or do not constitute the entirety of the dark matter.

\section{Summary}

We conducted $N$-body simulations of the formation of a MW-sized 
dark matter halo in CDM and WDM cosmologies. 
Such simulations are complicated by the formation of nonphysical small mass
halos due to the discreteness of the initial conditions 
but with sufficient resolution they are only important at small
scales and can be avoided with an appropriate circular velocity cut.

We studied the number of satellite halos as a function of
distance from the MW. 
The $4$~keV WDM simulation can adequately
reproduce the observed number of satellites at hundreds of kiloparsecs
while the $2$~keV simulation is slightly deficient and the $1$~keV severely deficient.
Our high resolution simulations followed the formation of two MW-sized halos. 
Numerical simulations of MW-sized halos show significant variance
in the number of satellites, an effect that can
be easily quantified using published studies and was incorporated in our results.
We calculated the number of satellites in the inner $50$~kpc, corrected for the effects of 
numerical destruction, and accounted for the variance by conservatively adopting 
a $30\%$ ($1\sigma$) intrinsic 
scatter in the number of satellites in addition to a scatter from Poisson statistics.
We also accounted for the uncertainty in the number of observed MW satellites 
due to the survey area of the SDSS and 
derived a 
very conservative lower limit on the dark matter particle mass of $> 2.3$~keV ($95\%$ C.L.). 
This agrees with the earlier Lyman-$\alpha$ forest modeling work of Viel et
al.~\cite{vie2006} that $m_{WDM} > 2$~keV but the two methods are
independent and almost certainly are subject to different systematic
errors if any exist. Their latest work \cite{vie2008} raised the limit
to $m_{WDM} > 4$~keV but problems with the derived IGM temperature and mean 
density may indicate the SDSS spectra they used
may suffer a systematic error \cite{vie2009}.

Our lower limit of $2.3$~keV for a thermal dark matter particle scales
to lower limits of $13.3$, $8.9$, $3.0$~keV ($95\%$ C.L.) for DW, SF, and Higgs decay produced sterile neutrinos. 
Sterile neutrinos, if they exist, are expected to decay into X-rays and
active neutrinos. Observations of the unresolved cosmic X-ray
background and X-ray observations of dark matter halos on scales from dwarf galaxies to clusters 
set upper limits below our lower limit and the limits of Lyman-$\alpha$ forest modeling. 
These limits are derived under many assumptions and, in general, the constraints apply 
to a parameter space of $m_s$, $\theta$, and $\Omega_s$. 

Our constraint is a
conservative lower limit since we only correct the number of SDSS
dwarfs for sky completeness. An analysis that takes into account the
surface brightness limits of the observational data
may allow tighter constraints; however, the analysis would be
somewhat model dependent. 
We have also not included the effects on subhalos of baryonic structures in the inner MW halo such as a disk. 
The presence of a disk could lead to greater subhalo destruction due to increased dynamical friction
and tidal heating. By increasing the subhalo destruction rate in the inner halo, disks would increase our lower bounds 
on the dark matter particle mass. The assumption of no disk is a conservative one and an analysis that includes a disk may allow tighter constraints.

We have demonstrated how $N$-body simulations of the MW and its
satellites can set limits on the dark matter particle mass comparable
to and independent of complementary methods such as modeling the Lyman-$\alpha$
forest. These limits are helped greatly by the discovery of many new
MW satellites in the SDSS. There may still be a population of low luminosity, low 
surface brightness dwarf galaxies undetectable by the SDSS~\cite{ric2005, ric2008, ric2010, bov2009}. 
Future surveys like LSST, DES, PanSTARRS,
and SkyMapper have the potential to discover many more MW satellites
and further improve constraints on the mass of the dark matter
particle.
Better constraints will result from the smaller uncertainty in the number of observed satellites achieved by improving the sky coverage and reducing luminosity corrections. In addition, the existence of a yet unknown population of even fainter satellites is not unlikely.

\begin{acknowledgments}
It is our pleasure to thank Alexander Kusenko, Kev Abazajian, Signe Riemer-S{\o}rensen, Oleg Ruchayskiy, Maxim Khlopov, H. J. de Vega, N. G. Sanchez, and Salucci Paolo for helpful discussions, comments, and suggestions. 
Emil Polisensky acknowledges support under the Edison Memorial Graduate Training Program
at the Naval Research Laboratory.
\end{acknowledgments}

\bibliographystyle{apsrev4-1}
\bibliography{MWsatsPRD.v7}

\providecommand{\noopsort}[1]{}\providecommand{\singleletter}[1]{#1}%
\begin{thebibliography}{100}%
\makeatletter
\providecommand \@ifxundefined [1]{%
 \ifx #1\undefined \expandafter \@firstoftwo
 \else \expandafter \@secondoftwo
\fi
}%
\providecommand \@ifnum [1]{%
 \ifnum #1\expandafter \@firstoftwo
 \else \expandafter \@secondoftwo
\fi
}%
\providecommand \enquote [1]{``#1''}%
\providecommand \bibnamefont  [1]{#1}%
\providecommand \bibfnamefont [1]{#1}%
\providecommand \citenamefont [1]{#1}%
\providecommand\href[0]{\@sanitize\@href}%
\providecommand\@href[1]{\endgroup\@@startlink{#1}\endgroup\@@href}%
\providecommand\@@href[1]{#1\@@endlink}%
\providecommand \@sanitize [0]{\begingroup\catcode`\&12\catcode`\#12\relax}%
\@ifxundefined \pdfoutput {\@firstoftwo}{%
 \@ifnum{\z@=\pdfoutput}{\@firstoftwo}{\@secondoftwo}%
}{%
 \providecommand\@@startlink[1]{\leavevmode\special{html:<a href="#1">}}%
 \providecommand\@@endlink[0]{\special{html:</a>}}%
}{%
 \providecommand\@@startlink[1]{%
  \leavevmode
  \pdfstartlink
   attr{/Border[0 0 1 ]/H/I/C[0 1 1]}%
   user{/Subtype/Link/A<</Type/Action/S/URI/URI(#1)>>}%
  \relax
 }%
 \providecommand\@@endlink[0]{\pdfendlink}%
}%
\providecommand \url  [0]{\begingroup\@sanitize \@url }%
\providecommand \@url [1]{\endgroup\@href {#1}{\urlprefix}}%
\providecommand \urlprefix [0]{URL }%
\providecommand \Eprint[0]{\href }%
\@ifxundefined \urlstyle {%
  \providecommand \doi [1]{doi:\discretionary{}{}{}#1}%
}{%
  \providecommand \doi [0]{doi:\discretionary{}{}{}\begingroup
  \urlstyle{rm}\Url }%
}%
\providecommand \doibase [0]{http://dx.doi.org/}%
\providecommand \Doi[1]{\href{\doibase#1}}%
\providecommand \bibAnnote [3]{%
  \BibitemShut{#1}%
  \begin{quotation}\noindent
    \textsc{Key:}\ #2\\\textsc{Annotation:}\ #3%
  \end{quotation}%
}%
\providecommand \bibAnnoteFile [2]{%
  \IfFileExists{#2}{\bibAnnote {#1} {#2} {\input{#2}}}{}%
}%
\providecommand \typeout [0]{\immediate \write \m@ne }%
\providecommand \selectlanguage [0]{\@gobble}%
\providecommand \bibinfo [0]{\@secondoftwo}%
\providecommand \bibfield [0]{\@secondoftwo}%
\providecommand \translation [1]{[#1]}%
\providecommand \BibitemOpen[0]{}%
\providecommand \bibitemStop [0]{}%
\providecommand \bibitemNoStop [0]{.\EOS\space}%
\providecommand \EOS [0]{\spacefactor3000\relax}%
\providecommand \BibitemShut [1]{\csname bibitem#1\endcsname}%
\bibitem{kly1999}%
  \BibitemOpen
  \bibfield{author}{%
  \bibinfo {author} {\bibfnamefont{A.}~\bibnamefont{{Klypin}}}, \bibinfo
  {author} {\bibfnamefont{A.~V.}\ \bibnamefont{{Kravtsov}}}, \bibinfo {author}
  {\bibfnamefont{O.}~\bibnamefont{{Valenzuela}}},\ and\ \bibinfo {author}
  {\bibfnamefont{F.}~\bibnamefont{{Prada}}},\ }%
  \bibfield{journal}{%
  \Doi{10.1086/307643}{\bibinfo {journal} {\apj}}\ }%
  \textbf{\bibinfo {volume} {522}},\ \bibinfo {pages} {82} (\bibinfo {month}
  {Sep.}\ \bibinfo {year} {1999}),\
  \Eprint{http://arxiv.org/abs/arXiv:astro-ph/9901240}{arXiv:astro-ph/9901240}%
  \bibAnnoteFile{NoStop}{kly1999}%
\bibitem{moo1999}%
  \BibitemOpen
  \bibfield{author}{%
  \bibinfo {author} {\bibfnamefont{B.}~\bibnamefont{{Moore}}}, \bibinfo
  {author} {\bibfnamefont{S.}~\bibnamefont{{Ghigna}}}, \bibinfo {author}
  {\bibfnamefont{F.}~\bibnamefont{{Governato}}}, \bibinfo {author}
  {\bibfnamefont{G.}~\bibnamefont{{Lake}}}, \bibinfo {author}
  {\bibfnamefont{T.}~\bibnamefont{{Quinn}}}, \bibinfo {author}
  {\bibfnamefont{J.}~\bibnamefont{{Stadel}}},\ and\ \bibinfo {author}
  {\bibfnamefont{P.}~\bibnamefont{{Tozzi}}},\ }%
  \bibfield{journal}{%
  \Doi{10.1086/312287}{\bibinfo {journal} {Astrophys. J. Lett.}}\ }%
  \textbf{\bibinfo {volume} {524}},\ \bibinfo {pages} {L19} (\bibinfo {month}
  {Oct.}\ \bibinfo {year} {1999}),\
  \Eprint{http://arxiv.org/abs/arXiv:astro-ph/9907411}{arXiv:astro-ph/9907411}%
  \bibAnnoteFile{NoStop}{moo1999}%
\bibitem{efs1992}%
  \BibitemOpen
  \bibfield{author}{%
  \bibinfo {author} {\bibfnamefont{G.}~\bibnamefont{{Efstathiou}}},\ }%
  \bibfield{journal}{%
  \bibinfo {journal} {Mon. Not. R. Astron. Soc.}\ }%
  \textbf{\bibinfo {volume} {256}},\ \bibinfo {pages} {43P} (\bibinfo {month}
  {May}\ \bibinfo {year} {1992})%
  \bibAnnoteFile{NoStop}{efs1992}%
\bibitem{tho1996}%
  \BibitemOpen
  \bibfield{author}{%
  \bibinfo {author} {\bibfnamefont{A.~A.}\ \bibnamefont{{Thoul}}}\ and\
  \bibinfo {author} {\bibfnamefont{D.~H.}\ \bibnamefont{{Weinberg}}},\ }%
  \bibfield{journal}{%
  \Doi{10.1086/177446}{\bibinfo {journal} {\apj}}\ }%
  \textbf{\bibinfo {volume} {465}},\ \bibinfo {pages} {608} (\bibinfo {month}
  {Jul.}\ \bibinfo {year} {1996}),\
  \Eprint{http://arxiv.org/abs/arXiv:astro-ph/9510154}{arXiv:astro-ph/9510154}%
  \bibAnnoteFile{NoStop}{tho1996}%
\bibitem{bul2001}%
  \BibitemOpen
  \bibfield{author}{%
  \bibinfo {author} {\bibfnamefont{J.~S.}\ \bibnamefont{{Bullock}}}, \bibinfo
  {author} {\bibfnamefont{A.~V.}\ \bibnamefont{{Kravtsov}}},\ and\ \bibinfo
  {author} {\bibfnamefont{D.~H.}\ \bibnamefont{{Weinberg}}},\ }%
  \bibfield{journal}{%
  \Doi{10.1086/318681}{\bibinfo {journal} {\apj}}\ }%
  \textbf{\bibinfo {volume} {548}},\ \bibinfo {pages} {33} (\bibinfo {month}
  {Feb.}\ \bibinfo {year} {2001}),\
  \Eprint{http://arxiv.org/abs/arXiv:astro-ph/0007295}{arXiv:astro-ph/0007295}%
  \bibAnnoteFile{NoStop}{bul2001}%
\bibitem{ric2004}%
  \BibitemOpen
  \bibfield{author}{%
  \bibinfo {author} {\bibfnamefont{M.}~\bibnamefont{{Ricotti}}}\ and\ \bibinfo
  {author} {\bibfnamefont{J.~P.}\ \bibnamefont{{Ostriker}}},\ }%
  \bibfield{journal}{%
  \Doi{10.1111/j.1365-2966.2004.07942.x}{\bibinfo {journal} {Mon. Not. R.
  Astron. Soc.}}\ }%
  \textbf{\bibinfo {volume} {352}},\ \bibinfo {pages} {547} (\bibinfo {month}
  {Aug.}\ \bibinfo {year} {2004}),\
  \Eprint{http://arxiv.org/abs/arXiv:astro-ph/0311003}{arXiv:astro-ph/0311003}%
  \bibAnnoteFile{NoStop}{ric2004}%
\bibitem{ric2005A}%
  \BibitemOpen
  \bibfield{author}{%
  \bibinfo {author} {\bibfnamefont{M.}~\bibnamefont{{Ricotti}}}, \bibinfo
  {author} {\bibfnamefont{J.~P.}\ \bibnamefont{{Ostriker}}},\ and\ \bibinfo
  {author} {\bibfnamefont{N.~Y.}\ \bibnamefont{{Gnedin}}},\ }%
  \bibfield{journal}{%
  \Doi{10.1111/j.1365-2966.2004.08623.x}{\bibinfo {journal} {Mon. Not. R.
  Astron. Soc.}}\ }%
  \textbf{\bibinfo {volume} {357}},\ \bibinfo {pages} {207} (\bibinfo {month}
  {Feb.}\ \bibinfo {year} {2005}),\
  \Eprint{http://arxiv.org/abs/arXiv:astro-ph/0404318}{arXiv:astro-ph/0404318}%
  \bibAnnoteFile{NoStop}{ric2005A}%
\bibitem{col2000}%
  \BibitemOpen
  \bibfield{author}{%
  \bibinfo {author} {\bibfnamefont{P.}~\bibnamefont{{Col{\'{\i}}n}}}, \bibinfo
  {author} {\bibfnamefont{V.}~\bibnamefont{{Avila-Reese}}},\ and\ \bibinfo
  {author} {\bibfnamefont{O.}~\bibnamefont{{Valenzuela}}},\ }%
  \bibfield{journal}{%
  \Doi{10.1086/317057}{\bibinfo {journal} {\apj}}\ }%
  \textbf{\bibinfo {volume} {542}},\ \bibinfo {pages} {622} (\bibinfo {month}
  {Oct.}\ \bibinfo {year} {2000}),\
  \Eprint{http://arxiv.org/abs/arXiv:astro-ph/0004115}{arXiv:astro-ph/0004115}%
  \bibAnnoteFile{NoStop}{col2000}%
\bibitem{avi2001}%
  \BibitemOpen
  \bibfield{author}{%
  \bibinfo {author} {\bibfnamefont{V.}~\bibnamefont{{Avila-Reese}}}, \bibinfo
  {author} {\bibfnamefont{P.}~\bibnamefont{{Col{\'{\i}}n}}}, \bibinfo {author}
  {\bibfnamefont{O.}~\bibnamefont{{Valenzuela}}}, \bibinfo {author}
  {\bibfnamefont{E.}~\bibnamefont{{D'Onghia}}},\ and\ \bibinfo {author}
  {\bibfnamefont{C.}~\bibnamefont{{Firmani}}},\ }%
  \bibfield{journal}{%
  \Doi{10.1086/322411}{\bibinfo {journal} {\apj}}\ }%
  \textbf{\bibinfo {volume} {559}},\ \bibinfo {pages} {516} (\bibinfo {month}
  {Oct.}\ \bibinfo {year} {2001}),\
  \Eprint{http://arxiv.org/abs/arXiv:astro-ph/0010525}{arXiv:astro-ph/0010525}%
  \bibAnnoteFile{NoStop}{avi2001}%
\bibitem{bod2001}%
  \BibitemOpen
  \bibfield{author}{%
  \bibinfo {author} {\bibfnamefont{P.}~\bibnamefont{{Bode}}}, \bibinfo {author}
  {\bibfnamefont{J.~P.}\ \bibnamefont{{Ostriker}}},\ and\ \bibinfo {author}
  {\bibfnamefont{N.}~\bibnamefont{{Turok}}},\ }%
  \bibfield{journal}{%
  \Doi{10.1086/321541}{\bibinfo {journal} {\apj}}\ }%
  \textbf{\bibinfo {volume} {556}},\ \bibinfo {pages} {93} (\bibinfo {month}
  {Jul.}\ \bibinfo {year} {2001}),\
  \Eprint{http://arxiv.org/abs/arXiv:astro-ph/0010389}{arXiv:astro-ph/0010389}%
  \bibAnnoteFile{NoStop}{bod2001}%
\bibitem{kne2002}%
  \BibitemOpen
  \bibfield{author}{%
  \bibinfo {author} {\bibfnamefont{A.}~\bibnamefont{{Knebe}}}, \bibinfo
  {author} {\bibfnamefont{J.~E.~G.}\ \bibnamefont{{Devriendt}}}, \bibinfo
  {author} {\bibfnamefont{A.}~\bibnamefont{{Mahmood}}},\ and\ \bibinfo {author}
  {\bibfnamefont{J.}~\bibnamefont{{Silk}}},\ }%
  \bibfield{journal}{%
  \Doi{10.1046/j.1365-8711.2002.05017.x}{\bibinfo {journal} {Mon. Not. R.
  Astron. Soc.}}\ }%
  \textbf{\bibinfo {volume} {329}},\ \bibinfo {pages} {813} (\bibinfo {month}
  {Feb.}\ \bibinfo {year} {2002}),\
  \Eprint{http://arxiv.org/abs/arXiv:astro-ph/0105316}{arXiv:astro-ph/0105316}%
  \bibAnnoteFile{NoStop}{kne2002}%
\bibitem{kne2003}%
  \BibitemOpen
  \bibfield{author}{%
  \bibinfo {author} {\bibfnamefont{A.}~\bibnamefont{{Knebe}}}, \bibinfo
  {author} {\bibfnamefont{J.~E.~G.}\ \bibnamefont{{Devriendt}}}, \bibinfo
  {author} {\bibfnamefont{B.~K.}\ \bibnamefont{{Gibson}}},\ and\ \bibinfo
  {author} {\bibfnamefont{J.}~\bibnamefont{{Silk}}},\ }%
  \bibfield{journal}{%
  \Doi{10.1046/j.1365-2966.2003.07044.x}{\bibinfo {journal} {Mon. Not. R.
  Astron. Soc.}}\ }%
  \textbf{\bibinfo {volume} {345}},\ \bibinfo {pages} {1285} (\bibinfo {month}
  {Nov.}\ \bibinfo {year} {2003}),\
  \Eprint{http://arxiv.org/abs/arXiv:astro-ph/0302443}{arXiv:astro-ph/0302443}%
  \bibAnnoteFile{NoStop}{kne2003}%
\bibitem{zen2003}%
  \BibitemOpen
  \bibfield{author}{%
  \bibinfo {author} {\bibfnamefont{A.~R.}\ \bibnamefont{{Zentner}}}\ and\
  \bibinfo {author} {\bibfnamefont{J.~S.}\ \bibnamefont{{Bullock}}},\ }%
  \bibfield{journal}{%
  \Doi{10.1086/378797}{\bibinfo {journal} {\apj}}\ }%
  \textbf{\bibinfo {volume} {598}},\ \bibinfo {pages} {49} (\bibinfo {month}
  {Nov.}\ \bibinfo {year} {2003}),\
  \Eprint{http://arxiv.org/abs/arXiv:astro-ph/0304292}{arXiv:astro-ph/0304292}%
  \bibAnnoteFile{NoStop}{zen2003}%
\bibitem{mac2009}%
  \BibitemOpen
  \bibfield{author}{%
  \bibinfo {author} {\bibfnamefont{A.~V.}\ \bibnamefont{{Maccio'}}}\ and\
  \bibinfo {author} {\bibfnamefont{F.}~\bibnamefont{{Fontanot}}},\ }%
  \bibfield{journal}{%
  \bibinfo {journal} {ArXiv e-prints}}%
   (\bibinfo {month} {Oct.}\ \bibinfo {year} {2009}),\
  \Eprint{http://arxiv.org/abs/0910.2460}{arXiv:0910.2460}%
  \bibAnnoteFile{NoStop}{mac2009}%
\bibitem{van2001}%
  \BibitemOpen
  \bibfield{author}{%
  \bibinfo {author} {\bibfnamefont{F.~C.}\ \bibnamefont{{van den Bosch}}}\ and\
  \bibinfo {author} {\bibfnamefont{R.~A.}\ \bibnamefont{{Swaters}}},\ }%
  \bibfield{journal}{%
  \Doi{10.1046/j.1365-8711.2001.04456.x}{\bibinfo {journal} {Mon. Not. R.
  Astron. Soc.}}\ }%
  \textbf{\bibinfo {volume} {325}},\ \bibinfo {pages} {1017} (\bibinfo {month}
  {Aug.}\ \bibinfo {year} {2001}),\
  \Eprint{http://arxiv.org/abs/arXiv:astro-ph/0006048}{arXiv:astro-ph/0006048}%
  \bibAnnoteFile{NoStop}{van2001}%
\bibitem{swa2003}%
  \BibitemOpen
  \bibfield{author}{%
  \bibinfo {author} {\bibfnamefont{R.~A.}\ \bibnamefont{{Swaters}}}, \bibinfo
  {author} {\bibfnamefont{B.~F.}\ \bibnamefont{{Madore}}}, \bibinfo {author}
  {\bibfnamefont{F.~C.}\ \bibnamefont{{van den Bosch}}},\ and\ \bibinfo
  {author} {\bibfnamefont{M.}~\bibnamefont{{Balcells}}},\ }%
  \bibfield{journal}{%
  \Doi{10.1086/345426}{\bibinfo {journal} {\apj}}\ }%
  \textbf{\bibinfo {volume} {583}},\ \bibinfo {pages} {732} (\bibinfo {month}
  {Feb.}\ \bibinfo {year} {2003}),\
  \Eprint{http://arxiv.org/abs/arXiv:astro-ph/0210152}{arXiv:astro-ph/0210152}%
  \bibAnnoteFile{NoStop}{swa2003}%
\bibitem{wel2003}%
  \BibitemOpen
  \bibfield{author}{%
  \bibinfo {author} {\bibfnamefont{D.~T.~F.}\ \bibnamefont{{Weldrake}}},
  \bibinfo {author} {\bibfnamefont{W.~J.~G.}\ \bibnamefont{{de Blok}}},\ and\
  \bibinfo {author} {\bibfnamefont{F.}~\bibnamefont{{Walter}}},\ }%
  \bibfield{journal}{%
  \Doi{10.1046/j.1365-8711.2003.06170.x}{\bibinfo {journal} {Mon. Not. R.
  Astron. Soc.}}\ }%
  \textbf{\bibinfo {volume} {340}},\ \bibinfo {pages} {12} (\bibinfo {month}
  {Mar.}\ \bibinfo {year} {2003}),\
  \Eprint{http://arxiv.org/abs/arXiv:astro-ph/0210568}{arXiv:astro-ph/0210568}%
  \bibAnnoteFile{NoStop}{wel2003}%
\bibitem{don2004}%
  \BibitemOpen
  \bibfield{author}{%
  \bibinfo {author} {\bibfnamefont{F.}~\bibnamefont{{Donato}}}, \bibinfo
  {author} {\bibfnamefont{G.}~\bibnamefont{{Gentile}}},\ and\ \bibinfo {author}
  {\bibfnamefont{P.}~\bibnamefont{{Salucci}}},\ }%
  \bibfield{journal}{%
  \Doi{10.1111/j.1365-2966.2004.08220.x}{\bibinfo {journal} {Mon. Not. R.
  Astron. Soc.}}\ }%
  \textbf{\bibinfo {volume} {353}},\ \bibinfo {pages} {L17} (\bibinfo {month}
  {Sep.}\ \bibinfo {year} {2004}),\
  \Eprint{http://arxiv.org/abs/arXiv:astro-ph/0403206}{arXiv:astro-ph/0403206}%
  \bibAnnoteFile{NoStop}{don2004}%
\bibitem{gen2005}%
  \BibitemOpen
  \bibfield{author}{%
  \bibinfo {author} {\bibfnamefont{G.}~\bibnamefont{{Gentile}}}, \bibinfo
  {author} {\bibfnamefont{A.}~\bibnamefont{{Burkert}}}, \bibinfo {author}
  {\bibfnamefont{P.}~\bibnamefont{{Salucci}}}, \bibinfo {author}
  {\bibfnamefont{U.}~\bibnamefont{{Klein}}},\ and\ \bibinfo {author}
  {\bibfnamefont{F.}~\bibnamefont{{Walter}}},\ }%
  \bibfield{journal}{%
  \Doi{10.1086/498939}{\bibinfo {journal} {Astrophys. J. Lett.}}\ }%
  \textbf{\bibinfo {volume} {634}},\ \bibinfo {pages} {L145} (\bibinfo {month}
  {Dec.}\ \bibinfo {year} {2005}),\
  \Eprint{http://arxiv.org/abs/arXiv:astro-ph/0506538}{arXiv:astro-ph/0506538}%
  \bibAnnoteFile{NoStop}{gen2005}%
\bibitem{sim2005}%
  \BibitemOpen
  \bibfield{author}{%
  \bibinfo {author} {\bibfnamefont{J.~D.}\ \bibnamefont{{Simon}}}, \bibinfo
  {author} {\bibfnamefont{A.~D.}\ \bibnamefont{{Bolatto}}}, \bibinfo {author}
  {\bibfnamefont{A.}~\bibnamefont{{Leroy}}}, \bibinfo {author}
  {\bibfnamefont{L.}~\bibnamefont{{Blitz}}},\ and\ \bibinfo {author}
  {\bibfnamefont{E.~L.}\ \bibnamefont{{Gates}}},\ }%
  \bibfield{journal}{%
  \Doi{10.1086/427684}{\bibinfo {journal} {\apj}}\ }%
  \textbf{\bibinfo {volume} {621}},\ \bibinfo {pages} {757} (\bibinfo {month}
  {Mar.}\ \bibinfo {year} {2005}),\
  \Eprint{http://arxiv.org/abs/arXiv:astro-ph/0412035}{arXiv:astro-ph/0412035}%
  \bibAnnoteFile{NoStop}{sim2005}%
\bibitem{gen2007}%
  \BibitemOpen
  \bibfield{author}{%
  \bibinfo {author} {\bibfnamefont{G.}~\bibnamefont{{Gentile}}}, \bibinfo
  {author} {\bibfnamefont{P.}~\bibnamefont{{Salucci}}}, \bibinfo {author}
  {\bibfnamefont{U.}~\bibnamefont{{Klein}}},\ and\ \bibinfo {author}
  {\bibfnamefont{G.~L.}\ \bibnamefont{{Granato}}},\ }%
  \bibfield{journal}{%
  \Doi{10.1111/j.1365-2966.2006.11283.x}{\bibinfo {journal} {Mon. Not. R.
  Astron. Soc.}}\ }%
  \textbf{\bibinfo {volume} {375}},\ \bibinfo {pages} {199} (\bibinfo {month}
  {Feb.}\ \bibinfo {year} {2007}),\
  \Eprint{http://arxiv.org/abs/arXiv:astro-ph/0611355}{arXiv:astro-ph/0611355}%
  \bibAnnoteFile{NoStop}{gen2007}%
\bibitem{sal2007}%
  \BibitemOpen
  \bibfield{author}{%
  \bibinfo {author} {\bibfnamefont{P.}~\bibnamefont{{Salucci}}}, \bibinfo
  {author} {\bibfnamefont{A.}~\bibnamefont{{Lapi}}}, \bibinfo {author}
  {\bibfnamefont{C.}~\bibnamefont{{Tonini}}}, \bibinfo {author}
  {\bibfnamefont{G.}~\bibnamefont{{Gentile}}}, \bibinfo {author}
  {\bibfnamefont{I.}~\bibnamefont{{Yegorova}}},\ and\ \bibinfo {author}
  {\bibfnamefont{U.}~\bibnamefont{{Klein}}},\ }%
  \bibfield{journal}{%
  \Doi{10.1111/j.1365-2966.2007.11696.x}{\bibinfo {journal} {Mon. Not. R.
  Astron. Soc.}}\ }%
  \textbf{\bibinfo {volume} {378}},\ \bibinfo {pages} {41} (\bibinfo {month}
  {Jun.}\ \bibinfo {year} {2007}),\
  \Eprint{http://arxiv.org/abs/arXiv:astro-ph/0703115}{arXiv:astro-ph/0703115}%
  \bibAnnoteFile{NoStop}{sal2007}%
\bibitem{kuz2010}%
  \BibitemOpen
  \bibfield{author}{%
  \bibinfo {author} {\bibfnamefont{R.}~\bibnamefont{{Kuzio de Naray}}},
  \bibinfo {author} {\bibfnamefont{G.~D.}\ \bibnamefont{{Martinez}}}, \bibinfo
  {author} {\bibfnamefont{J.~S.}\ \bibnamefont{{Bullock}}},\ and\ \bibinfo
  {author} {\bibfnamefont{M.}~\bibnamefont{{Kaplinghat}}},\ }%
  \bibfield{journal}{%
  \Doi{10.1088/2041-8205/710/2/L161}{\bibinfo {journal} {Astrophys. J. Lett.}}\
  }%
  \textbf{\bibinfo {volume} {710}},\ \bibinfo {pages} {L161} (\bibinfo {month}
  {Feb.}\ \bibinfo {year} {2010}),\
  \Eprint{http://arxiv.org/abs/0912.3518}{arXiv:0912.3518 [astro-ph.CO]}%
  \bibAnnoteFile{NoStop}{kuz2010}%
\bibitem{mel2007}%
  \BibitemOpen
  \bibfield{author}{%
  \bibinfo {author} {\bibfnamefont{A.~L.}\ \bibnamefont{{Melott}}},\ }%
  \bibfield{journal}{%
  \bibinfo {journal} {ArXiv e-prints}}%
   (\bibinfo {month} {Sep.}\ \bibinfo {year} {2007}),\
  \Eprint{http://arxiv.org/abs/0709.0745}{arXiv:0709.0745}%
  \bibAnnoteFile{NoStop}{mel2007}%
\bibitem{wan2007}%
  \BibitemOpen
  \bibfield{author}{%
  \bibinfo {author} {\bibfnamefont{J.}~\bibnamefont{{Wang}}}\ and\ \bibinfo
  {author} {\bibfnamefont{S.~D.~M.}\ \bibnamefont{{White}}},\ }%
  \bibfield{journal}{%
  \Doi{10.1111/j.1365-2966.2007.12053.x}{\bibinfo {journal} {Mon. Not. R.
  Astron. Soc.}}\ }%
  \textbf{\bibinfo {volume} {380}},\ \bibinfo {pages} {93} (\bibinfo {month}
  {Sep.}\ \bibinfo {year} {2007}),\
  \Eprint{http://arxiv.org/abs/arXiv:astro-ph/0702575}{arXiv:astro-ph/0702575}%
  \bibAnnoteFile{NoStop}{wan2007}%
\bibitem{cas1998}%
  \BibitemOpen
  \bibfield{author}{%
  \bibinfo {author} {\bibfnamefont{F.~J.}\ \bibnamefont{{Castander}}},\ }%
  \bibfield{journal}{%
  \Doi{10.1023/A:1002196414003}{\bibinfo {journal} {Astrophys. Space Sci.}}\ }%
  \textbf{\bibinfo {volume} {263}},\ \bibinfo {pages} {91} (\bibinfo {month}
  {Jun.}\ \bibinfo {year} {1998})%
  \bibAnnoteFile{NoStop}{cas1998}%
\bibitem{spr2005}%
  \BibitemOpen
  \bibfield{author}{%
  \bibinfo {author} {\bibfnamefont{V.}~\bibnamefont{{Springel}}},\ }%
  \bibfield{journal}{%
  \Doi{10.1111/j.1365-2966.2005.09655.x}{\bibinfo {journal} {Mon. Not. R.
  Astron. Soc.}}\ }%
  \textbf{\bibinfo {volume} {364}},\ \bibinfo {pages} {1105} (\bibinfo {month}
  {Dec.}\ \bibinfo {year} {2005}),\
  \Eprint{http://arxiv.org/abs/arXiv:astro-ph/0505010}{arXiv:astro-ph/0505010}%
  \bibAnnoteFile{NoStop}{spr2005}%
\bibitem{spe2007}%
  \BibitemOpen
  \bibfield{author}{%
  \bibinfo {author} {\bibfnamefont{D.~N.}\ \bibnamefont{{Spergel}}}, \bibinfo
  {author} {\bibfnamefont{R.}~\bibnamefont{{Bean}}}, \bibinfo {author}
  {\bibfnamefont{O.}~\bibnamefont{{Dor{\'e}}}}, \bibinfo {author}
  {\bibfnamefont{M.~R.}\ \bibnamefont{{Nolta}}}, \bibinfo {author}
  {\bibfnamefont{C.~L.}\ \bibnamefont{{Bennett}}}, \bibinfo {author}
  {\bibfnamefont{J.}~\bibnamefont{{Dunkley}}}, \bibinfo {author}
  {\bibfnamefont{G.}~\bibnamefont{{Hinshaw}}}, \bibinfo {author}
  {\bibfnamefont{N.}~\bibnamefont{{Jarosik}}}, \bibinfo {author}
  {\bibfnamefont{E.}~\bibnamefont{{Komatsu}}}, \bibinfo {author}
  {\bibfnamefont{L.}~\bibnamefont{{Page}}}, \bibinfo {author}
  {\bibfnamefont{H.~V.}\ \bibnamefont{{Peiris}}}, \bibinfo {author}
  {\bibfnamefont{L.}~\bibnamefont{{Verde}}}, \bibinfo {author}
  {\bibfnamefont{M.}~\bibnamefont{{Halpern}}}, \bibinfo {author}
  {\bibfnamefont{R.~S.}\ \bibnamefont{{Hill}}}, \bibinfo {author}
  {\bibfnamefont{A.}~\bibnamefont{{Kogut}}}, \bibinfo {author}
  {\bibfnamefont{M.}~\bibnamefont{{Limon}}}, \bibinfo {author}
  {\bibfnamefont{S.~S.}\ \bibnamefont{{Meyer}}}, \bibinfo {author}
  {\bibfnamefont{N.}~\bibnamefont{{Odegard}}}, \bibinfo {author}
  {\bibfnamefont{G.~S.}\ \bibnamefont{{Tucker}}}, \bibinfo {author}
  {\bibfnamefont{J.~L.}\ \bibnamefont{{Weiland}}}, \bibinfo {author}
  {\bibfnamefont{E.}~\bibnamefont{{Wollack}}},\ and\ \bibinfo {author}
  {\bibfnamefont{E.~L.}\ \bibnamefont{{Wright}}},\ }%
  \bibfield{journal}{%
  \Doi{10.1086/513700}{\bibinfo {journal} {Astrophys. J. Suppl.}}\ }%
  \textbf{\bibinfo {volume} {170}},\ \bibinfo {pages} {377} (\bibinfo {month}
  {Jun.}\ \bibinfo {year} {2007}),\
  \Eprint{http://arxiv.org/abs/arXiv:astro-ph/0603449}{arXiv:astro-ph/0603449}%
  \bibAnnoteFile{NoStop}{spe2007}%
\bibitem{die2008}%
  \BibitemOpen
  \bibfield{author}{%
  \bibinfo {author} {\bibfnamefont{J.}~\bibnamefont{{Diemand}}}, \bibinfo
  {author} {\bibfnamefont{M.}~\bibnamefont{{Kuhlen}}}, \bibinfo {author}
  {\bibfnamefont{P.}~\bibnamefont{{Madau}}}, \bibinfo {author}
  {\bibfnamefont{M.}~\bibnamefont{{Zemp}}}, \bibinfo {author}
  {\bibfnamefont{B.}~\bibnamefont{{Moore}}}, \bibinfo {author}
  {\bibfnamefont{D.}~\bibnamefont{{Potter}}},\ and\ \bibinfo {author}
  {\bibfnamefont{J.}~\bibnamefont{{Stadel}}},\ }%
  \bibfield{journal}{%
  \Doi{10.1038/nature07153}{\bibinfo {journal} {\nat}}\ }%
  \textbf{\bibinfo {volume} {454}},\ \bibinfo {pages} {735} (\bibinfo {month}
  {Aug.}\ \bibinfo {year} {2008}),\
  \Eprint{http://arxiv.org/abs/0805.1244}{arXiv:0805.1244}%
  \bibAnnoteFile{NoStop}{die2008}%
\bibitem{ber2001}%
  \BibitemOpen
  \bibfield{author}{%
  \bibinfo {author} {\bibfnamefont{E.}~\bibnamefont{{Bertschinger}}},\ }%
  \bibfield{journal}{%
  \Doi{10.1086/322526}{\bibinfo {journal} {Astrophys. J. Suppl.}}\ }%
  \textbf{\bibinfo {volume} {137}},\ \bibinfo {pages} {1} (\bibinfo {month}
  {Nov.}\ \bibinfo {year} {2001}),\
  \Eprint{http://arxiv.org/abs/arXiv:astro-ph/0103301}{arXiv:astro-ph/0103301}%
  \bibAnnoteFile{NoStop}{ber2001}%
\bibitem{bbks}%
  \BibitemOpen
  \bibfield{author}{%
  \bibinfo {author} {\bibfnamefont{J.~M.}\ \bibnamefont{{Bardeen}}}, \bibinfo
  {author} {\bibfnamefont{J.~R.}\ \bibnamefont{{Bond}}}, \bibinfo {author}
  {\bibfnamefont{N.}~\bibnamefont{{Kaiser}}},\ and\ \bibinfo {author}
  {\bibfnamefont{A.~S.}\ \bibnamefont{{Szalay}}},\ }%
  \bibfield{journal}{%
  \Doi{10.1086/164143}{\bibinfo {journal} {\apj}}\ }%
  \textbf{\bibinfo {volume} {304}},\ \bibinfo {pages} {15} (\bibinfo {month}
  {May}\ \bibinfo {year} {1986})%
  \bibAnnoteFile{NoStop}{bbks}%
\bibitem{eis1997}%
  \BibitemOpen
  \bibfield{author}{%
  \bibinfo {author} {\bibfnamefont{D.~J.}\ \bibnamefont{{Eisenstein}}}\ and\
  \bibinfo {author} {\bibfnamefont{W.}~\bibnamefont{{Hu}}},\ }%
  \bibfield{journal}{%
  \Doi{10.1086/305424}{\bibinfo {journal} {\apj}}\ }%
  \textbf{\bibinfo {volume} {496}},\ \bibinfo {pages} {605} (\bibinfo {month}
  {Mar.}\ \bibinfo {year} {1998}),\
  \Eprint{http://arxiv.org/abs/arXiv:astro-ph/9709112}{arXiv:astro-ph/9709112}%
  \bibAnnoteFile{NoStop}{eis1997}%
\bibitem{gor2008}%
  \BibitemOpen
  \bibfield{author}{%
  \bibinfo {author} {\bibfnamefont{D.}~\bibnamefont{{Gorbunov}}}, \bibinfo
  {author} {\bibfnamefont{A.}~\bibnamefont{{Khmelnitsky}}},\ and\ \bibinfo
  {author} {\bibfnamefont{V.}~\bibnamefont{{Rubakov}}},\ }%
  \bibfield{journal}{%
  \Doi{10.1088/1126-6708/2008/12/055}{\bibinfo {journal} {Journal of High
  Energy Physics}}\ }%
  \textbf{\bibinfo {volume} {12}},\ \bibinfo {pages} {55} (\bibinfo {month}
  {Dec.}\ \bibinfo {year} {2008}),\
  \Eprint{http://arxiv.org/abs/0805.2836}{arXiv:0805.2836 [hep-ph]}%
  \bibAnnoteFile{NoStop}{gor2008}%
\bibitem{kus2009}%
  \BibitemOpen
  \bibfield{author}{%
  \bibinfo {author} {\bibfnamefont{A.}~\bibnamefont{{Kusenko}}},\ }%
  \bibfield{journal}{%
  \Doi{10.1016/j.physrep.2009.07.004}{\bibinfo {journal} {Physics Reports}}\ }%
  \textbf{\bibinfo {volume} {481}},\ \bibinfo {pages} {1} (\bibinfo {month}
  {Sep.}\ \bibinfo {year} {2009}),\
  \Eprint{http://arxiv.org/abs/0906.2968}{arXiv:0906.2968}%
  \bibAnnoteFile{NoStop}{kus2009}%
\bibitem{gni2010a}%
  \BibitemOpen
  \bibfield{author}{%
  \bibinfo {author} {\bibfnamefont{S.}~\bibnamefont{{Gninenko}}},\ }%
  \bibfield{journal}{%
  \bibinfo {journal} {ArXiv e-prints}}%
   (\bibinfo {month} {Sep.}\ \bibinfo {year} {2010}),\
  \Eprint{http://arxiv.org/abs/1009.5536}{arXiv:1009.5536 [hep-ph]}%
  \bibAnnoteFile{NoStop}{gni2010a}%
\bibitem{gni2010b}%
  \BibitemOpen
  \bibfield{author}{%
  \bibinfo {author} {\bibfnamefont{S.~N.}\ \bibnamefont{{Gninenko}}}\ and\
  \bibinfo {author} {\bibfnamefont{D.~S.}\ \bibnamefont{{Gorbunov}}},\ }%
  \bibfield{journal}{%
  \Doi{10.1103/PhysRevD.81.075013}{\bibinfo {journal} {\prd}}\ }%
  \textbf{\bibinfo {volume} {81}},\ \bibinfo {pages} {075013} (\bibinfo {month}
  {Apr.}\ \bibinfo {year} {2010}),\
  \Eprint{http://arxiv.org/abs/0907.4666}{arXiv:0907.4666 [hep-ph]}%
  \bibAnnoteFile{NoStop}{gni2010b}%
\bibitem{kar2009}%
  \BibitemOpen
  \bibfield{author}{%
  \bibinfo {author} {\bibfnamefont{G.}~\bibnamefont{{Karagiorgi}}}, \bibinfo
  {author} {\bibfnamefont{Z.}~\bibnamefont{{Djurcic}}}, \bibinfo {author}
  {\bibfnamefont{J.~M.}\ \bibnamefont{{Conrad}}}, \bibinfo {author}
  {\bibfnamefont{M.~H.}\ \bibnamefont{{Shaevitz}}},\ and\ \bibinfo {author}
  {\bibfnamefont{M.}~\bibnamefont{{Sorel}}},\ }%
  \bibfield{journal}{%
  \Doi{10.1103/PhysRevD.80.073001}{\bibinfo {journal} {\prd}}\ }%
  \textbf{\bibinfo {volume} {80}},\ \bibinfo {pages} {073001} (\bibinfo {month}
  {Oct.}\ \bibinfo {year} {2009}),\
  \Eprint{http://arxiv.org/abs/0906.1997}{arXiv:0906.1997 [hep-ph]}%
  \bibAnnoteFile{NoStop}{kar2009}%
\bibitem{sor2004}%
  \BibitemOpen
  \bibfield{author}{%
  \bibinfo {author} {\bibfnamefont{M.}~\bibnamefont{{Sorel}}}, \bibinfo
  {author} {\bibfnamefont{J.~M.}\ \bibnamefont{{Conrad}}},\ and\ \bibinfo
  {author} {\bibfnamefont{M.~H.}\ \bibnamefont{{Shaevitz}}},\ }%
  \bibfield{journal}{%
  \Doi{10.1103/PhysRevD.70.073004}{\bibinfo {journal} {\prd}}\ }%
  \textbf{\bibinfo {volume} {70}},\ \bibinfo {pages} {073004} (\bibinfo {month}
  {Oct.}\ \bibinfo {year} {2004}),\
  \Eprint{http://arxiv.org/abs/arXiv:hep-ph/0305255}{arXiv:hep-ph/0305255}%
  \bibAnnoteFile{NoStop}{sor2004}%
\bibitem{mel2009}%
  \BibitemOpen
  \bibfield{author}{%
  \bibinfo {author} {\bibfnamefont{A.}~\bibnamefont{{Melchiorri}}}, \bibinfo
  {author} {\bibfnamefont{O.}~\bibnamefont{{Mena}}}, \bibinfo {author}
  {\bibfnamefont{S.}~\bibnamefont{{Palomares-Ruiz}}}, \bibinfo {author}
  {\bibfnamefont{S.}~\bibnamefont{{Pascoli}}}, \bibinfo {author}
  {\bibfnamefont{A.}~\bibnamefont{{Slosar}}},\ and\ \bibinfo {author}
  {\bibfnamefont{M.}~\bibnamefont{{Sorel}}},\ }%
  \bibfield{journal}{%
  \Doi{10.1088/1475-7516/2009/01/036}{\bibinfo {journal} {J. Cosmo. and
  Astropart. Phys.}}\ }%
  \textbf{\bibinfo {volume} {1}},\ \bibinfo {pages} {36} (\bibinfo {month}
  {Jan.}\ \bibinfo {year} {2009}),\
  \Eprint{http://arxiv.org/abs/0810.5133}{arXiv:0810.5133 [hep-ph]}%
  \bibAnnoteFile{NoStop}{mel2009}%
\bibitem{mal2007}%
  \BibitemOpen
  \bibfield{author}{%
  \bibinfo {author} {\bibfnamefont{M.}~\bibnamefont{{Maltoni}}}\ and\ \bibinfo
  {author} {\bibfnamefont{T.}~\bibnamefont{{Schwetz}}},\ }%
  \bibfield{journal}{%
  \Doi{10.1103/PhysRevD.76.093005}{\bibinfo {journal} {\prd}}\ }%
  \textbf{\bibinfo {volume} {76}},\ \bibinfo {pages} {093005} (\bibinfo {month}
  {Nov.}\ \bibinfo {year} {2007}),\
  \Eprint{http://arxiv.org/abs/0705.0107}{arXiv:0705.0107 [hep-ph]}%
  \bibAnnoteFile{NoStop}{mal2007}%
\bibitem{pas2005}%
  \BibitemOpen
  \bibfield{author}{%
  \bibinfo {author} {\bibfnamefont{H.}~\bibnamefont{{P{\"a}s}}}, \bibinfo
  {author} {\bibfnamefont{S.}~\bibnamefont{{Pakvasa}}},\ and\ \bibinfo {author}
  {\bibfnamefont{T.~J.}\ \bibnamefont{{Weiler}}},\ }%
  \bibfield{journal}{%
  \Doi{10.1103/PhysRevD.72.095017}{\bibinfo {journal} {\prd}}\ }%
  \textbf{\bibinfo {volume} {72}},\ \bibinfo {pages} {095017} (\bibinfo {month}
  {Nov.}\ \bibinfo {year} {2005}),\
  \Eprint{http://arxiv.org/abs/arXiv:hep-ph/0504096}{arXiv:hep-ph/0504096}%
  \bibAnnoteFile{NoStop}{pas2005}%
\bibitem{akh2010}%
  \BibitemOpen
  \bibfield{author}{%
  \bibinfo {author} {\bibfnamefont{E.}~\bibnamefont{{Akhmedov}}}\ and\ \bibinfo
  {author} {\bibfnamefont{T.}~\bibnamefont{{Schwetz}}},\ }%
  \bibfield{journal}{%
  \Doi{10.1007/JHEP10(2010)115}{\bibinfo {journal} {Journal of High Energy
  Physics}}\ }%
  \textbf{\bibinfo {volume} {10}},\ \bibinfo {pages} {115} (\bibinfo {month}
  {Oct.}\ \bibinfo {year} {2010}),\
  \Eprint{http://arxiv.org/abs/1007.4171}{arXiv:1007.4171 [hep-ph]}%
  \bibAnnoteFile{NoStop}{akh2010}%
\bibitem{ath1995}%
  \BibitemOpen
  \bibfield{author}{%
  \bibinfo {author} {\bibfnamefont{C.}~\bibnamefont{{Athanassopoulos}}},
  \bibinfo {author} {\bibfnamefont{L.~B.}\ \bibnamefont{{Auerbach}}}, \bibinfo
  {author} {\bibfnamefont{D.~A.}\ \bibnamefont{{Bauer}}}, \bibinfo {author}
  {\bibfnamefont{R.~D.}\ \bibnamefont{{Bolton}}}, \bibinfo {author}
  {\bibfnamefont{B.}~\bibnamefont{{Boyd}}}, \bibinfo {author}
  {\bibfnamefont{R.~L.}\ \bibnamefont{{Burman}}}, \bibinfo {author}
  {\bibfnamefont{D.~O.}\ \bibnamefont{{Caldwell}}}, \bibinfo {author}
  {\bibfnamefont{I.}~\bibnamefont{{Cohen}}}, \bibinfo {author}
  {\bibfnamefont{B.~D.}\ \bibnamefont{{Dieterle}}}, \bibinfo {author}
  {\bibfnamefont{J.~B.}\ \bibnamefont{{Donahue}}}, \bibinfo {author}
  {\bibfnamefont{A.~M.}\ \bibnamefont{{Eisner}}}, \bibinfo {author}
  {\bibfnamefont{A.}~\bibnamefont{{Fazely}}}, \bibinfo {author}
  {\bibfnamefont{F.~J.}\ \bibnamefont{{Federspiel}}}, \bibinfo {author}
  {\bibfnamefont{G.~T.}\ \bibnamefont{{Garvey}}}, \bibinfo {author}
  {\bibfnamefont{M.}~\bibnamefont{{Gray}}},\ and\ \bibinfo {author}
  {\bibnamefont{{et al.}}},\ }%
  \bibfield{journal}{%
  \Doi{10.1103/PhysRevLett.75.2650}{\bibinfo {journal} {Physical Review
  Letters}}\ }%
  \textbf{\bibinfo {volume} {75}},\ \bibinfo {pages} {2650} (\bibinfo {month}
  {Oct.}\ \bibinfo {year} {1995}),\
  \Eprint{http://arxiv.org/abs/arXiv:nucl-ex/9504002}{arXiv:nucl-ex/9504002}%
  \bibAnnoteFile{NoStop}{ath1995}%
\bibitem{ath1996}%
  \BibitemOpen
  \bibfield{author}{%
  \bibinfo {author} {\bibfnamefont{C.}~\bibnamefont{{Athanassopoulos}}},
  \bibinfo {author} {\bibfnamefont{L.~B.}\ \bibnamefont{{Auerbach}}}, \bibinfo
  {author} {\bibfnamefont{R.~L.}\ \bibnamefont{{Burman}}}, \bibinfo {author}
  {\bibfnamefont{I.}~\bibnamefont{{Cohen}}}, \bibinfo {author}
  {\bibfnamefont{D.~O.}\ \bibnamefont{{Caldwell}}}, \bibinfo {author}
  {\bibfnamefont{B.~D.}\ \bibnamefont{{Dieterle}}}, \bibinfo {author}
  {\bibfnamefont{J.~B.}\ \bibnamefont{{Donahue}}}, \bibinfo {author}
  {\bibfnamefont{A.~M.}\ \bibnamefont{{Eisner}}}, \bibinfo {author}
  {\bibfnamefont{A.}~\bibnamefont{{Fazely}}}, \bibinfo {author}
  {\bibfnamefont{F.~J.}\ \bibnamefont{{Federspiel}}}, \bibinfo {author}
  {\bibfnamefont{G.~T.}\ \bibnamefont{{Garvey}}}, \bibinfo {author}
  {\bibfnamefont{M.}~\bibnamefont{{Gray}}}, \bibinfo {author}
  {\bibfnamefont{R.~M.}\ \bibnamefont{{Gunasingha}}}, \bibinfo {author}
  {\bibfnamefont{R.}~\bibnamefont{{Imlay}}}, \bibinfo {author}
  {\bibfnamefont{K.}~\bibnamefont{{Johnston}}}, \bibinfo {author}
  {\bibfnamefont{H.~J.}\ \bibnamefont{{Kim}}}, \bibinfo {author}
  {\bibfnamefont{W.~C.}\ \bibnamefont{{Louis}}}, \bibinfo {author}
  {\bibfnamefont{R.}~\bibnamefont{{Majkic}}}, \bibinfo {author}
  {\bibfnamefont{J.}~\bibnamefont{{Margulies}}}, \bibinfo {author}
  {\bibfnamefont{K.}~\bibnamefont{{McIlhany}}}, \bibinfo {author}
  {\bibfnamefont{W.}~\bibnamefont{{Metcalf}}}, \bibinfo {author}
  {\bibfnamefont{G.~B.}\ \bibnamefont{{Mills}}}, \bibinfo {author}
  {\bibfnamefont{R.~A.}\ \bibnamefont{{Reeder}}}, \bibinfo {author}
  {\bibfnamefont{V.}~\bibnamefont{{Sandberg}}}, \bibinfo {author}
  {\bibfnamefont{D.}~\bibnamefont{{Smith}}}, \bibinfo {author}
  {\bibfnamefont{I.}~\bibnamefont{{Stancu}}}, \bibinfo {author}
  {\bibfnamefont{W.}~\bibnamefont{{Strossman}}}, \bibinfo {author}
  {\bibfnamefont{R.}~\bibnamefont{{Tayloe}}}, \bibinfo {author}
  {\bibfnamefont{G.~J.}\ \bibnamefont{{Vandalen}}}, \bibinfo {author}
  {\bibfnamefont{W.}~\bibnamefont{{Vernon}}}, \bibinfo {author}
  {\bibfnamefont{N.}~\bibnamefont{{Wadia}}}, \bibinfo {author}
  {\bibfnamefont{J.}~\bibnamefont{{Waltz}}}, \bibinfo {author}
  {\bibfnamefont{Y.}~\bibnamefont{{Wang}}}, \bibinfo {author}
  {\bibfnamefont{D.~H.}\ \bibnamefont{{White}}}, \bibinfo {author}
  {\bibfnamefont{D.}~\bibnamefont{{Works}}}, \bibinfo {author}
  {\bibfnamefont{Y.}~\bibnamefont{{Xiao}}},\ and\ \bibinfo {author}
  {\bibfnamefont{S.}~\bibnamefont{{Yellin}}},\ }%
  \bibfield{journal}{%
  \Doi{10.1103/PhysRevLett.77.3082}{\bibinfo {journal} {Physical Review
  Letters}}\ }%
  \textbf{\bibinfo {volume} {77}},\ \bibinfo {pages} {3082} (\bibinfo {month}
  {Oct.}\ \bibinfo {year} {1996}),\
  \Eprint{http://arxiv.org/abs/arXiv:nucl-ex/9605003}{arXiv:nucl-ex/9605003}%
  \bibAnnoteFile{NoStop}{ath1996}%
\bibitem{ath1998a}%
  \BibitemOpen
  \bibfield{author}{%
  \bibinfo {author} {\bibfnamefont{C.}~\bibnamefont{{Athanassopoulos}}},
  \bibinfo {author} {\bibfnamefont{L.~B.}\ \bibnamefont{{Auerbach}}}, \bibinfo
  {author} {\bibfnamefont{R.~L.}\ \bibnamefont{{Burman}}}, \bibinfo {author}
  {\bibfnamefont{D.~O.}\ \bibnamefont{{Caldwell}}}, \bibinfo {author}
  {\bibfnamefont{E.~D.}\ \bibnamefont{{Church}}}, \bibinfo {author}
  {\bibfnamefont{I.}~\bibnamefont{{Cohen}}}, \bibinfo {author}
  {\bibfnamefont{J.~B.}\ \bibnamefont{{Donahue}}}, \bibinfo {author}
  {\bibfnamefont{A.}~\bibnamefont{{Fazely}}}, \bibinfo {author}
  {\bibfnamefont{F.~J.}\ \bibnamefont{{Federspiel}}}, \bibinfo {author}
  {\bibfnamefont{G.~T.}\ \bibnamefont{{Garvey}}}, \bibinfo {author}
  {\bibfnamefont{R.~M.}\ \bibnamefont{{Gunasingha}}}, \bibinfo {author}
  {\bibfnamefont{R.}~\bibnamefont{{Imlay}}}, \bibinfo {author}
  {\bibfnamefont{K.}~\bibnamefont{{Johnston}}}, \bibinfo {author}
  {\bibfnamefont{H.~J.}\ \bibnamefont{{Kim}}}, \bibinfo {author}
  {\bibfnamefont{W.~C.}\ \bibnamefont{{Louis}}}, \bibinfo {author}
  {\bibfnamefont{R.}~\bibnamefont{{Majkic}}}, \bibinfo {author}
  {\bibfnamefont{K.}~\bibnamefont{{McIlhany}}}, \bibinfo {author}
  {\bibfnamefont{G.~B.}\ \bibnamefont{{Mills}}}, \bibinfo {author}
  {\bibfnamefont{R.~A.}\ \bibnamefont{{Reeder}}}, \bibinfo {author}
  {\bibfnamefont{V.}~\bibnamefont{{Sandberg}}}, \bibinfo {author}
  {\bibfnamefont{D.}~\bibnamefont{{Smith}}}, \bibinfo {author}
  {\bibfnamefont{I.}~\bibnamefont{{Stancu}}}, \bibinfo {author}
  {\bibfnamefont{W.}~\bibnamefont{{Strossman}}}, \bibinfo {author}
  {\bibfnamefont{R.}~\bibnamefont{{Tayloe}}}, \bibinfo {author}
  {\bibfnamefont{G.~J.}\ \bibnamefont{{Vandalen}}}, \bibinfo {author}
  {\bibfnamefont{W.}~\bibnamefont{{Vernon}}}, \bibinfo {author}
  {\bibfnamefont{N.}~\bibnamefont{{Wadia}}}, \bibinfo {author}
  {\bibfnamefont{J.}~\bibnamefont{{Waltz}}}, \bibinfo {author}
  {\bibfnamefont{D.~H.}\ \bibnamefont{{White}}}, \bibinfo {author}
  {\bibfnamefont{D.}~\bibnamefont{{Works}}}, \bibinfo {author}
  {\bibfnamefont{Y.}~\bibnamefont{{Xiao}}},\ and\ \bibinfo {author}
  {\bibfnamefont{S.}~\bibnamefont{{Yellin}}},\ }%
  \bibfield{journal}{%
  \Doi{10.1103/PhysRevLett.81.1774}{\bibinfo {journal} {Physical Review
  Letters}}\ }%
  \textbf{\bibinfo {volume} {81}},\ \bibinfo {pages} {1774} (\bibinfo {month}
  {Aug.}\ \bibinfo {year} {1998}),\
  \Eprint{http://arxiv.org/abs/arXiv:nucl-ex/9709006}{arXiv:nucl-ex/9709006}%
  \bibAnnoteFile{NoStop}{ath1998a}%
\bibitem{ath1998b}%
  \BibitemOpen
  \bibfield{author}{%
  \bibinfo {author} {\bibfnamefont{C.}~\bibnamefont{{Athanassopoulos}}},
  \bibinfo {author} {\bibfnamefont{L.~B.}\ \bibnamefont{{Auerbach}}}, \bibinfo
  {author} {\bibfnamefont{R.~L.}\ \bibnamefont{{Burman}}}, \bibinfo {author}
  {\bibfnamefont{D.~O.}\ \bibnamefont{{Caldwell}}}, \bibinfo {author}
  {\bibfnamefont{E.~D.}\ \bibnamefont{{Church}}}, \bibinfo {author}
  {\bibfnamefont{I.}~\bibnamefont{{Cohen}}}, \bibinfo {author}
  {\bibfnamefont{J.~B.}\ \bibnamefont{{Donahue}}}, \bibinfo {author}
  {\bibfnamefont{A.}~\bibnamefont{{Fazely}}}, \bibinfo {author}
  {\bibfnamefont{F.~J.}\ \bibnamefont{{Federspiel}}}, \bibinfo {author}
  {\bibfnamefont{G.~T.}\ \bibnamefont{{Garvey}}}, \bibinfo {author}
  {\bibfnamefont{R.~M.}\ \bibnamefont{{Gunasingha}}}, \bibinfo {author}
  {\bibfnamefont{R.}~\bibnamefont{{Imlay}}}, \bibinfo {author}
  {\bibfnamefont{K.}~\bibnamefont{{Johnston}}}, \bibinfo {author}
  {\bibfnamefont{H.~J.}\ \bibnamefont{{Kim}}}, \bibinfo {author}
  {\bibfnamefont{W.~C.}\ \bibnamefont{{Louis}}}, \bibinfo {author}
  {\bibfnamefont{R.}~\bibnamefont{{Majkic}}}, \bibinfo {author}
  {\bibfnamefont{K.}~\bibnamefont{{McIlhany}}}, \bibinfo {author}
  {\bibfnamefont{W.}~\bibnamefont{{Metcalf}}}, \bibinfo {author}
  {\bibfnamefont{G.~B.}\ \bibnamefont{{Mills}}}, \bibinfo {author}
  {\bibfnamefont{R.~A.}\ \bibnamefont{{Reeder}}}, \bibinfo {author}
  {\bibfnamefont{V.}~\bibnamefont{{Sandberg}}}, \bibinfo {author}
  {\bibfnamefont{D.}~\bibnamefont{{Smith}}}, \bibinfo {author}
  {\bibfnamefont{I.}~\bibnamefont{{Stancu}}}, \bibinfo {author}
  {\bibfnamefont{W.}~\bibnamefont{{Strossman}}}, \bibinfo {author}
  {\bibfnamefont{R.}~\bibnamefont{{Tayloe}}}, \bibinfo {author}
  {\bibfnamefont{G.~J.}\ \bibnamefont{{Vandalen}}}, \bibinfo {author}
  {\bibfnamefont{W.}~\bibnamefont{{Vernon}}}, \bibinfo {author}
  {\bibfnamefont{N.}~\bibnamefont{{Wadia}}}, \bibinfo {author}
  {\bibfnamefont{J.}~\bibnamefont{{Waltz}}}, \bibinfo {author}
  {\bibfnamefont{D.~H.}\ \bibnamefont{{White}}}, \bibinfo {author}
  {\bibfnamefont{D.}~\bibnamefont{{Works}}}, \bibinfo {author}
  {\bibfnamefont{Y.}~\bibnamefont{{Xiao}}},\ and\ \bibinfo {author}
  {\bibfnamefont{S.}~\bibnamefont{{Yellin}}},\ }%
  \bibfield{journal}{%
  \Doi{10.1103/PhysRevC.58.2489}{\bibinfo {journal} {\prc}}\ }%
  \textbf{\bibinfo {volume} {58}},\ \bibinfo {pages} {2489} (\bibinfo {month}
  {Oct.}\ \bibinfo {year} {1998}),\
  \Eprint{http://arxiv.org/abs/arXiv:nucl-ex/9706006}{arXiv:nucl-ex/9706006}%
  \bibAnnoteFile{NoStop}{ath1998b}%
\bibitem{aa2007}%
  \BibitemOpen
  \bibfield{author}{%
  \bibinfo {author} {\bibfnamefont{A.~A.}\ \bibnamefont{{Aguilar-Arevalo}}},
  \bibinfo {author} {\bibfnamefont{A.~O.}\ \bibnamefont{{Bazarko}}}, \bibinfo
  {author} {\bibfnamefont{S.~J.}\ \bibnamefont{{Brice}}}, \bibinfo {author}
  {\bibfnamefont{B.~C.}\ \bibnamefont{{Brown}}}, \bibinfo {author}
  {\bibfnamefont{L.}~\bibnamefont{{Bugel}}}, \bibinfo {author}
  {\bibfnamefont{J.}~\bibnamefont{{Cao}}}, \bibinfo {author}
  {\bibfnamefont{L.}~\bibnamefont{{Coney}}}, \bibinfo {author}
  {\bibfnamefont{J.~M.}\ \bibnamefont{{Conrad}}}, \bibinfo {author}
  {\bibfnamefont{D.~C.}\ \bibnamefont{{Cox}}}, \bibinfo {author}
  {\bibfnamefont{A.}~\bibnamefont{{Curioni}}}, \bibinfo {author}
  {\bibfnamefont{Z.}~\bibnamefont{{Djurcic}}}, \bibinfo {author}
  {\bibfnamefont{D.~A.}\ \bibnamefont{{Finley}}}, \bibinfo {author}
  {\bibfnamefont{B.~T.}\ \bibnamefont{{Fleming}}}, \bibinfo {author}
  {\bibfnamefont{R.}~\bibnamefont{{Ford}}}, \bibinfo {author}
  {\bibfnamefont{F.~G.}\ \bibnamefont{{Garcia}}}, \bibinfo {author}
  {\bibfnamefont{G.~T.}\ \bibnamefont{{Garvey}}}, \bibinfo {author}
  {\bibfnamefont{C.}~\bibnamefont{{Green}}}, \bibinfo {author}
  {\bibfnamefont{J.~A.}\ \bibnamefont{{Green}}}, \bibinfo {author}
  {\bibfnamefont{T.~L.}\ \bibnamefont{{Hart}}}, \bibinfo {author}
  {\bibfnamefont{E.}~\bibnamefont{{Hawker}}}, \bibinfo {author}
  {\bibfnamefont{R.}~\bibnamefont{{Imlay}}}, \bibinfo {author}
  {\bibfnamefont{R.~A.}\ \bibnamefont{{Johnson}}}, \bibinfo {author}
  {\bibfnamefont{P.}~\bibnamefont{{Kasper}}}, \bibinfo {author}
  {\bibfnamefont{T.}~\bibnamefont{{Katori}}}, \bibinfo {author}
  {\bibfnamefont{T.}~\bibnamefont{{Kobilarcik}}}, \bibinfo {author}
  {\bibfnamefont{I.}~\bibnamefont{{Kourbanis}}}, \bibinfo {author}
  {\bibfnamefont{S.}~\bibnamefont{{Koutsoliotas}}}, \bibinfo {author}
  {\bibfnamefont{E.~M.}\ \bibnamefont{{Laird}}}, \bibinfo {author}
  {\bibfnamefont{J.~M.}\ \bibnamefont{{Link}}}, \bibinfo {author}
  {\bibfnamefont{Y.}~\bibnamefont{{Liu}}}, \bibinfo {author}
  {\bibfnamefont{Y.}~\bibnamefont{{Liu}}}, \bibinfo {author}
  {\bibfnamefont{W.~C.}\ \bibnamefont{{Louis}}}, \bibinfo {author}
  {\bibfnamefont{K.~B.~M.}\ \bibnamefont{{Mahn}}}, \bibinfo {author}
  {\bibfnamefont{W.}~\bibnamefont{{Marsh}}}, \bibinfo {author}
  {\bibfnamefont{P.~S.}\ \bibnamefont{{Martin}}}, \bibinfo {author}
  {\bibfnamefont{G.}~\bibnamefont{{McGregor}}}, \bibinfo {author}
  {\bibfnamefont{W.}~\bibnamefont{{Metcalf}}}, \bibinfo {author}
  {\bibfnamefont{P.~D.}\ \bibnamefont{{Meyers}}}, \bibinfo {author}
  {\bibfnamefont{F.}~\bibnamefont{{Mills}}}, \bibinfo {author}
  {\bibfnamefont{G.~B.}\ \bibnamefont{{Mills}}}, \bibinfo {author}
  {\bibfnamefont{J.}~\bibnamefont{{Monroe}}}, \bibinfo {author}
  {\bibfnamefont{C.~D.}\ \bibnamefont{{Moore}}}, \bibinfo {author}
  {\bibfnamefont{R.~H.}\ \bibnamefont{{Nelson}}}, \bibinfo {author}
  {\bibfnamefont{P.}~\bibnamefont{{Nienaber}}}, \bibinfo {author}
  {\bibfnamefont{S.}~\bibnamefont{{Ouedraogo}}}, \bibinfo {author}
  {\bibfnamefont{R.~B.}\ \bibnamefont{{Patterson}}}, \bibinfo {author}
  {\bibfnamefont{D.}~\bibnamefont{{Perevalov}}}, \bibinfo {author}
  {\bibfnamefont{C.~C.}\ \bibnamefont{{Polly}}}, \bibinfo {author}
  {\bibfnamefont{E.}~\bibnamefont{{Prebys}}}, \bibinfo {author}
  {\bibfnamefont{J.~L.}\ \bibnamefont{{Raaf}}}, \bibinfo {author}
  {\bibfnamefont{H.}~\bibnamefont{{Ray}}}, \bibinfo {author}
  {\bibfnamefont{B.~P.}\ \bibnamefont{{Roe}}}, \bibinfo {author}
  {\bibfnamefont{A.~D.}\ \bibnamefont{{Russell}}}, \bibinfo {author}
  {\bibfnamefont{V.}~\bibnamefont{{Sandberg}}}, \bibinfo {author}
  {\bibfnamefont{R.}~\bibnamefont{{Schirato}}}, \bibinfo {author}
  {\bibfnamefont{D.}~\bibnamefont{{Schmitz}}}, \bibinfo {author}
  {\bibfnamefont{M.~H.}\ \bibnamefont{{Shaevitz}}}, \bibinfo {author}
  {\bibfnamefont{F.~C.}\ \bibnamefont{{Shoemaker}}}, \bibinfo {author}
  {\bibfnamefont{D.}~\bibnamefont{{Smith}}}, \bibinfo {author}
  {\bibfnamefont{M.}~\bibnamefont{{Sorel}}}, \bibinfo {author}
  {\bibfnamefont{P.}~\bibnamefont{{Spentzouris}}}, \bibinfo {author}
  {\bibfnamefont{I.}~\bibnamefont{{Stancu}}}, \bibinfo {author}
  {\bibfnamefont{R.~J.}\ \bibnamefont{{Stefanski}}}, \bibinfo {author}
  {\bibfnamefont{M.}~\bibnamefont{{Sung}}}, \bibinfo {author}
  {\bibfnamefont{H.~A.}\ \bibnamefont{{Tanaka}}}, \bibinfo {author}
  {\bibfnamefont{R.}~\bibnamefont{{Tayloe}}}, \bibinfo {author}
  {\bibfnamefont{M.}~\bibnamefont{{Tzanov}}}, \bibinfo {author}
  {\bibfnamefont{R.}~\bibnamefont{{van de Water}}}, \bibinfo {author}
  {\bibfnamefont{M.~O.}\ \bibnamefont{{Wascko}}}, \bibinfo {author}
  {\bibfnamefont{D.~H.}\ \bibnamefont{{White}}}, \bibinfo {author}
  {\bibfnamefont{M.~J.}\ \bibnamefont{{Wilking}}}, \bibinfo {author}
  {\bibfnamefont{H.~J.}\ \bibnamefont{{Yang}}}, \bibinfo {author}
  {\bibfnamefont{G.~P.}\ \bibnamefont{{Zeller}}},\ and\ \bibinfo {author}
  {\bibfnamefont{E.~D.}\ \bibnamefont{{Zimmerman}}},\ }%
  \bibfield{journal}{%
  \Doi{10.1103/PhysRevLett.98.231801}{\bibinfo {journal} {Physical Review
  Letters}}\ }%
  \textbf{\bibinfo {volume} {98}},\ \bibinfo {pages} {231801} (\bibinfo {month}
  {Jun.}\ \bibinfo {year} {2007}),\
  \Eprint{http://arxiv.org/abs/0704.1500}{arXiv:0704.1500 [hep-ex]}%
  \bibAnnoteFile{NoStop}{aa2007}%
\bibitem{aa2009}%
  \BibitemOpen
  \bibfield{author}{%
  \bibinfo {author} {\bibfnamefont{A.~A.}\ \bibnamefont{{Aguilar-Arevalo}}},
  \bibinfo {author} {\bibfnamefont{C.~E.}\ \bibnamefont{{Anderson}}}, \bibinfo
  {author} {\bibfnamefont{A.~O.}\ \bibnamefont{{Bazarko}}}, \bibinfo {author}
  {\bibfnamefont{S.~J.}\ \bibnamefont{{Brice}}}, \bibinfo {author}
  {\bibfnamefont{B.~C.}\ \bibnamefont{{Brown}}}, \bibinfo {author}
  {\bibfnamefont{L.}~\bibnamefont{{Bugel}}}, \bibinfo {author}
  {\bibfnamefont{J.}~\bibnamefont{{Cao}}}, \bibinfo {author}
  {\bibfnamefont{L.}~\bibnamefont{{Coney}}}, \bibinfo {author}
  {\bibfnamefont{J.~M.}\ \bibnamefont{{Conrad}}}, \bibinfo {author}
  {\bibfnamefont{D.~C.}\ \bibnamefont{{Cox}}}, \bibinfo {author}
  {\bibfnamefont{A.}~\bibnamefont{{Curioni}}}, \bibinfo {author}
  {\bibfnamefont{Z.}~\bibnamefont{{Djurcic}}}, \bibinfo {author}
  {\bibfnamefont{D.~A.}\ \bibnamefont{{Finley}}}, \bibinfo {author}
  {\bibfnamefont{B.~T.}\ \bibnamefont{{Fleming}}}, \bibinfo {author}
  {\bibfnamefont{R.}~\bibnamefont{{Ford}}}, \bibinfo {author}
  {\bibfnamefont{F.~G.}\ \bibnamefont{{Garcia}}}, \bibinfo {author}
  {\bibfnamefont{G.~T.}\ \bibnamefont{{Garvey}}}, \bibinfo {author}
  {\bibfnamefont{C.}~\bibnamefont{{Green}}}, \bibinfo {author}
  {\bibfnamefont{J.~A.}\ \bibnamefont{{Green}}}, \bibinfo {author}
  {\bibfnamefont{T.~L.}\ \bibnamefont{{Hart}}}, \bibinfo {author}
  {\bibfnamefont{E.}~\bibnamefont{{Hawker}}}, \bibinfo {author}
  {\bibfnamefont{R.}~\bibnamefont{{Imlay}}}, \bibinfo {author}
  {\bibfnamefont{R.~A.}\ \bibnamefont{{Johnson}}}, \bibinfo {author}
  {\bibfnamefont{G.}~\bibnamefont{{Karagiorgi}}}, \bibinfo {author}
  {\bibfnamefont{P.}~\bibnamefont{{Kasper}}}, \bibinfo {author}
  {\bibfnamefont{T.}~\bibnamefont{{Katori}}}, \bibinfo {author}
  {\bibfnamefont{T.}~\bibnamefont{{Kobilarcik}}}, \bibinfo {author}
  {\bibfnamefont{I.}~\bibnamefont{{Kourbanis}}}, \bibinfo {author}
  {\bibfnamefont{S.}~\bibnamefont{{Koutsoliotas}}}, \bibinfo {author}
  {\bibfnamefont{E.~M.}\ \bibnamefont{{Laird}}}, \bibinfo {author}
  {\bibfnamefont{S.~K.}\ \bibnamefont{{Linden}}}, \bibinfo {author}
  {\bibfnamefont{J.~M.}\ \bibnamefont{{Link}}}, \bibinfo {author}
  {\bibfnamefont{Y.}~\bibnamefont{{Liu}}}, \bibinfo {author}
  {\bibfnamefont{Y.}~\bibnamefont{{Liu}}}, \bibinfo {author}
  {\bibfnamefont{W.~C.}\ \bibnamefont{{Louis}}}, \bibinfo {author}
  {\bibfnamefont{K.~B.~M.}\ \bibnamefont{{Mahn}}}, \bibinfo {author}
  {\bibfnamefont{W.}~\bibnamefont{{Marsh}}}, \bibinfo {author}
  {\bibfnamefont{G.}~\bibnamefont{{McGregor}}}, \bibinfo {author}
  {\bibfnamefont{W.}~\bibnamefont{{Metcalf}}}, \bibinfo {author}
  {\bibfnamefont{P.~D.}\ \bibnamefont{{Meyers}}}, \bibinfo {author}
  {\bibfnamefont{F.}~\bibnamefont{{Mills}}}, \bibinfo {author}
  {\bibfnamefont{G.~B.}\ \bibnamefont{{Mills}}}, \bibinfo {author}
  {\bibfnamefont{J.}~\bibnamefont{{Monroe}}}, \bibinfo {author}
  {\bibfnamefont{C.~D.}\ \bibnamefont{{Moore}}}, \bibinfo {author}
  {\bibfnamefont{R.~H.}\ \bibnamefont{{Nelson}}}, \bibinfo {author}
  {\bibfnamefont{V.~T.}\ \bibnamefont{{Nguyen}}}, \bibinfo {author}
  {\bibfnamefont{P.}~\bibnamefont{{Nienaber}}}, \bibinfo {author}
  {\bibfnamefont{J.~A.}\ \bibnamefont{{Nowak}}}, \bibinfo {author}
  {\bibfnamefont{S.}~\bibnamefont{{Ouedraogo}}}, \bibinfo {author}
  {\bibfnamefont{R.~B.}\ \bibnamefont{{Patterson}}}, \bibinfo {author}
  {\bibfnamefont{D.}~\bibnamefont{{Perevalov}}}, \bibinfo {author}
  {\bibfnamefont{C.~C.}\ \bibnamefont{{Polly}}}, \bibinfo {author}
  {\bibfnamefont{E.}~\bibnamefont{{Prebys}}}, \bibinfo {author}
  {\bibfnamefont{J.~L.}\ \bibnamefont{{Raaf}}}, \bibinfo {author}
  {\bibfnamefont{H.}~\bibnamefont{{Ray}}}, \bibinfo {author}
  {\bibfnamefont{B.~P.}\ \bibnamefont{{Roe}}}, \bibinfo {author}
  {\bibfnamefont{A.~D.}\ \bibnamefont{{Russell}}}, \bibinfo {author}
  {\bibfnamefont{V.}~\bibnamefont{{Sandberg}}}, \bibinfo {author}
  {\bibfnamefont{R.}~\bibnamefont{{Schirato}}}, \bibinfo {author}
  {\bibfnamefont{D.}~\bibnamefont{{Schmitz}}}, \bibinfo {author}
  {\bibfnamefont{M.~H.}\ \bibnamefont{{Shaevitz}}}, \bibinfo {author}
  {\bibfnamefont{F.~C.}\ \bibnamefont{{Shoemaker}}}, \bibinfo {author}
  {\bibfnamefont{D.}~\bibnamefont{{Smith}}}, \bibinfo {author}
  {\bibfnamefont{M.}~\bibnamefont{{Sodeberg}}}, \bibinfo {author}
  {\bibfnamefont{M.}~\bibnamefont{{Sorel}}}, \bibinfo {author}
  {\bibfnamefont{P.}~\bibnamefont{{Spentzouris}}}, \bibinfo {author}
  {\bibfnamefont{I.}~\bibnamefont{{Stancu}}}, \bibinfo {author}
  {\bibfnamefont{R.~J.}\ \bibnamefont{{Stefanski}}}, \bibinfo {author}
  {\bibfnamefont{M.}~\bibnamefont{{Sung}}}, \bibinfo {author}
  {\bibfnamefont{H.~A.}\ \bibnamefont{{Tanaka}}}, \bibinfo {author}
  {\bibfnamefont{R.}~\bibnamefont{{Tayloe}}}, \bibinfo {author}
  {\bibfnamefont{M.}~\bibnamefont{{Tzanov}}}, \bibinfo {author}
  {\bibfnamefont{R.}~\bibnamefont{{van de Water}}}, \bibinfo {author}
  {\bibfnamefont{M.~O.}\ \bibnamefont{{Wascko}}}, \bibinfo {author}
  {\bibfnamefont{D.~H.}\ \bibnamefont{{White}}}, \bibinfo {author}
  {\bibfnamefont{M.~J.}\ \bibnamefont{{Wilking}}}, \bibinfo {author}
  {\bibfnamefont{H.~J.}\ \bibnamefont{{Yang}}}, \bibinfo {author}
  {\bibfnamefont{G.~P.}\ \bibnamefont{{Zeller}}},\ and\ \bibinfo {author}
  {\bibfnamefont{E.~D.}\ \bibnamefont{{Zimmerman}}},\ }%
  \bibfield{journal}{%
  \Doi{10.1103/PhysRevLett.102.101802}{\bibinfo {journal} {Physical Review
  Letters}}\ }%
  \textbf{\bibinfo {volume} {102}},\ \bibinfo {pages} {101802} (\bibinfo
  {month} {Mar.}\ \bibinfo {year} {2009}),\
  \Eprint{http://arxiv.org/abs/0812.2243}{arXiv:0812.2243 [hep-ex]}%
  \bibAnnoteFile{NoStop}{aa2009}%
\bibitem{aa2010}%
  \BibitemOpen
  \bibfield{author}{%
  \bibinfo {author} {\bibfnamefont{A.~A.}\ \bibnamefont{{Aguilar-Arevalo}}},
  \bibinfo {author} {\bibfnamefont{C.~E.}\ \bibnamefont{{Anderson}}}, \bibinfo
  {author} {\bibfnamefont{S.~J.}\ \bibnamefont{{Brice}}}, \bibinfo {author}
  {\bibfnamefont{B.~C.}\ \bibnamefont{{Brown}}}, \bibinfo {author}
  {\bibfnamefont{L.}~\bibnamefont{{Bugel}}}, \bibinfo {author}
  {\bibfnamefont{J.~M.}\ \bibnamefont{{Conrad}}}, \bibinfo {author}
  {\bibfnamefont{R.}~\bibnamefont{{Dharmapalan}}}, \bibinfo {author}
  {\bibfnamefont{Z.}~\bibnamefont{{Djurcic}}}, \bibinfo {author}
  {\bibfnamefont{B.~T.}\ \bibnamefont{{Fleming}}}, \bibinfo {author}
  {\bibfnamefont{R.}~\bibnamefont{{Ford}}}, \bibinfo {author}
  {\bibfnamefont{F.~G.}\ \bibnamefont{{Garcia}}}, \bibinfo {author}
  {\bibfnamefont{G.~T.}\ \bibnamefont{{Garvey}}}, \bibinfo {author}
  {\bibfnamefont{J.}~\bibnamefont{{Mirabal}}}, \bibinfo {author}
  {\bibfnamefont{J.}~\bibnamefont{{Grange}}}, \bibinfo {author}
  {\bibfnamefont{J.~A.}\ \bibnamefont{{Green}}}, \bibinfo {author}
  {\bibfnamefont{R.}~\bibnamefont{{Imlay}}}, \bibinfo {author}
  {\bibfnamefont{R.~A.}\ \bibnamefont{{Johnson}}}, \bibinfo {author}
  {\bibfnamefont{G.}~\bibnamefont{{Karagiorgi}}}, \bibinfo {author}
  {\bibfnamefont{T.}~\bibnamefont{{Katori}}}, \bibinfo {author}
  {\bibfnamefont{T.}~\bibnamefont{{Kobilarcik}}}, \bibinfo {author}
  {\bibfnamefont{S.~K.}\ \bibnamefont{{Linden}}}, \bibinfo {author}
  {\bibfnamefont{W.~C.}\ \bibnamefont{{Louis}}}, \bibinfo {author}
  {\bibfnamefont{K.~B.~M.}\ \bibnamefont{{Mahn}}}, \bibinfo {author}
  {\bibfnamefont{W.}~\bibnamefont{{Marsh}}}, \bibinfo {author}
  {\bibfnamefont{C.}~\bibnamefont{{Mauger}}}, \bibinfo {author}
  {\bibfnamefont{W.}~\bibnamefont{{Metcalf}}}, \bibinfo {author}
  {\bibfnamefont{G.~B.}\ \bibnamefont{{Mills}}}, \bibinfo {author}
  {\bibfnamefont{C.~D.}\ \bibnamefont{{Moore}}}, \bibinfo {author}
  {\bibfnamefont{J.}~\bibnamefont{{Mousseau}}}, \bibinfo {author}
  {\bibfnamefont{R.~H.}\ \bibnamefont{{Nelson}}}, \bibinfo {author}
  {\bibfnamefont{V.}~\bibnamefont{{Nguyen}}}, \bibinfo {author}
  {\bibfnamefont{P.}~\bibnamefont{{Nienaber}}}, \bibinfo {author}
  {\bibfnamefont{J.~A.}\ \bibnamefont{{Nowak}}}, \bibinfo {author}
  {\bibfnamefont{B.}~\bibnamefont{{Osmanov}}}, \bibinfo {author}
  {\bibfnamefont{Z.}~\bibnamefont{{Pavlovic}}}, \bibinfo {author}
  {\bibfnamefont{D.}~\bibnamefont{{Perevalov}}}, \bibinfo {author}
  {\bibfnamefont{C.~C.}\ \bibnamefont{{Polly}}}, \bibinfo {author}
  {\bibfnamefont{H.}~\bibnamefont{{Ray}}}, \bibinfo {author}
  {\bibfnamefont{B.~P.}\ \bibnamefont{{Roe}}}, \bibinfo {author}
  {\bibfnamefont{A.~D.}\ \bibnamefont{{Russell}}}, \bibinfo {author}
  {\bibfnamefont{R.}~\bibnamefont{{Schirato}}}, \bibinfo {author}
  {\bibfnamefont{M.~H.}\ \bibnamefont{{Shaevitz}}}, \bibinfo {author}
  {\bibfnamefont{M.}~\bibnamefont{{Sorel}}}, \bibinfo {author}
  {\bibfnamefont{J.}~\bibnamefont{{Spitz}}}, \bibinfo {author}
  {\bibfnamefont{I.}~\bibnamefont{{Stancu}}}, \bibinfo {author}
  {\bibfnamefont{R.~J.}\ \bibnamefont{{Stefanski}}}, \bibinfo {author}
  {\bibfnamefont{R.}~\bibnamefont{{Tayloe}}}, \bibinfo {author}
  {\bibfnamefont{M.}~\bibnamefont{{Tzanov}}}, \bibinfo {author}
  {\bibfnamefont{R.~G.}\ \bibnamefont{{van de Water}}}, \bibinfo {author}
  {\bibfnamefont{M.~O.}\ \bibnamefont{{Wascko}}}, \bibinfo {author}
  {\bibfnamefont{D.~H.}\ \bibnamefont{{White}}}, \bibinfo {author}
  {\bibfnamefont{M.~J.}\ \bibnamefont{{Wilking}}}, \bibinfo {author}
  {\bibfnamefont{G.~P.}\ \bibnamefont{{Zeller}}},\ and\ \bibinfo {author}
  {\bibfnamefont{E.~D.}\ \bibnamefont{{Zimmerman}}},\ }%
  \bibfield{journal}{%
  \Doi{10.1103/PhysRevLett.105.181801}{\bibinfo {journal} {Physical Review
  Letters}}\ }%
  \textbf{\bibinfo {volume} {105}},\ \bibinfo {pages} {181801} (\bibinfo
  {month} {Oct.}\ \bibinfo {year} {2010})%
  \bibAnnoteFile{NoStop}{aa2010}%
\bibitem{dw1994}%
  \BibitemOpen
  \bibfield{author}{%
  \bibinfo {author} {\bibfnamefont{S.}~\bibnamefont{{Dodelson}}}\ and\ \bibinfo
  {author} {\bibfnamefont{L.~M.}\ \bibnamefont{{Widrow}}},\ }%
  \bibfield{journal}{%
  \Doi{10.1103/PhysRevLett.72.17}{\bibinfo {journal} {Physical Review
  Letters}}\ }%
  \textbf{\bibinfo {volume} {72}},\ \bibinfo {pages} {17} (\bibinfo {month}
  {Jan.}\ \bibinfo {year} {1994}),\
  \Eprint{http://arxiv.org/abs/arXiv:hep-ph/9303287}{arXiv:hep-ph/9303287}%
  \bibAnnoteFile{NoStop}{dw1994}%
\bibitem{aba2006}%
  \BibitemOpen
  \bibfield{author}{%
  \bibinfo {author} {\bibfnamefont{K.}~\bibnamefont{{Abazajian}}},\ }%
  \bibfield{journal}{%
  \Doi{10.1103/PhysRevD.73.063513}{\bibinfo {journal} {\prd}}\ }%
  \textbf{\bibinfo {volume} {73}},\ \bibinfo {pages} {063513} (\bibinfo {month}
  {Mar.}\ \bibinfo {year} {2006}),\
  \Eprint{http://arxiv.org/abs/arXiv:astro-ph/0512631}{arXiv:astro-ph/0512631}%
  \bibAnnoteFile{NoStop}{aba2006}%
\bibitem{vie2005}%
  \BibitemOpen
  \bibfield{author}{%
  \bibinfo {author} {\bibfnamefont{M.}~\bibnamefont{{Viel}}}, \bibinfo {author}
  {\bibfnamefont{J.}~\bibnamefont{{Lesgourgues}}}, \bibinfo {author}
  {\bibfnamefont{M.~G.}\ \bibnamefont{{Haehnelt}}}, \bibinfo {author}
  {\bibfnamefont{S.}~\bibnamefont{{Matarrese}}},\ and\ \bibinfo {author}
  {\bibfnamefont{A.}~\bibnamefont{{Riotto}}},\ }%
  \bibfield{journal}{%
  \Doi{10.1103/PhysRevD.71.063534}{\bibinfo {journal} {\prd}}\ }%
  \textbf{\bibinfo {volume} {71}},\ \bibinfo {pages} {063534} (\bibinfo {month}
  {Mar.}\ \bibinfo {year} {2005}),\
  \Eprint{http://arxiv.org/abs/arXiv:astro-ph/0501562}{arXiv:astro-ph/0501562}%
  \bibAnnoteFile{NoStop}{vie2005}%
\bibitem{sf1999}%
  \BibitemOpen
  \bibfield{author}{%
  \bibinfo {author} {\bibfnamefont{X.}~\bibnamefont{{Shi}}}\ and\ \bibinfo
  {author} {\bibfnamefont{G.~M.}\ \bibnamefont{{Fuller}}},\ }%
  \bibfield{journal}{%
  \Doi{10.1103/PhysRevLett.82.2832}{\bibinfo {journal} {Physical Review
  Letters}}\ }%
  \textbf{\bibinfo {volume} {82}},\ \bibinfo {pages} {2832} (\bibinfo {month}
  {Apr.}\ \bibinfo {year} {1999}),\
  \Eprint{http://arxiv.org/abs/arXiv:astro-ph/9810076}{arXiv:astro-ph/9810076}%
  \bibAnnoteFile{NoStop}{sf1999}%
\bibitem{kus2006}%
  \BibitemOpen
  \bibfield{author}{%
  \bibinfo {author} {\bibfnamefont{A.}~\bibnamefont{{Kusenko}}},\ }%
  \bibfield{journal}{%
  \Doi{10.1103/PhysRevLett.97.241301}{\bibinfo {journal} {Physical Review
  Letters}}\ }%
  \textbf{\bibinfo {volume} {97}},\ \bibinfo {pages} {241301} (\bibinfo {month}
  {Dec.}\ \bibinfo {year} {2006}),\
  \Eprint{http://arxiv.org/abs/arXiv:hep-ph/0609081}{arXiv:hep-ph/0609081}%
  \bibAnnoteFile{NoStop}{kus2006}%
\bibitem{kam2000}%
  \BibitemOpen
  \bibfield{author}{%
  \bibinfo {author} {\bibfnamefont{M.}~\bibnamefont{{Kamionkowski}}}\ and\
  \bibinfo {author} {\bibfnamefont{A.~R.}\ \bibnamefont{{Liddle}}},\ }%
  \bibfield{journal}{%
  \Doi{10.1103/PhysRevLett.84.4525}{\bibinfo {journal} {Physical Review
  Letters}}\ }%
  \textbf{\bibinfo {volume} {84}},\ \bibinfo {pages} {4525} (\bibinfo {month}
  {May}\ \bibinfo {year} {2000}),\
  \Eprint{http://arxiv.org/abs/arXiv:astro-ph/9911103}{arXiv:astro-ph/9911103}%
  \bibAnnoteFile{NoStop}{kam2000}%
\bibitem{sig2004}%
  \BibitemOpen
  \bibfield{author}{%
  \bibinfo {author} {\bibfnamefont{K.}~\bibnamefont{{Sigurdson}}}\ and\
  \bibinfo {author} {\bibfnamefont{M.}~\bibnamefont{{Kamionkowski}}},\ }%
  \bibfield{journal}{%
  \Doi{10.1103/PhysRevLett.92.171302}{\bibinfo {journal} {Physical Review
  Letters}}\ }%
  \textbf{\bibinfo {volume} {92}},\ \bibinfo {pages} {171302} (\bibinfo {month}
  {Apr.}\ \bibinfo {year} {2004}),\
  \Eprint{http://arxiv.org/abs/arXiv:astro-ph/0311486}{arXiv:astro-ph/0311486}%
  \bibAnnoteFile{NoStop}{sig2004}%
\bibitem{kap2005}%
  \BibitemOpen
  \bibfield{author}{%
  \bibinfo {author} {\bibfnamefont{M.}~\bibnamefont{{Kaplinghat}}},\ }%
  \bibfield{journal}{%
  \Doi{10.1103/PhysRevD.72.063510}{\bibinfo {journal} {\prd}}\ }%
  \textbf{\bibinfo {volume} {72}},\ \bibinfo {pages} {063510} (\bibinfo {month}
  {Sep.}\ \bibinfo {year} {2005}),\
  \Eprint{http://arxiv.org/abs/arXiv:astro-ph/0507300}{arXiv:astro-ph/0507300}%
  \bibAnnoteFile{NoStop}{kap2005}%
\bibitem{khl2005}%
  \BibitemOpen
  \bibfield{author}{%
  \bibinfo {author} {\bibfnamefont{M.~Y.}\ \bibnamefont{{Khlopov}}},\ }%
  \bibfield{journal}{%
  \bibinfo {journal} {ArXiv Astrophysics e-prints}}%
   (\bibinfo {month} {Nov.}\ \bibinfo {year} {2005}),\
  \Eprint{http://arxiv.org/abs/arXiv:astro-ph/0511796}{arXiv:astro-ph/0511796}%
  \bibAnnoteFile{NoStop}{khl2005}%
\bibitem{khl2006}%
  \BibitemOpen
  \bibfield{author}{%
  \bibinfo {author} {\bibfnamefont{M.~Y.}\ \bibnamefont{{Khlopov}}},\ }%
  \bibfield{journal}{%
  \bibinfo {journal} {ArXiv Astrophysics e-prints}}%
   (\bibinfo {month} {Jul.}\ \bibinfo {year} {2006}),\
  \Eprint{http://arxiv.org/abs/arXiv:astro-ph/0607048}{arXiv:astro-ph/0607048}%
  \bibAnnoteFile{NoStop}{khl2006}%
\bibitem{bel2006A}%
  \BibitemOpen
  \bibfield{author}{%
  \bibinfo {author} {\bibfnamefont{K.}~\bibnamefont{{Belotsky}}}, \bibinfo
  {author} {\bibfnamefont{M.}~\bibnamefont{{Khlopov}}},\ and\ \bibinfo {author}
  {\bibfnamefont{K.}~\bibnamefont{{Shibaev}}},\ }%
  in\ \emph{\bibinfo {booktitle} {Particle Physics at the Year of
  250$^{\mbox{th}}$ Anniversary of Moscow University}},\ \bibinfo {editor}
  {edited by\ \bibinfo {editor} {\bibnamefont{{A.~Studenikin}}}}\ (\bibinfo
  {year} {2006})\ pp.\ \bibinfo {pages} {180--+},\
  \Eprint{http://arxiv.org/abs/arXiv:astro-ph/0602261}{arXiv:astro-ph/0602261}%
  \bibAnnoteFile{NoStop}{bel2006A}%
\bibitem{bel2006B}%
  \BibitemOpen
  \bibfield{author}{%
  \bibinfo {author} {\bibfnamefont{K.~M.}\ \bibnamefont{{Belotsky}}}, \bibinfo
  {author} {\bibfnamefont{M.~Y.}\ \bibnamefont{{Khlopov}}},\ and\ \bibinfo
  {author} {\bibfnamefont{K.~I.}\ \bibnamefont{{Shibaev}}},\ }%
  \bibfield{journal}{%
  \bibinfo {journal} {Gravitation and Cosmology}\ }%
  \textbf{\bibinfo {volume} {12}},\ \bibinfo {pages} {93} (\bibinfo {month}
  {Jun.}\ \bibinfo {year} {2006}),\
  \Eprint{http://arxiv.org/abs/arXiv:astro-ph/0604518}{arXiv:astro-ph/0604518}%
  \bibAnnoteFile{NoStop}{bel2006B}%
\bibitem{khl2007}%
  \BibitemOpen
  \bibfield{author}{%
  \bibinfo {author} {\bibfnamefont{M.~Y.}\ \bibnamefont{{Khlopov}}}\ and\
  \bibinfo {author} {\bibfnamefont{C.}~\bibnamefont{{Kouvaris}}},\ }%
  \bibfield{journal}{%
  \Doi{10.1103/PhysRevD.77.065002}{\bibinfo {journal} {\prd}}\ }%
  \textbf{\bibinfo {volume} {77}},\ \bibinfo {pages} {065002} (\bibinfo {month}
  {Mar.}\ \bibinfo {year} {2008}),\
  \Eprint{http://arxiv.org/abs/0710.2189}{arXiv:0710.2189}%
  \bibAnnoteFile{NoStop}{khl2007}%
\bibitem{khl2008A}%
  \BibitemOpen
  \bibfield{author}{%
  \bibinfo {author} {\bibfnamefont{M.~Y.}\ \bibnamefont{{Khlopov}}}\ and\
  \bibinfo {author} {\bibfnamefont{C.}~\bibnamefont{{Kouvaris}}},\ }%
  \bibfield{journal}{%
  \Doi{10.1103/PhysRevD.78.065040}{\bibinfo {journal} {\prd}}\ }%
  \textbf{\bibinfo {volume} {78}},\ \bibinfo {pages} {065040} (\bibinfo {month}
  {Sep.}\ \bibinfo {year} {2008}),\
  \Eprint{http://arxiv.org/abs/0806.1191}{arXiv:0806.1191}%
  \bibAnnoteFile{NoStop}{khl2008A}%
\bibitem{khl2008B}%
  \BibitemOpen
  \bibfield{author}{%
  \bibinfo {author} {\bibfnamefont{M.~Y.}\ \bibnamefont{{Khlopov}}},\ }%
  \bibfield{journal}{%
  \bibinfo {journal} {ArXiv e-prints}}%
   (\bibinfo {month} {Jun.}\ \bibinfo {year} {2008}),\
  \Eprint{http://arxiv.org/abs/0806.3581}{arXiv:0806.3581}%
  \bibAnnoteFile{NoStop}{khl2008B}%
\bibitem{eis1998}%
  \BibitemOpen
  \bibfield{author}{%
  \bibinfo {author} {\bibfnamefont{D.~J.}\ \bibnamefont{{Eisenstein}}}\ and\
  \bibinfo {author} {\bibfnamefont{P.}~\bibnamefont{{Hut}}},\ }%
  \bibfield{journal}{%
  \Doi{10.1086/305535}{\bibinfo {journal} {\apj}}\ }%
  \textbf{\bibinfo {volume} {498}},\ \bibinfo {pages} {137} (\bibinfo {month}
  {May}\ \bibinfo {year} {1998}),\
  \Eprint{http://arxiv.org/abs/arXiv:astro-ph/9712200}{arXiv:astro-ph/9712200}%
  \bibAnnoteFile{NoStop}{eis1998}%
\bibitem{kno2009}%
  \BibitemOpen
  \bibfield{author}{%
  \bibinfo {author} {\bibfnamefont{S.~R.}\ \bibnamefont{{Knollmann}}}\ and\
  \bibinfo {author} {\bibfnamefont{A.}~\bibnamefont{{Knebe}}},\ }%
  \bibfield{journal}{%
  \Doi{10.1088/0067-0049/182/2/608}{\bibinfo {journal} {Astrophys. J. Suppl.}}\
  }%
  \textbf{\bibinfo {volume} {182}},\ \bibinfo {pages} {608} (\bibinfo {month}
  {Jun.}\ \bibinfo {year} {2009}),\
  \Eprint{http://arxiv.org/abs/0904.3662}{arXiv:0904.3662}%
  \bibAnnoteFile{NoStop}{kno2009}%
\bibitem{eke1996}%
  \BibitemOpen
  \bibfield{author}{%
  \bibinfo {author} {\bibfnamefont{V.~R.}\ \bibnamefont{{Eke}}}, \bibinfo
  {author} {\bibfnamefont{S.}~\bibnamefont{{Cole}}},\ and\ \bibinfo {author}
  {\bibfnamefont{C.~S.}\ \bibnamefont{{Frenk}}},\ }%
  \bibfield{journal}{%
  \bibinfo {journal} {Mon. Not. R. Astron. Soc.}\ }%
  \textbf{\bibinfo {volume} {282}},\ \bibinfo {pages} {263} (\bibinfo {month}
  {Sep.}\ \bibinfo {year} {1996}),\
  \Eprint{http://arxiv.org/abs/arXiv:astro-ph/9601088}{arXiv:astro-ph/9601088}%
  \bibAnnoteFile{NoStop}{eke1996}%
\bibitem{bry1998}%
  \BibitemOpen
  \bibfield{author}{%
  \bibinfo {author} {\bibfnamefont{G.~L.}\ \bibnamefont{{Bryan}}}\ and\
  \bibinfo {author} {\bibfnamefont{M.~L.}\ \bibnamefont{{Norman}}},\ }%
  \bibfield{journal}{%
  \Doi{10.1086/305262}{\bibinfo {journal} {\apj}}\ }%
  \textbf{\bibinfo {volume} {495}},\ \bibinfo {pages} {80} (\bibinfo {month}
  {Mar.}\ \bibinfo {year} {1998}),\
  \Eprint{http://arxiv.org/abs/arXiv:astro-ph/9710107}{arXiv:astro-ph/9710107}%
  \bibAnnoteFile{NoStop}{bry1998}%
\bibitem{pow2003}%
  \BibitemOpen
  \bibfield{author}{%
  \bibinfo {author} {\bibfnamefont{C.}~\bibnamefont{{Power}}}, \bibinfo
  {author} {\bibfnamefont{J.~F.}\ \bibnamefont{{Navarro}}}, \bibinfo {author}
  {\bibfnamefont{A.}~\bibnamefont{{Jenkins}}}, \bibinfo {author}
  {\bibfnamefont{C.~S.}\ \bibnamefont{{Frenk}}}, \bibinfo {author}
  {\bibfnamefont{S.~D.~M.}\ \bibnamefont{{White}}}, \bibinfo {author}
  {\bibfnamefont{V.}~\bibnamefont{{Springel}}}, \bibinfo {author}
  {\bibfnamefont{J.}~\bibnamefont{{Stadel}}},\ and\ \bibinfo {author}
  {\bibfnamefont{T.}~\bibnamefont{{Quinn}}},\ }%
  \bibfield{journal}{%
  \Doi{10.1046/j.1365-8711.2003.05925.x}{\bibinfo {journal} {Mon. Not. R.
  Astron. Soc.}}\ }%
  \textbf{\bibinfo {volume} {338}},\ \bibinfo {pages} {14} (\bibinfo {month}
  {Jan.}\ \bibinfo {year} {2003}),\
  \Eprint{http://arxiv.org/abs/arXiv:astro-ph/0201544}{arXiv:astro-ph/0201544}%
  \bibAnnoteFile{NoStop}{pow2003}%
\bibitem{ghi2000}%
  \BibitemOpen
  \bibfield{author}{%
  \bibinfo {author} {\bibfnamefont{S.}~\bibnamefont{{Ghigna}}}, \bibinfo
  {author} {\bibfnamefont{B.}~\bibnamefont{{Moore}}}, \bibinfo {author}
  {\bibfnamefont{F.}~\bibnamefont{{Governato}}}, \bibinfo {author}
  {\bibfnamefont{G.}~\bibnamefont{{Lake}}}, \bibinfo {author}
  {\bibfnamefont{T.}~\bibnamefont{{Quinn}}},\ and\ \bibinfo {author}
  {\bibfnamefont{J.}~\bibnamefont{{Stadel}}},\ }%
  \bibfield{journal}{%
  \Doi{10.1086/317221}{\bibinfo {journal} {\apj}}\ }%
  \textbf{\bibinfo {volume} {544}},\ \bibinfo {pages} {616} (\bibinfo {month}
  {Dec.}\ \bibinfo {year} {2000}),\
  \Eprint{http://arxiv.org/abs/arXiv:astro-ph/9910166}{arXiv:astro-ph/9910166}%
  \bibAnnoteFile{NoStop}{ghi2000}%
\bibitem{hel2002}%
  \BibitemOpen
  \bibfield{author}{%
  \bibinfo {author} {\bibfnamefont{A.}~\bibnamefont{{Helmi}}}, \bibinfo
  {author} {\bibfnamefont{S.~D.}\ \bibnamefont{{White}}},\ and\ \bibinfo
  {author} {\bibfnamefont{V.}~\bibnamefont{{Springel}}},\ }%
  \bibfield{journal}{%
  \Doi{10.1103/PhysRevD.66.063502}{\bibinfo {journal} {\prd}}\ }%
  \textbf{\bibinfo {volume} {66}},\ \bibinfo {pages} {063502} (\bibinfo {month}
  {Sep.}\ \bibinfo {year} {2002}),\
  \Eprint{http://arxiv.org/abs/arXiv:astro-ph/0201289}{arXiv:astro-ph/0201289}%
  \bibAnnoteFile{NoStop}{hel2002}%
\bibitem{gao2004}%
  \BibitemOpen
  \bibfield{author}{%
  \bibinfo {author} {\bibfnamefont{L.}~\bibnamefont{{Gao}}}, \bibinfo {author}
  {\bibfnamefont{S.~D.~M.}\ \bibnamefont{{White}}}, \bibinfo {author}
  {\bibfnamefont{A.}~\bibnamefont{{Jenkins}}}, \bibinfo {author}
  {\bibfnamefont{F.}~\bibnamefont{{Stoehr}}},\ and\ \bibinfo {author}
  {\bibfnamefont{V.}~\bibnamefont{{Springel}}},\ }%
  \bibfield{journal}{%
  \Doi{10.1111/j.1365-2966.2004.08360.x}{\bibinfo {journal} {Mon. Not. R.
  Astron. Soc.}}\ }%
  \textbf{\bibinfo {volume} {355}},\ \bibinfo {pages} {819} (\bibinfo {month}
  {Dec.}\ \bibinfo {year} {2004}),\
  \Eprint{http://arxiv.org/abs/arXiv:astro-ph/0404589}{arXiv:astro-ph/0404589}%
  \bibAnnoteFile{NoStop}{gao2004}%
\bibitem{del2004}%
  \BibitemOpen
  \bibfield{author}{%
  \bibinfo {author} {\bibfnamefont{G.}~\bibnamefont{{De Lucia}}}, \bibinfo
  {author} {\bibfnamefont{G.}~\bibnamefont{{Kauffmann}}}, \bibinfo {author}
  {\bibfnamefont{V.}~\bibnamefont{{Springel}}}, \bibinfo {author}
  {\bibfnamefont{S.~D.~M.}\ \bibnamefont{{White}}}, \bibinfo {author}
  {\bibfnamefont{B.}~\bibnamefont{{Lanzoni}}}, \bibinfo {author}
  {\bibfnamefont{F.}~\bibnamefont{{Stoehr}}}, \bibinfo {author}
  {\bibfnamefont{G.}~\bibnamefont{{Tormen}}},\ and\ \bibinfo {author}
  {\bibfnamefont{N.}~\bibnamefont{{Yoshida}}},\ }%
  \bibfield{journal}{%
  \Doi{10.1111/j.1365-2966.2004.07372.x}{\bibinfo {journal} {Mon. Not. R.
  Astron. Soc.}}\ }%
  \textbf{\bibinfo {volume} {348}},\ \bibinfo {pages} {333} (\bibinfo {month}
  {Feb.}\ \bibinfo {year} {2004}),\
  \Eprint{http://arxiv.org/abs/arXiv:astro-ph/0306205}{arXiv:astro-ph/0306205}%
  \bibAnnoteFile{NoStop}{del2004}%
\bibitem{van2005}%
  \BibitemOpen
  \bibfield{author}{%
  \bibinfo {author} {\bibfnamefont{F.~C.}\ \bibnamefont{{van den Bosch}}},
  \bibinfo {author} {\bibfnamefont{G.}~\bibnamefont{{Tormen}}},\ and\ \bibinfo
  {author} {\bibfnamefont{C.}~\bibnamefont{{Giocoli}}},\ }%
  \bibfield{journal}{%
  \Doi{10.1111/j.1365-2966.2005.08964.x}{\bibinfo {journal} {Mon. Not. R.
  Astron. Soc.}}\ }%
  \textbf{\bibinfo {volume} {359}},\ \bibinfo {pages} {1029} (\bibinfo {month}
  {May}\ \bibinfo {year} {2005}),\
  \Eprint{http://arxiv.org/abs/arXiv:astro-ph/0409201}{arXiv:astro-ph/0409201}%
  \bibAnnoteFile{NoStop}{van2005}%
\bibitem{die2007}%
  \BibitemOpen
  \bibfield{author}{%
  \bibinfo {author} {\bibfnamefont{J.}~\bibnamefont{{Diemand}}}, \bibinfo
  {author} {\bibfnamefont{M.}~\bibnamefont{{Kuhlen}}},\ and\ \bibinfo {author}
  {\bibfnamefont{P.}~\bibnamefont{{Madau}}},\ }%
  \bibfield{journal}{%
  \Doi{10.1086/510736}{\bibinfo {journal} {\apj}}\ }%
  \textbf{\bibinfo {volume} {657}},\ \bibinfo {pages} {262} (\bibinfo {month}
  {Mar.}\ \bibinfo {year} {2007}),\
  \Eprint{http://arxiv.org/abs/arXiv:astro-ph/0611370}{arXiv:astro-ph/0611370}%
  \bibAnnoteFile{NoStop}{die2007}%
\bibitem{gio2008}%
  \BibitemOpen
  \bibfield{author}{%
  \bibinfo {author} {\bibfnamefont{C.}~\bibnamefont{{Giocoli}}}, \bibinfo
  {author} {\bibfnamefont{G.}~\bibnamefont{{Tormen}}},\ and\ \bibinfo {author}
  {\bibfnamefont{F.~C.}\ \bibnamefont{{van den Bosch}}},\ }%
  \bibfield{journal}{%
  \Doi{10.1111/j.1365-2966.2008.13182.x}{\bibinfo {journal} {Mon. Not. R.
  Astron. Soc.}}\ }%
  \textbf{\bibinfo {volume} {386}},\ \bibinfo {pages} {2135} (\bibinfo {month}
  {Jun.}\ \bibinfo {year} {2008}),\
  \Eprint{http://arxiv.org/abs/0712.1563}{arXiv:0712.1563}%
  \bibAnnoteFile{NoStop}{gio2008}%
\bibitem{spr2008}%
  \BibitemOpen
  \bibfield{author}{%
  \bibinfo {author} {\bibfnamefont{V.}~\bibnamefont{{Springel}}}, \bibinfo
  {author} {\bibfnamefont{J.}~\bibnamefont{{Wang}}}, \bibinfo {author}
  {\bibfnamefont{M.}~\bibnamefont{{Vogelsberger}}}, \bibinfo {author}
  {\bibfnamefont{A.}~\bibnamefont{{Ludlow}}}, \bibinfo {author}
  {\bibfnamefont{A.}~\bibnamefont{{Jenkins}}}, \bibinfo {author}
  {\bibfnamefont{A.}~\bibnamefont{{Helmi}}}, \bibinfo {author}
  {\bibfnamefont{J.~F.}\ \bibnamefont{{Navarro}}}, \bibinfo {author}
  {\bibfnamefont{C.~S.}\ \bibnamefont{{Frenk}}},\ and\ \bibinfo {author}
  {\bibfnamefont{S.~D.~M.}\ \bibnamefont{{White}}},\ }%
  \bibfield{journal}{%
  \Doi{10.1111/j.1365-2966.2008.14066.x}{\bibinfo {journal} {Mon. Not. R.
  Astron. Soc.}}\ }%
  \textbf{\bibinfo {volume} {391}},\ \bibinfo {pages} {1685} (\bibinfo {month}
  {Dec.}\ \bibinfo {year} {2008}),\
  \Eprint{http://arxiv.org/abs/0809.0898}{arXiv:0809.0898}%
  \bibAnnoteFile{NoStop}{spr2008}%
\bibitem{ish2009}%
  \BibitemOpen
  \bibfield{author}{%
  \bibinfo {author} {\bibfnamefont{T.}~\bibnamefont{{Ishiyama}}}, \bibinfo
  {author} {\bibfnamefont{T.}~\bibnamefont{{Fukushige}}},\ and\ \bibinfo
  {author} {\bibfnamefont{J.}~\bibnamefont{{Makino}}},\ }%
  \bibfield{journal}{%
  \Doi{10.1088/0004-637X/696/2/2115}{\bibinfo {journal} {\apj}}\ }%
  \textbf{\bibinfo {volume} {696}},\ \bibinfo {pages} {2115} (\bibinfo {month}
  {May}\ \bibinfo {year} {2009}),\
  \Eprint{http://arxiv.org/abs/0812.0683}{arXiv:0812.0683}%
  \bibAnnoteFile{NoStop}{ish2009}%
\bibitem{kly2010}%
  \BibitemOpen
  \bibfield{author}{%
  \bibinfo {author} {\bibfnamefont{A.}~\bibnamefont{{Klypin}}}, \bibinfo
  {author} {\bibfnamefont{S.}~\bibnamefont{{Trujillo-Gomez}}},\ and\ \bibinfo
  {author} {\bibfnamefont{J.}~\bibnamefont{{Primack}}},\ }%
  \bibfield{journal}{%
  \bibinfo {journal} {ArXiv e-prints}}%
   (\bibinfo {month} {Feb.}\ \bibinfo {year} {2010}),\
  \Eprint{http://arxiv.org/abs/1002.3660}{arXiv:1002.3660 [astro-ph.CO]}%
  \bibAnnoteFile{NoStop}{kly2010}%
\bibitem{jar2010}%
  \BibitemOpen
  \bibfield{author}{%
  \bibinfo {author} {\bibfnamefont{N.}~\bibnamefont{{Jarosik}}}, \bibinfo
  {author} {\bibfnamefont{C.~L.}\ \bibnamefont{{Bennett}}}, \bibinfo {author}
  {\bibfnamefont{J.}~\bibnamefont{{Dunkley}}}, \bibinfo {author}
  {\bibfnamefont{B.}~\bibnamefont{{Gold}}}, \bibinfo {author}
  {\bibfnamefont{M.~R.}\ \bibnamefont{{Greason}}}, \bibinfo {author}
  {\bibfnamefont{M.}~\bibnamefont{{Halpern}}}, \bibinfo {author}
  {\bibfnamefont{R.~S.}\ \bibnamefont{{Hill}}}, \bibinfo {author}
  {\bibfnamefont{G.}~\bibnamefont{{Hinshaw}}}, \bibinfo {author}
  {\bibfnamefont{A.}~\bibnamefont{{Kogut}}}, \bibinfo {author}
  {\bibfnamefont{E.}~\bibnamefont{{Komatsu}}}, \bibinfo {author}
  {\bibfnamefont{D.}~\bibnamefont{{Larson}}}, \bibinfo {author}
  {\bibfnamefont{M.}~\bibnamefont{{Limon}}}, \bibinfo {author}
  {\bibfnamefont{S.~S.}\ \bibnamefont{{Meyer}}}, \bibinfo {author}
  {\bibfnamefont{M.~R.}\ \bibnamefont{{Nolta}}}, \bibinfo {author}
  {\bibfnamefont{N.}~\bibnamefont{{Odegard}}}, \bibinfo {author}
  {\bibfnamefont{L.}~\bibnamefont{{Page}}}, \bibinfo {author}
  {\bibfnamefont{K.~M.}\ \bibnamefont{{Smith}}}, \bibinfo {author}
  {\bibfnamefont{D.~N.}\ \bibnamefont{{Spergel}}}, \bibinfo {author}
  {\bibfnamefont{G.~S.}\ \bibnamefont{{Tucker}}}, \bibinfo {author}
  {\bibfnamefont{J.~L.}\ \bibnamefont{{Weiland}}}, \bibinfo {author}
  {\bibfnamefont{E.}~\bibnamefont{{Wollack}}},\ and\ \bibinfo {author}
  {\bibfnamefont{E.~L.}\ \bibnamefont{{Wright}}},\ }%
  \bibfield{journal}{%
  \bibinfo {journal} {ArXiv e-prints}}%
   (\bibinfo {month} {Jan.}\ \bibinfo {year} {2010}),\
  \Eprint{http://arxiv.org/abs/1001.4744}{arXiv:1001.4744 [astro-ph.CO]}%
  \bibAnnoteFile{NoStop}{jar2010}%
\bibitem{bar2001}%
  \BibitemOpen
  \bibfield{author}{%
  \bibinfo {author} {\bibfnamefont{R.}~\bibnamefont{{Barkana}}}, \bibinfo
  {author} {\bibfnamefont{Z.}~\bibnamefont{{Haiman}}},\ and\ \bibinfo {author}
  {\bibfnamefont{J.~P.}\ \bibnamefont{{Ostriker}}},\ }%
  \bibfield{journal}{%
  \Doi{10.1086/322393}{\bibinfo {journal} {\apj}}\ }%
  \textbf{\bibinfo {volume} {558}},\ \bibinfo {pages} {482} (\bibinfo {month}
  {Sep.}\ \bibinfo {year} {2001}),\
  \Eprint{http://arxiv.org/abs/arXiv:astro-ph/0102304}{arXiv:astro-ph/0102304}%
  \bibAnnoteFile{NoStop}{bar2001}%
\bibitem{wal2009}%
  \BibitemOpen
  \bibfield{author}{%
  \bibinfo {author} {\bibfnamefont{S.~M.}\ \bibnamefont{{Walsh}}}, \bibinfo
  {author} {\bibfnamefont{B.}~\bibnamefont{{Willman}}},\ and\ \bibinfo {author}
  {\bibfnamefont{H.}~\bibnamefont{{Jerjen}}},\ }%
  \bibfield{journal}{%
  \Doi{10.1088/0004-6256/137/1/450}{\bibinfo {journal} {Astron. J.}}\ }%
  \textbf{\bibinfo {volume} {137}},\ \bibinfo {pages} {450} (\bibinfo {month}
  {Jan.}\ \bibinfo {year} {2009}),\
  \Eprint{http://arxiv.org/abs/0807.3345}{arXiv:0807.3345}%
  \bibAnnoteFile{NoStop}{wal2009}%
\bibitem{tol2008}%
  \BibitemOpen
  \bibfield{author}{%
  \bibinfo {author} {\bibfnamefont{E.~J.}\ \bibnamefont{{Tollerud}}}, \bibinfo
  {author} {\bibfnamefont{J.~S.}\ \bibnamefont{{Bullock}}}, \bibinfo {author}
  {\bibfnamefont{L.~E.}\ \bibnamefont{{Strigari}}},\ and\ \bibinfo {author}
  {\bibfnamefont{B.}~\bibnamefont{{Willman}}},\ }%
  \bibfield{journal}{%
  \Doi{10.1086/592102}{\bibinfo {journal} {\apj}}\ }%
  \textbf{\bibinfo {volume} {688}},\ \bibinfo {pages} {277} (\bibinfo {month}
  {Nov.}\ \bibinfo {year} {2008}),\
  \Eprint{http://arxiv.org/abs/0806.4381}{arXiv:0806.4381}%
  \bibAnnoteFile{NoStop}{tol2008}%
\bibitem{kop2008}%
  \BibitemOpen
  \bibfield{author}{%
  \bibinfo {author} {\bibfnamefont{S.}~\bibnamefont{{Koposov}}}, \bibinfo
  {author} {\bibfnamefont{V.}~\bibnamefont{{Belokurov}}}, \bibinfo {author}
  {\bibfnamefont{N.~W.}\ \bibnamefont{{Evans}}}, \bibinfo {author}
  {\bibfnamefont{P.~C.}\ \bibnamefont{{Hewett}}}, \bibinfo {author}
  {\bibfnamefont{M.~J.}\ \bibnamefont{{Irwin}}}, \bibinfo {author}
  {\bibfnamefont{G.}~\bibnamefont{{Gilmore}}}, \bibinfo {author}
  {\bibfnamefont{D.~B.}\ \bibnamefont{{Zucker}}}, \bibinfo {author}
  {\bibfnamefont{H.}~\bibnamefont{{Rix}}}, \bibinfo {author}
  {\bibfnamefont{M.}~\bibnamefont{{Fellhauer}}}, \bibinfo {author}
  {\bibfnamefont{E.~F.}\ \bibnamefont{{Bell}}},\ and\ \bibinfo {author}
  {\bibfnamefont{E.~V.}\ \bibnamefont{{Glushkova}}},\ }%
  \bibfield{journal}{%
  \Doi{10.1086/589911}{\bibinfo {journal} {\apj}}\ }%
  \textbf{\bibinfo {volume} {686}},\ \bibinfo {pages} {279} (\bibinfo {month}
  {Oct.}\ \bibinfo {year} {2008}),\
  \Eprint{http://arxiv.org/abs/0706.2687}{arXiv:0706.2687}%
  \bibAnnoteFile{NoStop}{kop2008}%
\bibitem{mat1998}%
  \BibitemOpen
  \bibfield{author}{%
  \bibinfo {author} {\bibfnamefont{M.~L.}\ \bibnamefont{{Mateo}}},\ }%
  \bibfield{journal}{%
  \Doi{10.1146/annurev.astro.36.1.435}{\bibinfo {journal} {Ann. Rev. Astron.
  Astrophys.}}\ }%
  \textbf{\bibinfo {volume} {36}},\ \bibinfo {pages} {435} (\bibinfo {year}
  {1998}),\
  \Eprint{http://arxiv.org/abs/arXiv:astro-ph/9810070}{arXiv:astro-ph/9810070}%
  \bibAnnoteFile{NoStop}{mat1998}%
\bibitem{geh2009}%
  \BibitemOpen
  \bibfield{author}{%
  \bibinfo {author} {\bibfnamefont{M.}~\bibnamefont{{Geha}}}, \bibinfo {author}
  {\bibfnamefont{B.}~\bibnamefont{{Willman}}}, \bibinfo {author}
  {\bibfnamefont{J.~D.}\ \bibnamefont{{Simon}}}, \bibinfo {author}
  {\bibfnamefont{L.~E.}\ \bibnamefont{{Strigari}}}, \bibinfo {author}
  {\bibfnamefont{E.~N.}\ \bibnamefont{{Kirby}}}, \bibinfo {author}
  {\bibfnamefont{D.~R.}\ \bibnamefont{{Law}}},\ and\ \bibinfo {author}
  {\bibfnamefont{J.}~\bibnamefont{{Strader}}},\ }%
  \bibfield{journal}{%
  \Doi{10.1088/0004-637X/692/2/1464}{\bibinfo {journal} {\apj}}\ }%
  \textbf{\bibinfo {volume} {692}},\ \bibinfo {pages} {1464} (\bibinfo {month}
  {Feb.}\ \bibinfo {year} {2009}),\
  \Eprint{http://arxiv.org/abs/0809.2781}{arXiv:0809.2781}%
  \bibAnnoteFile{NoStop}{geh2009}%
\bibitem{mar2007}%
  \BibitemOpen
  \bibfield{author}{%
  \bibinfo {author} {\bibfnamefont{N.~F.}\ \bibnamefont{{Martin}}}, \bibinfo
  {author} {\bibfnamefont{R.~A.}\ \bibnamefont{{Ibata}}}, \bibinfo {author}
  {\bibfnamefont{S.~C.}\ \bibnamefont{{Chapman}}}, \bibinfo {author}
  {\bibfnamefont{M.}~\bibnamefont{{Irwin}}},\ and\ \bibinfo {author}
  {\bibfnamefont{G.~F.}\ \bibnamefont{{Lewis}}},\ }%
  \bibfield{journal}{%
  \Doi{10.1111/j.1365-2966.2007.12055.x}{\bibinfo {journal} {Mon. Not. R.
  Astron. Soc.}}\ }%
  \textbf{\bibinfo {volume} {380}},\ \bibinfo {pages} {281} (\bibinfo {month}
  {Sep.}\ \bibinfo {year} {2007}),\
  \Eprint{http://arxiv.org/abs/0705.4622}{arXiv:0705.4622}%
  \bibAnnoteFile{NoStop}{mar2007}%
\bibitem{sim2007}%
  \BibitemOpen
  \bibfield{author}{%
  \bibinfo {author} {\bibfnamefont{J.~D.}\ \bibnamefont{{Simon}}}\ and\
  \bibinfo {author} {\bibfnamefont{M.}~\bibnamefont{{Geha}}},\ }%
  \bibfield{journal}{%
  \Doi{10.1086/521816}{\bibinfo {journal} {\apj}}\ }%
  \textbf{\bibinfo {volume} {670}},\ \bibinfo {pages} {313} (\bibinfo {month}
  {Nov.}\ \bibinfo {year} {2007}),\
  \Eprint{http://arxiv.org/abs/0706.0516}{arXiv:0706.0516}%
  \bibAnnoteFile{NoStop}{sim2007}%
\bibitem{bel2009}%
  \BibitemOpen
  \bibfield{author}{%
  \bibinfo {author} {\bibfnamefont{V.}~\bibnamefont{{Belokurov}}}, \bibinfo
  {author} {\bibfnamefont{M.~G.}\ \bibnamefont{{Walker}}}, \bibinfo {author}
  {\bibfnamefont{N.~W.}\ \bibnamefont{{Evans}}}, \bibinfo {author}
  {\bibfnamefont{G.}~\bibnamefont{{Gilmore}}}, \bibinfo {author}
  {\bibfnamefont{M.~J.}\ \bibnamefont{{Irwin}}}, \bibinfo {author}
  {\bibfnamefont{M.}~\bibnamefont{{Mateo}}}, \bibinfo {author}
  {\bibfnamefont{L.}~\bibnamefont{{Mayer}}}, \bibinfo {author}
  {\bibfnamefont{E.}~\bibnamefont{{Olszewski}}}, \bibinfo {author}
  {\bibfnamefont{J.}~\bibnamefont{{Bechtold}}},\ and\ \bibinfo {author}
  {\bibfnamefont{T.}~\bibnamefont{{Pickering}}},\ }%
  \bibfield{journal}{%
  \Doi{10.1111/j.1365-2966.2009.15106.x}{\bibinfo {journal} {Mon. Not. R.
  Astron. Soc.}}\ }%
  \textbf{\bibinfo {volume} {397}},\ \bibinfo {pages} {1748} (\bibinfo {month}
  {Aug.}\ \bibinfo {year} {2009}),\
  \Eprint{http://arxiv.org/abs/0903.0818}{arXiv:0903.0818}%
  \bibAnnoteFile{NoStop}{bel2009}%
\bibitem{wil2005}%
  \BibitemOpen
  \bibfield{author}{%
  \bibinfo {author} {\bibfnamefont{B.}~\bibnamefont{{Willman}}}, \bibinfo
  {author} {\bibfnamefont{M.~R.}\ \bibnamefont{{Blanton}}}, \bibinfo {author}
  {\bibfnamefont{A.~A.}\ \bibnamefont{{West}}}, \bibinfo {author}
  {\bibfnamefont{J.~J.}\ \bibnamefont{{Dalcanton}}}, \bibinfo {author}
  {\bibfnamefont{D.~W.}\ \bibnamefont{{Hogg}}}, \bibinfo {author}
  {\bibfnamefont{D.~P.}\ \bibnamefont{{Schneider}}}, \bibinfo {author}
  {\bibfnamefont{N.}~\bibnamefont{{Wherry}}}, \bibinfo {author}
  {\bibfnamefont{B.}~\bibnamefont{{Yanny}}},\ and\ \bibinfo {author}
  {\bibfnamefont{J.}~\bibnamefont{{Brinkmann}}},\ }%
  \bibfield{journal}{%
  \Doi{10.1086/430214}{\bibinfo {journal} {Astron. J.}}\ }%
  \textbf{\bibinfo {volume} {129}},\ \bibinfo {pages} {2692} (\bibinfo {month}
  {Jun.}\ \bibinfo {year} {2005}),\
  \Eprint{http://arxiv.org/abs/arXiv:astro-ph/0410416}{arXiv:astro-ph/0410416}%
  \bibAnnoteFile{NoStop}{wil2005}%
\bibitem{bel2007}%
  \BibitemOpen
  \bibfield{author}{%
  \bibinfo {author} {\bibfnamefont{V.}~\bibnamefont{{Belokurov}}}, \bibinfo
  {author} {\bibfnamefont{D.~B.}\ \bibnamefont{{Zucker}}}, \bibinfo {author}
  {\bibfnamefont{N.~W.}\ \bibnamefont{{Evans}}}, \bibinfo {author}
  {\bibfnamefont{J.~T.}\ \bibnamefont{{Kleyna}}}, \bibinfo {author}
  {\bibfnamefont{S.}~\bibnamefont{{Koposov}}}, \bibinfo {author}
  {\bibfnamefont{S.~T.}\ \bibnamefont{{Hodgkin}}}, \bibinfo {author}
  {\bibfnamefont{M.~J.}\ \bibnamefont{{Irwin}}}, \bibinfo {author}
  {\bibfnamefont{G.}~\bibnamefont{{Gilmore}}}, \bibinfo {author}
  {\bibfnamefont{M.~I.}\ \bibnamefont{{Wilkinson}}}, \bibinfo {author}
  {\bibfnamefont{M.}~\bibnamefont{{Fellhauer}}}, \bibinfo {author}
  {\bibfnamefont{D.~M.}\ \bibnamefont{{Bramich}}}, \bibinfo {author}
  {\bibfnamefont{P.~C.}\ \bibnamefont{{Hewett}}}, \bibinfo {author}
  {\bibfnamefont{S.}~\bibnamefont{{Vidrih}}}, \bibinfo {author}
  {\bibfnamefont{J.~T.~A.}\ \bibnamefont{{De Jong}}}, \bibinfo {author}
  {\bibfnamefont{J.~A.}\ \bibnamefont{{Smith}}}, \bibinfo {author}
  {\bibfnamefont{H.}~\bibnamefont{{Rix}}}, \bibinfo {author}
  {\bibfnamefont{E.~F.}\ \bibnamefont{{Bell}}}, \bibinfo {author}
  {\bibfnamefont{R.~F.~G.}\ \bibnamefont{{Wyse}}}, \bibinfo {author}
  {\bibfnamefont{H.~J.}\ \bibnamefont{{Newberg}}}, \bibinfo {author}
  {\bibfnamefont{P.~A.}\ \bibnamefont{{Mayeur}}}, \bibinfo {author}
  {\bibfnamefont{B.}~\bibnamefont{{Yanny}}}, \bibinfo {author}
  {\bibfnamefont{C.~M.}\ \bibnamefont{{Rockosi}}}, \bibinfo {author}
  {\bibfnamefont{O.~Y.}\ \bibnamefont{{Gnedin}}}, \bibinfo {author}
  {\bibfnamefont{D.~P.}\ \bibnamefont{{Schneider}}}, \bibinfo {author}
  {\bibfnamefont{T.~C.}\ \bibnamefont{{Beers}}}, \bibinfo {author}
  {\bibfnamefont{J.~C.}\ \bibnamefont{{Barentine}}}, \bibinfo {author}
  {\bibfnamefont{H.}~\bibnamefont{{Brewington}}}, \bibinfo {author}
  {\bibfnamefont{J.}~\bibnamefont{{Brinkmann}}}, \bibinfo {author}
  {\bibfnamefont{M.}~\bibnamefont{{Harvanek}}}, \bibinfo {author}
  {\bibfnamefont{S.~J.}\ \bibnamefont{{Kleinman}}}, \bibinfo {author}
  {\bibfnamefont{J.}~\bibnamefont{{Krzesinski}}}, \bibinfo {author}
  {\bibfnamefont{D.}~\bibnamefont{{Long}}}, \bibinfo {author}
  {\bibfnamefont{A.}~\bibnamefont{{Nitta}}},\ and\ \bibinfo {author}
  {\bibfnamefont{S.~A.}\ \bibnamefont{{Snedden}}},\ }%
  \bibfield{journal}{%
  \Doi{10.1086/509718}{\bibinfo {journal} {\apj}}\ }%
  \textbf{\bibinfo {volume} {654}},\ \bibinfo {pages} {897} (\bibinfo {month}
  {Jan.}\ \bibinfo {year} {2007}),\
  \Eprint{http://arxiv.org/abs/arXiv:astro-ph/0608448}{arXiv:astro-ph/0608448}%
  \bibAnnoteFile{NoStop}{bel2007}%
\bibitem{wal2007}%
  \BibitemOpen
  \bibfield{author}{%
  \bibinfo {author} {\bibfnamefont{S.~M.}\ \bibnamefont{{Walsh}}}, \bibinfo
  {author} {\bibfnamefont{H.}~\bibnamefont{{Jerjen}}},\ and\ \bibinfo {author}
  {\bibfnamefont{B.}~\bibnamefont{{Willman}}},\ }%
  \bibfield{journal}{%
  \Doi{10.1086/519684}{\bibinfo {journal} {Astrophys. J. Lett.}}\ }%
  \textbf{\bibinfo {volume} {662}},\ \bibinfo {pages} {L83} (\bibinfo {month}
  {Jun.}\ \bibinfo {year} {2007}),\
  \Eprint{http://arxiv.org/abs/0705.1378}{arXiv:0705.1378}%
  \bibAnnoteFile{NoStop}{wal2007}%
\bibitem{wat2009}%
  \BibitemOpen
  \bibfield{author}{%
  \bibinfo {author} {\bibfnamefont{L.~L.}\ \bibnamefont{{Watkins}}}, \bibinfo
  {author} {\bibfnamefont{N.~W.}\ \bibnamefont{{Evans}}}, \bibinfo {author}
  {\bibfnamefont{V.}~\bibnamefont{{Belokurov}}}, \bibinfo {author}
  {\bibfnamefont{M.~C.}\ \bibnamefont{{Smith}}}, \bibinfo {author}
  {\bibfnamefont{P.~C.}\ \bibnamefont{{Hewett}}}, \bibinfo {author}
  {\bibfnamefont{D.~M.}\ \bibnamefont{{Bramich}}}, \bibinfo {author}
  {\bibfnamefont{G.~F.}\ \bibnamefont{{Gilmore}}}, \bibinfo {author}
  {\bibfnamefont{M.~J.}\ \bibnamefont{{Irwin}}}, \bibinfo {author}
  {\bibfnamefont{S.}~\bibnamefont{{Vidrih}}}, \bibinfo {author}
  {\bibfnamefont{{\L}.}~\bibnamefont{{Wyrzykowski}}},\ and\ \bibinfo {author}
  {\bibfnamefont{D.~B.}\ \bibnamefont{{Zucker}}},\ }%
  \bibfield{journal}{%
  \Doi{10.1111/j.1365-2966.2009.15242.x}{\bibinfo {journal} {Mon. Not. R.
  Astron. Soc.}}\ }%
  \textbf{\bibinfo {volume} {398}},\ \bibinfo {pages} {1757} (\bibinfo {month}
  {Oct.}\ \bibinfo {year} {2009}),\
  \Eprint{http://arxiv.org/abs/0906.0498}{arXiv:0906.0498}%
  \bibAnnoteFile{NoStop}{wat2009}%
\bibitem{kol2009}%
  \BibitemOpen
  \bibfield{author}{%
  \bibinfo {author} {\bibfnamefont{J.~A.}\ \bibnamefont{{Kollmeier}}}, \bibinfo
  {author} {\bibfnamefont{A.}~\bibnamefont{{Gould}}}, \bibinfo {author}
  {\bibfnamefont{S.}~\bibnamefont{{Shectman}}}, \bibinfo {author}
  {\bibfnamefont{I.~B.}\ \bibnamefont{{Thompson}}}, \bibinfo {author}
  {\bibfnamefont{G.~W.}\ \bibnamefont{{Preston}}}, \bibinfo {author}
  {\bibfnamefont{J.~D.}\ \bibnamefont{{Simon}}}, \bibinfo {author}
  {\bibfnamefont{J.~D.}\ \bibnamefont{{Crane}}}, \bibinfo {author}
  {\bibfnamefont{{\v Z}.}~\bibnamefont{{Ivezi{\'c}}}},\ and\ \bibinfo {author}
  {\bibfnamefont{B.}~\bibnamefont{{Sesar}}},\ }%
  \bibfield{journal}{%
  \Doi{10.1088/0004-637X/705/2/L158}{\bibinfo {journal} {Astrophys. J. Lett.}}\
  }%
  \textbf{\bibinfo {volume} {705}},\ \bibinfo {pages} {L158} (\bibinfo {month}
  {Nov.}\ \bibinfo {year} {2009}),\
  \Eprint{http://arxiv.org/abs/0908.1381}{arXiv:0908.1381}%
  \bibAnnoteFile{NoStop}{kol2009}%
\bibitem{dej2009}%
  \BibitemOpen
  \bibfield{author}{%
  \bibinfo {author} {\bibfnamefont{J.~T.~A.}\ \bibnamefont{{de Jong}}},
  \bibinfo {author} {\bibfnamefont{N.~F.}\ \bibnamefont{{Martin}}}, \bibinfo
  {author} {\bibfnamefont{H.}~\bibnamefont{{Rix}}}, \bibinfo {author}
  {\bibfnamefont{K.~W.}\ \bibnamefont{{Smith}}}, \bibinfo {author}
  {\bibfnamefont{S.}~\bibnamefont{{Jin}}},\ and\ \bibinfo {author}
  {\bibfnamefont{A.~V.}\ \bibnamefont{{Maccio'}}},\ }%
  \bibfield{journal}{%
  \bibinfo {journal} {ArXiv e-prints}}%
   (\bibinfo {month} {Dec.}\ \bibinfo {year} {2009}),\
  \Eprint{http://arxiv.org/abs/0912.3251}{arXiv:0912.3251}%
  \bibAnnoteFile{NoStop}{dej2009}%
\bibitem{bel2008}%
  \BibitemOpen
  \bibfield{author}{%
  \bibinfo {author} {\bibfnamefont{V.}~\bibnamefont{{Belokurov}}}, \bibinfo
  {author} {\bibfnamefont{M.~G.}\ \bibnamefont{{Walker}}}, \bibinfo {author}
  {\bibfnamefont{N.~W.}\ \bibnamefont{{Evans}}}, \bibinfo {author}
  {\bibfnamefont{D.~C.}\ \bibnamefont{{Faria}}}, \bibinfo {author}
  {\bibfnamefont{G.}~\bibnamefont{{Gilmore}}}, \bibinfo {author}
  {\bibfnamefont{M.~J.}\ \bibnamefont{{Irwin}}}, \bibinfo {author}
  {\bibfnamefont{S.}~\bibnamefont{{Koposov}}}, \bibinfo {author}
  {\bibfnamefont{M.}~\bibnamefont{{Mateo}}}, \bibinfo {author}
  {\bibfnamefont{E.}~\bibnamefont{{Olszewski}}},\ and\ \bibinfo {author}
  {\bibfnamefont{D.~B.}\ \bibnamefont{{Zucker}}},\ }%
  \bibfield{journal}{%
  \Doi{10.1086/592962}{\bibinfo {journal} {Astrophys. J. Lett.}}\ }%
  \textbf{\bibinfo {volume} {686}},\ \bibinfo {pages} {L83} (\bibinfo {month}
  {Oct.}\ \bibinfo {year} {2008}),\
  \Eprint{http://arxiv.org/abs/0807.2831}{arXiv:0807.2831}%
  \bibAnnoteFile{NoStop}{bel2008}%
\bibitem{bel2010}%
  \BibitemOpen
  \bibfield{author}{%
  \bibinfo {author} {\bibfnamefont{V.}~\bibnamefont{{Belokurov}}}, \bibinfo
  {author} {\bibfnamefont{M.~G.}\ \bibnamefont{{Walker}}}, \bibinfo {author}
  {\bibfnamefont{N.~W.}\ \bibnamefont{{Evans}}}, \bibinfo {author}
  {\bibfnamefont{G.}~\bibnamefont{{Gilmore}}}, \bibinfo {author}
  {\bibfnamefont{M.~J.}\ \bibnamefont{{Irwin}}}, \bibinfo {author}
  {\bibfnamefont{D.}~\bibnamefont{{Just}}}, \bibinfo {author}
  {\bibfnamefont{S.}~\bibnamefont{{Koposov}}}, \bibinfo {author}
  {\bibfnamefont{M.}~\bibnamefont{{Mateo}}}, \bibinfo {author}
  {\bibfnamefont{E.}~\bibnamefont{{Olszewski}}}, \bibinfo {author}
  {\bibfnamefont{L.}~\bibnamefont{{Watkins}}},\ and\ \bibinfo {author}
  {\bibfnamefont{L.}~\bibnamefont{{Wyrzykowski}}},\ }%
  \bibfield{journal}{%
  \Doi{10.1088/2041-8205/712/1/L103}{\bibinfo {journal} {Astrophys. J. Lett.}}\
  }%
  \textbf{\bibinfo {volume} {712}},\ \bibinfo {pages} {L103} (\bibinfo {month}
  {Mar.}\ \bibinfo {year} {2010}),\
  \Eprint{http://arxiv.org/abs/1002.0504}{arXiv:1002.0504}%
  \bibAnnoteFile{NoStop}{bel2010}%
\bibitem{zuc2006}%
  \BibitemOpen
  \bibfield{author}{%
  \bibinfo {author} {\bibfnamefont{D.~B.}\ \bibnamefont{{Zucker}}}, \bibinfo
  {author} {\bibfnamefont{V.}~\bibnamefont{{Belokurov}}}, \bibinfo {author}
  {\bibfnamefont{N.~W.}\ \bibnamefont{{Evans}}}, \bibinfo {author}
  {\bibfnamefont{M.~I.}\ \bibnamefont{{Wilkinson}}}, \bibinfo {author}
  {\bibfnamefont{M.~J.}\ \bibnamefont{{Irwin}}}, \bibinfo {author}
  {\bibfnamefont{T.}~\bibnamefont{{Sivarani}}}, \bibinfo {author}
  {\bibfnamefont{S.}~\bibnamefont{{Hodgkin}}}, \bibinfo {author}
  {\bibfnamefont{D.~M.}\ \bibnamefont{{Bramich}}}, \bibinfo {author}
  {\bibfnamefont{J.~M.}\ \bibnamefont{{Irwin}}}, \bibinfo {author}
  {\bibfnamefont{G.}~\bibnamefont{{Gilmore}}}, \bibinfo {author}
  {\bibfnamefont{B.}~\bibnamefont{{Willman}}}, \bibinfo {author}
  {\bibfnamefont{S.}~\bibnamefont{{Vidrih}}}, \bibinfo {author}
  {\bibfnamefont{M.}~\bibnamefont{{Fellhauer}}}, \bibinfo {author}
  {\bibfnamefont{P.~C.}\ \bibnamefont{{Hewett}}}, \bibinfo {author}
  {\bibfnamefont{T.~C.}\ \bibnamefont{{Beers}}}, \bibinfo {author}
  {\bibfnamefont{E.~F.}\ \bibnamefont{{Bell}}}, \bibinfo {author}
  {\bibfnamefont{E.~K.}\ \bibnamefont{{Grebel}}}, \bibinfo {author}
  {\bibfnamefont{D.~P.}\ \bibnamefont{{Schneider}}}, \bibinfo {author}
  {\bibfnamefont{H.~J.}\ \bibnamefont{{Newberg}}}, \bibinfo {author}
  {\bibfnamefont{R.~F.~G.}\ \bibnamefont{{Wyse}}}, \bibinfo {author}
  {\bibfnamefont{C.~M.}\ \bibnamefont{{Rockosi}}}, \bibinfo {author}
  {\bibfnamefont{B.}~\bibnamefont{{Yanny}}}, \bibinfo {author}
  {\bibfnamefont{R.}~\bibnamefont{{Lupton}}}, \bibinfo {author}
  {\bibfnamefont{J.~A.}\ \bibnamefont{{Smith}}}, \bibinfo {author}
  {\bibfnamefont{J.~C.}\ \bibnamefont{{Barentine}}}, \bibinfo {author}
  {\bibfnamefont{H.}~\bibnamefont{{Brewington}}}, \bibinfo {author}
  {\bibfnamefont{J.}~\bibnamefont{{Brinkmann}}}, \bibinfo {author}
  {\bibfnamefont{M.}~\bibnamefont{{Harvanek}}}, \bibinfo {author}
  {\bibfnamefont{S.~J.}\ \bibnamefont{{Kleinman}}}, \bibinfo {author}
  {\bibfnamefont{J.}~\bibnamefont{{Krzesinski}}}, \bibinfo {author}
  {\bibfnamefont{D.}~\bibnamefont{{Long}}}, \bibinfo {author}
  {\bibfnamefont{A.}~\bibnamefont{{Nitta}}},\ and\ \bibinfo {author}
  {\bibfnamefont{S.~A.}\ \bibnamefont{{Snedden}}},\ }%
  \bibfield{journal}{%
  \Doi{10.1086/505216}{\bibinfo {journal} {Astrophys. J. Lett.}}\ }%
  \textbf{\bibinfo {volume} {643}},\ \bibinfo {pages} {L103} (\bibinfo {month}
  {Jun.}\ \bibinfo {year} {2006}),\
  \Eprint{http://arxiv.org/abs/arXiv:astro-ph/0604354}{arXiv:astro-ph/0604354}%
  \bibAnnoteFile{NoStop}{zuc2006}%
\bibitem{irw2007}%
  \BibitemOpen
  \bibfield{author}{%
  \bibinfo {author} {\bibfnamefont{M.~J.}\ \bibnamefont{{Irwin}}}, \bibinfo
  {author} {\bibfnamefont{V.}~\bibnamefont{{Belokurov}}}, \bibinfo {author}
  {\bibfnamefont{N.~W.}\ \bibnamefont{{Evans}}}, \bibinfo {author}
  {\bibfnamefont{E.~V.}\ \bibnamefont{{Ryan-Weber}}}, \bibinfo {author}
  {\bibfnamefont{J.~T.~A.}\ \bibnamefont{{de Jong}}}, \bibinfo {author}
  {\bibfnamefont{S.}~\bibnamefont{{Koposov}}}, \bibinfo {author}
  {\bibfnamefont{D.~B.}\ \bibnamefont{{Zucker}}}, \bibinfo {author}
  {\bibfnamefont{S.~T.}\ \bibnamefont{{Hodgkin}}}, \bibinfo {author}
  {\bibfnamefont{G.}~\bibnamefont{{Gilmore}}}, \bibinfo {author}
  {\bibfnamefont{P.}~\bibnamefont{{Prema}}}, \bibinfo {author}
  {\bibfnamefont{L.}~\bibnamefont{{Hebb}}}, \bibinfo {author}
  {\bibfnamefont{A.}~\bibnamefont{{Begum}}}, \bibinfo {author}
  {\bibfnamefont{M.}~\bibnamefont{{Fellhauer}}}, \bibinfo {author}
  {\bibfnamefont{P.~C.}\ \bibnamefont{{Hewett}}}, \bibinfo {author}
  {\bibfnamefont{R.~C.}\ \bibnamefont{{Kennicutt}}, \bibfnamefont{Jr.}},
  \bibinfo {author} {\bibfnamefont{M.~I.}\ \bibnamefont{{Wilkinson}}}, \bibinfo
  {author} {\bibfnamefont{D.~M.}\ \bibnamefont{{Bramich}}}, \bibinfo {author}
  {\bibfnamefont{S.}~\bibnamefont{{Vidrih}}}, \bibinfo {author}
  {\bibfnamefont{H.}~\bibnamefont{{Rix}}}, \bibinfo {author}
  {\bibfnamefont{T.~C.}\ \bibnamefont{{Beers}}}, \bibinfo {author}
  {\bibfnamefont{J.~C.}\ \bibnamefont{{Barentine}}}, \bibinfo {author}
  {\bibfnamefont{H.}~\bibnamefont{{Brewington}}}, \bibinfo {author}
  {\bibfnamefont{M.}~\bibnamefont{{Harvanek}}}, \bibinfo {author}
  {\bibfnamefont{J.}~\bibnamefont{{Krzesinski}}}, \bibinfo {author}
  {\bibfnamefont{D.}~\bibnamefont{{Long}}}, \bibinfo {author}
  {\bibfnamefont{A.}~\bibnamefont{{Nitta}}},\ and\ \bibinfo {author}
  {\bibfnamefont{S.~A.}\ \bibnamefont{{Snedden}}},\ }%
  \bibfield{journal}{%
  \Doi{10.1086/512183}{\bibinfo {journal} {Astrophys. J. Lett.}}\ }%
  \textbf{\bibinfo {volume} {656}},\ \bibinfo {pages} {L13} (\bibinfo {month}
  {Feb.}\ \bibinfo {year} {2007}),\
  \Eprint{http://arxiv.org/abs/arXiv:astro-ph/0701154}{arXiv:astro-ph/0701154}%
  \bibAnnoteFile{NoStop}{irw2007}%
\bibitem{ric2005}%
  \BibitemOpen
  \bibfield{author}{%
  \bibinfo {author} {\bibfnamefont{M.}~\bibnamefont{{Ricotti}}}\ and\ \bibinfo
  {author} {\bibfnamefont{N.~Y.}\ \bibnamefont{{Gnedin}}},\ }%
  \bibfield{journal}{%
  \Doi{10.1086/431415}{\bibinfo {journal} {\apj}}\ }%
  \textbf{\bibinfo {volume} {629}},\ \bibinfo {pages} {259} (\bibinfo {month}
  {Aug.}\ \bibinfo {year} {2005}),\
  \Eprint{http://arxiv.org/abs/arXiv:astro-ph/0408563}{arXiv:astro-ph/0408563}%
  \bibAnnoteFile{NoStop}{ric2005}%
\bibitem{ric2008}%
  \BibitemOpen
  \bibfield{author}{%
  \bibinfo {author} {\bibfnamefont{M.}~\bibnamefont{{Ricotti}}}, \bibinfo
  {author} {\bibfnamefont{N.~Y.}\ \bibnamefont{{Gnedin}}},\ and\ \bibinfo
  {author} {\bibfnamefont{J.~M.}\ \bibnamefont{{Shull}}},\ }%
  \bibfield{journal}{%
  \Doi{10.1086/590901}{\bibinfo {journal} {\apj}}\ }%
  \textbf{\bibinfo {volume} {685}},\ \bibinfo {pages} {21} (\bibinfo {month}
  {Sep.}\ \bibinfo {year} {2008}),\
  \Eprint{http://arxiv.org/abs/0802.2715}{arXiv:0802.2715}%
  \bibAnnoteFile{NoStop}{ric2008}%
\bibitem{ric2010}%
  \BibitemOpen
  \bibfield{author}{%
  \bibinfo {author} {\bibfnamefont{M.}~\bibnamefont{{Ricotti}}},\ }%
  \bibfield{journal}{%
  \bibinfo {journal} {Advances in Astronomy}\ }%
  \textbf{\bibinfo {volume} {2010}} (\bibinfo {year} {2010}),\ \doi{\bibinfo
  {doi} {10.1155/2010/271592}},\
  \Eprint{http://arxiv.org/abs/0911.2792}{arXiv:0911.2792}%
  \bibAnnoteFile{NoStop}{ric2010}%
\bibitem{bov2009}%
  \BibitemOpen
  \bibfield{author}{%
  \bibinfo {author} {\bibfnamefont{M.~S.}\ \bibnamefont{{Bovill}}}\ and\
  \bibinfo {author} {\bibfnamefont{M.}~\bibnamefont{{Ricotti}}},\ }%
  \bibfield{journal}{%
  \Doi{10.1088/0004-637X/693/2/1859}{\bibinfo {journal} {\apj}}\ }%
  \textbf{\bibinfo {volume} {693}},\ \bibinfo {pages} {1859} (\bibinfo {month}
  {Mar.}\ \bibinfo {year} {2009}),\
  \Eprint{http://arxiv.org/abs/0806.2340}{arXiv:0806.2340}%
  \bibAnnoteFile{NoStop}{bov2009}%
\bibitem{sie2008}%
  \BibitemOpen
  \bibfield{author}{%
  \bibinfo {author} {\bibfnamefont{M.~H.}\ \bibnamefont{{Siegel}}}, \bibinfo
  {author} {\bibfnamefont{M.~D.}\ \bibnamefont{{Shetrone}}},\ and\ \bibinfo
  {author} {\bibfnamefont{M.}~\bibnamefont{{Irwin}}},\ }%
  \bibfield{journal}{%
  \Doi{10.1088/0004-6256/135/6/2084}{\bibinfo {journal} {Astron. J.}}\ }%
  \textbf{\bibinfo {volume} {135}},\ \bibinfo {pages} {2084} (\bibinfo {month}
  {Jun.}\ \bibinfo {year} {2008}),\
  \Eprint{http://arxiv.org/abs/0803.2489}{arXiv:0803.2489}%
  \bibAnnoteFile{NoStop}{sie2008}%
\bibitem{loe2010}%
  \BibitemOpen
  \bibfield{author}{%
  \bibinfo {author} {\bibfnamefont{M.}~\bibnamefont{{Loewenstein}}}\ and\
  \bibinfo {author} {\bibfnamefont{A.}~\bibnamefont{{Kusenko}}},\ }%
  \bibfield{journal}{%
  \Doi{10.1088/0004-637X/714/1/652}{\bibinfo {journal} {\apj}}\ }%
  \textbf{\bibinfo {volume} {714}},\ \bibinfo {pages} {652} (\bibinfo {month}
  {May}\ \bibinfo {year} {2010}),\
  \Eprint{http://arxiv.org/abs/0912.0552}{arXiv:0912.0552 [astro-ph.HE]}%
  \bibAnnoteFile{NoStop}{loe2010}%
\bibitem{kus2010}%
  \BibitemOpen
  \bibfield{author}{%
  \bibinfo {author} {\bibfnamefont{A.}~\bibnamefont{{Kusenko}}}\ and\ \bibinfo
  {author} {\bibfnamefont{M.}~\bibnamefont{{Loewenstein}}},\ }%
  \bibfield{journal}{%
  \bibinfo {journal} {ArXiv e-prints}}%
   (\bibinfo {month} {Jan.}\ \bibinfo {year} {2010}),\
  \Eprint{http://arxiv.org/abs/1001.4055}{arXiv:1001.4055 [astro-ph.CO]}%
  \bibAnnoteFile{NoStop}{kus2010}%
\bibitem{boy2010}%
  \BibitemOpen
  \bibfield{author}{%
  \bibinfo {author} {\bibfnamefont{A.}~\bibnamefont{{Boyarsky}}}, \bibinfo
  {author} {\bibfnamefont{O.}~\bibnamefont{{Ruchayskiy}}}, \bibinfo {author}
  {\bibfnamefont{D.}~\bibnamefont{{Iakubovskyi}}}, \bibinfo {author}
  {\bibfnamefont{M.~G.}\ \bibnamefont{{Walker}}}, \bibinfo {author}
  {\bibfnamefont{S.}~\bibnamefont{{Riemer-S{\o}rensen}}},\ and\ \bibinfo
  {author} {\bibfnamefont{S.~H.}\ \bibnamefont{{Hansen}}},\ }%
  \bibfield{journal}{%
  \Doi{10.1111/j.1365-2966.2010.17004.x}{\bibinfo {journal} {Mon. Not. R.
  Astron. Soc.}}\ }%
  \textbf{\bibinfo {volume} {407}},\ \bibinfo {pages} {1188} (\bibinfo {month}
  {Sep.}\ \bibinfo {year} {2010}),\
  \Eprint{http://arxiv.org/abs/1001.0644}{arXiv:1001.0644 [astro-ph.CO]}%
  \bibAnnoteFile{NoStop}{boy2010}%
\bibitem{wol2009}%
  \BibitemOpen
  \bibfield{author}{%
  \bibinfo {author} {\bibfnamefont{J.}~\bibnamefont{{Wolf}}}, \bibinfo {author}
  {\bibfnamefont{G.~D.}\ \bibnamefont{{Martinez}}}, \bibinfo {author}
  {\bibfnamefont{J.~S.}\ \bibnamefont{{Bullock}}}, \bibinfo {author}
  {\bibfnamefont{M.}~\bibnamefont{{Kaplinghat}}}, \bibinfo {author}
  {\bibfnamefont{M.}~\bibnamefont{{Geha}}}, \bibinfo {author}
  {\bibfnamefont{R.~R.}\ \bibnamefont{{Munoz}}}, \bibinfo {author}
  {\bibfnamefont{J.~D.}\ \bibnamefont{{Simon}}},\ and\ \bibinfo {author}
  {\bibfnamefont{F.~F.}\ \bibnamefont{{Avedo}}},\ }%
  \bibfield{journal}{%
  \bibinfo {journal} {ArXiv e-prints}}%
   (\bibinfo {month} {Aug.}\ \bibinfo {year} {2009}),\
  \Eprint{http://arxiv.org/abs/0908.2995}{arXiv:0908.2995}%
  \bibAnnoteFile{NoStop}{wol2009}%
\bibitem{vie2006}%
  \BibitemOpen
  \bibfield{author}{%
  \bibinfo {author} {\bibfnamefont{M.}~\bibnamefont{{Viel}}}, \bibinfo {author}
  {\bibfnamefont{J.}~\bibnamefont{{Lesgourgues}}}, \bibinfo {author}
  {\bibfnamefont{M.~G.}\ \bibnamefont{{Haehnelt}}}, \bibinfo {author}
  {\bibfnamefont{S.}~\bibnamefont{{Matarrese}}},\ and\ \bibinfo {author}
  {\bibfnamefont{A.}~\bibnamefont{{Riotto}}},\ }%
  \bibfield{journal}{%
  \Doi{10.1103/PhysRevLett.97.071301}{\bibinfo {journal} {Physical Review
  Letters}}\ }%
  \textbf{\bibinfo {volume} {97}},\ \bibinfo {pages} {071301} (\bibinfo {month}
  {Aug.}\ \bibinfo {year} {2006}),\
  \Eprint{http://arxiv.org/abs/arXiv:astro-ph/0605706}{arXiv:astro-ph/0605706}%
  \bibAnnoteFile{NoStop}{vie2006}%
\bibitem{vie2008}%
  \BibitemOpen
  \bibfield{author}{%
  \bibinfo {author} {\bibfnamefont{M.}~\bibnamefont{{Viel}}}, \bibinfo {author}
  {\bibfnamefont{G.~D.}\ \bibnamefont{{Becker}}}, \bibinfo {author}
  {\bibfnamefont{J.~S.}\ \bibnamefont{{Bolton}}}, \bibinfo {author}
  {\bibfnamefont{M.~G.}\ \bibnamefont{{Haehnelt}}}, \bibinfo {author}
  {\bibfnamefont{M.}~\bibnamefont{{Rauch}}},\ and\ \bibinfo {author}
  {\bibfnamefont{W.~L.~W.}\ \bibnamefont{{Sargent}}},\ }%
  \bibfield{journal}{%
  \Doi{10.1103/PhysRevLett.100.041304}{\bibinfo {journal} {Physical Review
  Letters}}\ }%
  \textbf{\bibinfo {volume} {100}},\ \bibinfo {pages} {041304} (\bibinfo
  {month} {Feb.}\ \bibinfo {year} {2008}),\
  \Eprint{http://arxiv.org/abs/0709.0131}{arXiv:0709.0131}%
  \bibAnnoteFile{NoStop}{vie2008}%
\bibitem{mcd2006}%
  \BibitemOpen
  \bibfield{author}{%
  \bibinfo {author} {\bibfnamefont{P.}~\bibnamefont{{McDonald}}}, \bibinfo
  {author} {\bibfnamefont{U.}~\bibnamefont{{Seljak}}}, \bibinfo {author}
  {\bibfnamefont{S.}~\bibnamefont{{Burles}}}, \bibinfo {author}
  {\bibfnamefont{D.~J.}\ \bibnamefont{{Schlegel}}}, \bibinfo {author}
  {\bibfnamefont{D.~H.}\ \bibnamefont{{Weinberg}}}, \bibinfo {author}
  {\bibfnamefont{R.}~\bibnamefont{{Cen}}}, \bibinfo {author}
  {\bibfnamefont{D.}~\bibnamefont{{Shih}}}, \bibinfo {author}
  {\bibfnamefont{J.}~\bibnamefont{{Schaye}}}, \bibinfo {author}
  {\bibfnamefont{D.~P.}\ \bibnamefont{{Schneider}}}, \bibinfo {author}
  {\bibfnamefont{N.~A.}\ \bibnamefont{{Bahcall}}}, \bibinfo {author}
  {\bibfnamefont{J.~W.}\ \bibnamefont{{Briggs}}}, \bibinfo {author}
  {\bibfnamefont{J.}~\bibnamefont{{Brinkmann}}}, \bibinfo {author}
  {\bibfnamefont{R.~J.}\ \bibnamefont{{Brunner}}}, \bibinfo {author}
  {\bibfnamefont{M.}~\bibnamefont{{Fukugita}}}, \bibinfo {author}
  {\bibfnamefont{J.~E.}\ \bibnamefont{{Gunn}}}, \bibinfo {author}
  {\bibfnamefont{{\v Z}.}~\bibnamefont{{Ivezi{\'c}}}}, \bibinfo {author}
  {\bibfnamefont{S.}~\bibnamefont{{Kent}}}, \bibinfo {author}
  {\bibfnamefont{R.~H.}\ \bibnamefont{{Lupton}}},\ and\ \bibinfo {author}
  {\bibfnamefont{D.~E.}\ \bibnamefont{{Vanden Berk}}},\ }%
  \bibfield{journal}{%
  \Doi{10.1086/444361}{\bibinfo {journal} {Astrophys. J. Suppl.}}\ }%
  \textbf{\bibinfo {volume} {163}},\ \bibinfo {pages} {80} (\bibinfo {month}
  {Mar.}\ \bibinfo {year} {2006}),\
  \Eprint{http://arxiv.org/abs/arXiv:astro-ph/0405013}{arXiv:astro-ph/0405013}%
  \bibAnnoteFile{NoStop}{mcd2006}%
\bibitem{sel2006}%
  \BibitemOpen
  \bibfield{author}{%
  \bibinfo {author} {\bibfnamefont{U.}~\bibnamefont{{Seljak}}}, \bibinfo
  {author} {\bibfnamefont{A.}~\bibnamefont{{Makarov}}}, \bibinfo {author}
  {\bibfnamefont{P.}~\bibnamefont{{McDonald}}},\ and\ \bibinfo {author}
  {\bibfnamefont{H.}~\bibnamefont{{Trac}}},\ }%
  \bibfield{journal}{%
  \Doi{10.1103/PhysRevLett.97.191303}{\bibinfo {journal} {Physical Review
  Letters}}\ }%
  \textbf{\bibinfo {volume} {97}},\ \bibinfo {pages} {191303} (\bibinfo {month}
  {Nov.}\ \bibinfo {year} {2006}),\
  \Eprint{http://arxiv.org/abs/arXiv:astro-ph/0602430}{arXiv:astro-ph/0602430}%
  \bibAnnoteFile{NoStop}{sel2006}%
\bibitem{vie2009}%
  \BibitemOpen
  \bibfield{author}{%
  \bibinfo {author} {\bibfnamefont{M.}~\bibnamefont{{Viel}}}, \bibinfo {author}
  {\bibfnamefont{J.~S.}\ \bibnamefont{{Bolton}}},\ and\ \bibinfo {author}
  {\bibfnamefont{M.~G.}\ \bibnamefont{{Haehnelt}}},\ }%
  \bibfield{journal}{%
  \Doi{10.1111/j.1745-3933.2009.00720.x}{\bibinfo {journal} {Mon. Not. R.
  Astron. Soc.}}\ }%
  \textbf{\bibinfo {volume} {399}},\ \bibinfo {pages} {L39} (\bibinfo {month}
  {Oct.}\ \bibinfo {year} {2009}),\
  \Eprint{http://arxiv.org/abs/0907.2927}{arXiv:0907.2927}%
  \bibAnnoteFile{NoStop}{vie2009}%
\bibitem{ric2000}%
  \BibitemOpen
  \bibfield{author}{%
  \bibinfo {author} {\bibfnamefont{M.}~\bibnamefont{{Ricotti}}}, \bibinfo
  {author} {\bibfnamefont{N.~Y.}\ \bibnamefont{{Gnedin}}},\ and\ \bibinfo
  {author} {\bibfnamefont{J.~M.}\ \bibnamefont{{Shull}}},\ }%
  \bibfield{journal}{%
  \Doi{10.1086/308733}{\bibinfo {journal} {\apj}}\ }%
  \textbf{\bibinfo {volume} {534}},\ \bibinfo {pages} {41} (\bibinfo {month}
  {May}\ \bibinfo {year} {2000}),\
  \Eprint{http://arxiv.org/abs/arXiv:astro-ph/9906413}{arXiv:astro-ph/9906413}%
  \bibAnnoteFile{NoStop}{ric2000}%
\bibitem{sch2000}%
  \BibitemOpen
  \bibfield{author}{%
  \bibinfo {author} {\bibfnamefont{J.}~\bibnamefont{{Schaye}}}, \bibinfo
  {author} {\bibfnamefont{T.}~\bibnamefont{{Theuns}}}, \bibinfo {author}
  {\bibfnamefont{M.}~\bibnamefont{{Rauch}}}, \bibinfo {author}
  {\bibfnamefont{G.}~\bibnamefont{{Efstathiou}}},\ and\ \bibinfo {author}
  {\bibfnamefont{W.~L.~W.}\ \bibnamefont{{Sargent}}},\ }%
  \bibfield{journal}{%
  \Doi{10.1046/j.1365-8711.2000.03815.x}{\bibinfo {journal} {Mon. Not. R.
  Astron. Soc.}}\ }%
  \textbf{\bibinfo {volume} {318}},\ \bibinfo {pages} {817} (\bibinfo {month}
  {Nov.}\ \bibinfo {year} {2000}),\
  \Eprint{http://arxiv.org/abs/arXiv:astro-ph/9912432}{arXiv:astro-ph/9912432}%
  \bibAnnoteFile{NoStop}{sch2000}%
\bibitem{boy2006A}%
  \BibitemOpen
  \bibfield{author}{%
  \bibinfo {author} {\bibfnamefont{A.}~\bibnamefont{{Boyarsky}}}, \bibinfo
  {author} {\bibfnamefont{A.}~\bibnamefont{{Neronov}}}, \bibinfo {author}
  {\bibfnamefont{O.}~\bibnamefont{{Ruchayskiy}}},\ and\ \bibinfo {author}
  {\bibfnamefont{M.}~\bibnamefont{{Shaposhnikov}}},\ }%
  \bibfield{journal}{%
  \Doi{10.1111/j.1365-2966.2006.10458.x}{\bibinfo {journal} {Mon. Not. R.
  Astron. Soc.}}\ }%
  \textbf{\bibinfo {volume} {370}},\ \bibinfo {pages} {213} (\bibinfo {month}
  {Jul.}\ \bibinfo {year} {2006}),\
  \Eprint{http://arxiv.org/abs/arXiv:astro-ph/0512509}{arXiv:astro-ph/0512509}%
  \bibAnnoteFile{NoStop}{boy2006A}%
\bibitem{aba2006A}%
  \BibitemOpen
  \bibfield{author}{%
  \bibinfo {author} {\bibfnamefont{K.}~\bibnamefont{{Abazajian}}}\ and\
  \bibinfo {author} {\bibfnamefont{S.~M.}\ \bibnamefont{{Koushiappas}}},\ }%
  \bibfield{journal}{%
  \Doi{10.1103/PhysRevD.74.023527}{\bibinfo {journal} {\prd}}\ }%
  \textbf{\bibinfo {volume} {74}},\ \bibinfo {pages} {023527} (\bibinfo {month}
  {Jul.}\ \bibinfo {year} {2006}),\
  \Eprint{http://arxiv.org/abs/arXiv:astro-ph/0605271}{arXiv:astro-ph/0605271}%
  \bibAnnoteFile{NoStop}{aba2006A}%
\bibitem{boy2006}%
  \BibitemOpen
  \bibfield{author}{%
  \bibinfo {author} {\bibfnamefont{A.}~\bibnamefont{{Boyarsky}}}, \bibinfo
  {author} {\bibfnamefont{A.}~\bibnamefont{{Neronov}}}, \bibinfo {author}
  {\bibfnamefont{O.}~\bibnamefont{{Ruchayskiy}}},\ and\ \bibinfo {author}
  {\bibfnamefont{M.}~\bibnamefont{{Shaposhnikov}}},\ }%
  \bibfield{journal}{%
  \Doi{10.1103/PhysRevD.74.103506}{\bibinfo {journal} {\prd}}\ }%
  \textbf{\bibinfo {volume} {74}},\ \bibinfo {pages} {103506} (\bibinfo {month}
  {Nov.}\ \bibinfo {year} {2006}),\
  \Eprint{http://arxiv.org/abs/arXiv:astro-ph/0603368}{arXiv:astro-ph/0603368}%
  \bibAnnoteFile{NoStop}{boy2006}%
\bibitem{rie2007}%
  \BibitemOpen
  \bibfield{author}{%
  \bibinfo {author} {\bibfnamefont{S.}~\bibnamefont{{Riemer-Sorensen}}},
  \bibinfo {author} {\bibfnamefont{K.}~\bibnamefont{{Pedersen}}}, \bibinfo
  {author} {\bibfnamefont{S.~H.}\ \bibnamefont{{Hansen}}},\ and\ \bibinfo
  {author} {\bibfnamefont{H.}~\bibnamefont{{Dahle}}},\ }%
  \bibfield{journal}{%
  \Doi{10.1103/PhysRevD.76.043524}{\bibinfo {journal} {\prd}}\ }%
  \textbf{\bibinfo {volume} {76}},\ \bibinfo {pages} {043524} (\bibinfo {month}
  {Aug.}\ \bibinfo {year} {2007}),\
  \Eprint{http://arxiv.org/abs/arXiv:astro-ph/0610034}{arXiv:astro-ph/0610034}%
  \bibAnnoteFile{NoStop}{rie2007}%
\bibitem{boy2008}%
  \BibitemOpen
  \bibfield{author}{%
  \bibinfo {author} {\bibfnamefont{A.}~\bibnamefont{{Boyarsky}}}, \bibinfo
  {author} {\bibfnamefont{O.}~\bibnamefont{{Ruchayskiy}}},\ and\ \bibinfo
  {author} {\bibfnamefont{M.}~\bibnamefont{{Markevitch}}},\ }%
  \bibfield{journal}{%
  \Doi{10.1086/524397}{\bibinfo {journal} {\apj}}\ }%
  \textbf{\bibinfo {volume} {673}},\ \bibinfo {pages} {752} (\bibinfo {month}
  {Feb.}\ \bibinfo {year} {2008}),\
  \Eprint{http://arxiv.org/abs/arXiv:astro-ph/0611168}{arXiv:astro-ph/0611168}%
  \bibAnnoteFile{NoStop}{boy2008}%
\bibitem{wat2006}%
  \BibitemOpen
  \bibfield{author}{%
  \bibinfo {author} {\bibfnamefont{C.~R.}\ \bibnamefont{{Watson}}}, \bibinfo
  {author} {\bibfnamefont{J.~F.}\ \bibnamefont{{Beacom}}}, \bibinfo {author}
  {\bibfnamefont{H.}~\bibnamefont{{Y{\"u}ksel}}},\ and\ \bibinfo {author}
  {\bibfnamefont{T.~P.}\ \bibnamefont{{Walker}}},\ }%
  \bibfield{journal}{%
  \Doi{10.1103/PhysRevD.74.033009}{\bibinfo {journal} {\prd}}\ }%
  \textbf{\bibinfo {volume} {74}},\ \bibinfo {pages} {033009} (\bibinfo {month}
  {Aug.}\ \bibinfo {year} {2006}),\
  \Eprint{http://arxiv.org/abs/arXiv:astro-ph/0605424}{arXiv:astro-ph/0605424}%
  \bibAnnoteFile{NoStop}{wat2006}%
\bibitem{boy2007}%
  \BibitemOpen
  \bibfield{author}{%
  \bibinfo {author} {\bibfnamefont{A.}~\bibnamefont{{Boyarsky}}}, \bibinfo
  {author} {\bibfnamefont{J.}~\bibnamefont{{Nevalainen}}},\ and\ \bibinfo
  {author} {\bibfnamefont{O.}~\bibnamefont{{Ruchayskiy}}},\ }%
  \bibfield{journal}{%
  \Doi{10.1051/0004-6361:20066774}{\bibinfo {journal} {A\&A}}\ }%
  \textbf{\bibinfo {volume} {471}},\ \bibinfo {pages} {51} (\bibinfo {month}
  {Aug.}\ \bibinfo {year} {2007}),\
  \Eprint{http://arxiv.org/abs/arXiv:astro-ph/0610961}{arXiv:astro-ph/0610961}%
  \bibAnnoteFile{NoStop}{boy2007}%
\bibitem{rie2009}%
  \BibitemOpen
  \bibfield{author}{%
  \bibinfo {author} {\bibfnamefont{S.}~\bibnamefont{{Riemer-S{\o}rensen}}}\
  and\ \bibinfo {author} {\bibfnamefont{S.~H.}\ \bibnamefont{{Hansen}}},\ }%
  \bibfield{journal}{%
  \Doi{10.1051/0004-6361/200912430}{\bibinfo {journal} {A\&A}}\ }%
  \textbf{\bibinfo {volume} {500}},\ \bibinfo {pages} {L37} (\bibinfo {month}
  {Jun.}\ \bibinfo {year} {2009})%
  \bibAnnoteFile{NoStop}{rie2009}%
\bibitem{boy2009}%
  \BibitemOpen
  \bibfield{author}{%
  \bibinfo {author} {\bibfnamefont{A.}~\bibnamefont{{Boyarsky}}}, \bibinfo
  {author} {\bibfnamefont{J.}~\bibnamefont{{Lesgourgues}}}, \bibinfo {author}
  {\bibfnamefont{O.}~\bibnamefont{{Ruchayskiy}}},\ and\ \bibinfo {author}
  {\bibfnamefont{M.}~\bibnamefont{{Viel}}},\ }%
  \bibfield{journal}{%
  \Doi{10.1103/PhysRevLett.102.201304}{\bibinfo {journal} {Physical Review
  Letters}}\ }%
  \textbf{\bibinfo {volume} {102}},\ \bibinfo {pages} {201304} (\bibinfo
  {month} {May}\ \bibinfo {year} {2009}),\
  \Eprint{http://arxiv.org/abs/0812.3256}{arXiv:0812.3256 [hep-ph]}%
  \bibAnnoteFile{NoStop}{boy2009}%
\bibitem{loe2009}%
  \BibitemOpen
  \bibfield{author}{%
  \bibinfo {author} {\bibfnamefont{M.}~\bibnamefont{{Loewenstein}}}, \bibinfo
  {author} {\bibfnamefont{A.}~\bibnamefont{{Kusenko}}},\ and\ \bibinfo {author}
  {\bibfnamefont{P.~L.}\ \bibnamefont{{Biermann}}},\ }%
  \bibfield{journal}{%
  \Doi{10.1088/0004-637X/700/1/426}{\bibinfo {journal} {\apj}}\ }%
  \textbf{\bibinfo {volume} {700}},\ \bibinfo {pages} {426} (\bibinfo {month}
  {Jul.}\ \bibinfo {year} {2009}),\
  \Eprint{http://arxiv.org/abs/0812.2710}{arXiv:0812.2710}%
  \bibAnnoteFile{NoStop}{loe2009}%
\bibitem{rie2006}%
  \BibitemOpen
  \bibfield{author}{%
  \bibinfo {author} {\bibfnamefont{S.}~\bibnamefont{{Riemer-S{\o}rensen}}},
  \bibinfo {author} {\bibfnamefont{S.~H.}\ \bibnamefont{{Hansen}}},\ and\
  \bibinfo {author} {\bibfnamefont{K.}~\bibnamefont{{Pedersen}}},\ }%
  \bibfield{journal}{%
  \Doi{10.1086/505330}{\bibinfo {journal} {Astrophys. J. Lett.}}\ }%
  \textbf{\bibinfo {volume} {644}},\ \bibinfo {pages} {L33} (\bibinfo {month}
  {Jun.}\ \bibinfo {year} {2006}),\
  \Eprint{http://arxiv.org/abs/arXiv:astro-ph/0603661}{arXiv:astro-ph/0603661}%
  \bibAnnoteFile{NoStop}{rie2006}%
\bibitem{aba2007}%
  \BibitemOpen
  \bibfield{author}{%
  \bibinfo {author} {\bibfnamefont{K.~N.}\ \bibnamefont{{Abazajian}}}, \bibinfo
  {author} {\bibfnamefont{M.}~\bibnamefont{{Markevitch}}}, \bibinfo {author}
  {\bibfnamefont{S.~M.}\ \bibnamefont{{Koushiappas}}},\ and\ \bibinfo {author}
  {\bibfnamefont{R.~C.}\ \bibnamefont{{Hickox}}},\ }%
  \bibfield{journal}{%
  \Doi{10.1103/PhysRevD.75.063511}{\bibinfo {journal} {\prd}}\ }%
  \textbf{\bibinfo {volume} {75}},\ \bibinfo {pages} {063511} (\bibinfo {month}
  {Mar.}\ \bibinfo {year} {2007}),\
  \Eprint{http://arxiv.org/abs/arXiv:astro-ph/0611144}{arXiv:astro-ph/0611144}%
  \bibAnnoteFile{NoStop}{aba2007}%
\bibitem{asa2006}%
  \BibitemOpen
  \bibfield{author}{%
  \bibinfo {author} {\bibfnamefont{T.}~\bibnamefont{{Asaka}}}, \bibinfo
  {author} {\bibfnamefont{M.}~\bibnamefont{{Shaposhnikov}}},\ and\ \bibinfo
  {author} {\bibfnamefont{A.}~\bibnamefont{{Kusenko}}},\ }%
  \bibfield{journal}{%
  \Doi{10.1016/j.physletb.2006.05.067}{\bibinfo {journal} {Physics Letters B}}\
  }%
  \textbf{\bibinfo {volume} {638}},\ \bibinfo {pages} {401} (\bibinfo {month}
  {Jul.}\ \bibinfo {year} {2006}),\
  \Eprint{http://arxiv.org/abs/arXiv:hep-ph/0602150}{arXiv:hep-ph/0602150}%
  \bibAnnoteFile{NoStop}{asa2006}%
\end{thebibliography}%

\appendix*
\section{}

\begin{figure*}[!hbt]
\includegraphics*[scale=0.35,angle=270]{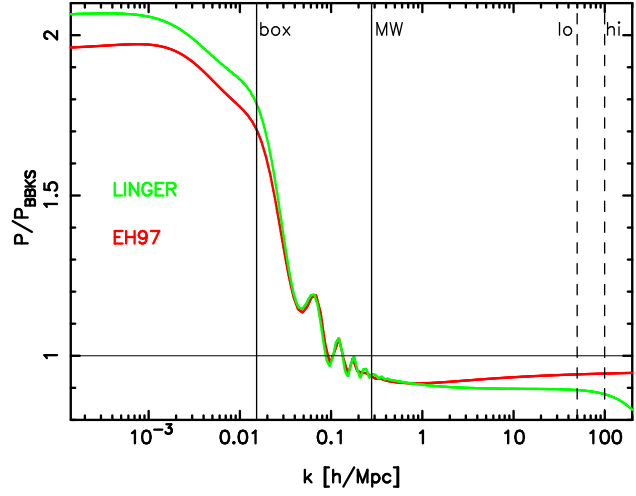}
\caption{Comparison of CDM power spectra calculated from the fitting
  formula of Eisenstein and Hu (EH97) and from the LINGER software normalized by BBKS. 
  On scales $k>0.1$~$h$/Mpc ($<
  14$~Mpc) the power spectra are nearly identical. The `MW' vertical line
  is the diameter of a spherical region with density $\Omega_m
  \rho_{c}$ enclosing a Milky Way-sized mass $2 \times
  10^{12} M_{\odot}$ ($\sim 5$~Mpc). This scale is well within the
  range where the power spectra are nearly equal.\label{figPScomp}}
\end{figure*}
We used the BBKS formula for the CDM transfer function when generating
the initial conditions for our high resolution simulations. This formula assumes a
baryon density of zero. 
Eisenstein and Hu~\cite{eis1997} calculated
transfer functions for CDM cosmologies that include baryon physics. 
We plot the power spectra for the fitting formula of Eisenstein and Hu and the
spectrum calculated with the LINGER software
(using $\Omega_b=0.04$) normalized by BBKS in Figure~\ref{figPScomp}.
With $\Omega_m=0.238$ a Milky Way-sized halo
with mass $\sim 2 \times 10^{12} M_{\odot}$ would form from a
spherical region with diameter $4.8$~Mpc ($k = 0.28$~$h$/Mpc); this is plotted along with the 
scale of the simulation box ($90$~Mpc) as solid vertical lines. Dashed vertical lines
show the cell size in the refinement region of the low and high resolution simulations.
Figure~\ref{figPScomp} shows
that, for a fixed value of $\sigma_8$, BBKS underestimates power on
scales $k \lesssim 0.1$ but the power spectra are nearly identical for
scales $\lesssim 14$~Mpc with BBKS slightly power overabundant by $\sim10\%$. 
The \textit{set C} halos showed subhalo abundance variations much greater than $10\%$
and the BBKS power overabundance is much less than the $30\%$ ($1\sigma$) intrinsic scatter in subhalo abundance for MW-sized halos that we adopt.

\begin{figure*}[!hbt]
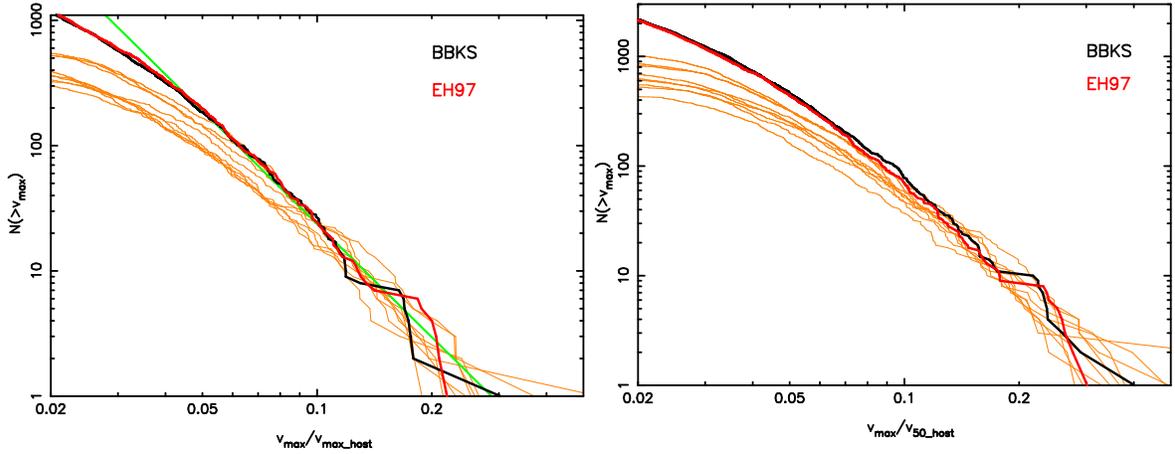

\includegraphics*[scale=0.32,angle=270]{fig13a.eps}
\includegraphics*[scale=0.32,angle=270]{fig13b.eps}
\caption{Subhalo velocity function 
  comparison for CDM high resolution {\it set B} simulations using
  fitting formula from BBKS and Eisenstein and Hu (\textit{thick lines}) and the 
  LINGER using \textit{set C} simulations (\textit{thin lines}). 
  (\textit{left}) Subhalos within $R_{100}$ and velocities normalized by $v_{max}$ of their host. The straight sloped line is the fitting formula from the Bolshoi simulation. (\textit{right}) Subhalos within $R_{50}$, normalized by $v_{50}$ of their host. The subhalo abundances between the BBKS and EH97 simulations are in good agreement and within the scatter of the \textit{set C} halos.\label{figPScomp2}}
\end{figure*}
To check if BBKS might affect the number of satellites, we reran the
{\it set B} high resolution CDM
using initial conditions generated from the
formula of Eisenstein and Hu. The panels of Figure~\ref{figPScomp2} compare the velocity
functions of satellites and show good agreement between the simulations and within the scatter of the \textit{set C} simulations. Based on this 
and the agreement between the BBKS and \textit{set C} simulations seen in Sec.~\ref{sec:3}, we
conclude that the use of BBKS has not introduced a systematic error
into our results.

\end{document}